\documentclass[aps,prb,twocolumn,superscriptaddress,nofootinbib]{revtex4-2}
\usepackage{amssymb}
\usepackage{amsmath}
\usepackage{mathrsfs}
\usepackage{mathtools}
\mathtoolsset{showonlyrefs=true}
\usepackage{bm}
\usepackage{physics}
\usepackage{enumitem}
\usepackage[svgnames]{xcolor}
\usepackage{float}
\usepackage{tikz}
\usepackage{graphicx}
\usepackage[
    setpagesize=false,
    bookmarks=false,
    colorlinks=true,
    allcolors=magenta
]{hyperref} 
\usepackage[titletoc]{appendix}

\begin{document}

\title{Gate Operations for Superconducting Qubits and Non-Markovianity}
\author{Kiyoto Nakamura}
\email{kiyoto.nakamura@uni-ulm.de}
\author{Joachim Ankerhold}
\affiliation{Institute for Complex Quantum Systems and IQST, Ulm University, D-89069 Ulm, Germany}

\date{\today}

\begin{abstract}
    While the accuracy of  qubit operations has been greatly improved in the last decade, further development is demanded to achieve the ultimate goal: a fault-tolerant quantum computer that can solve real-world problems more efficiently than classical computers. With growing fidelities even subtle effects of environmental noise such as qubit--reservoir correlations and non-Markovian dynamics turn into the focus for both circuit design and control. To guide progress, we disclose, in a numerically rigorous manner, a comprehensive picture of the single-qubit dynamics in  presence of a broad class of noise sources and for entire sequences of gate operations.
    Thermal reservoirs ranging from Ohmic to deep  $1/f^{\varepsilon}$-like sub-Ohmic behavior are considered to imitate realistic scenarios for superconducting qubits. 
    Apart from dynamical features, fidelities of the qubit performance over entire sequences are analyzed as a figure of merit.
    The relevance of retarded feedback and long-range qubit--reservoir correlations is demonstrated on a quantitative level, thus, providing a deeper understanding of the limitations of performances for current devices and guiding the design of future ones.
\end{abstract}

\maketitle
\section{Introduction}
The last decade has witnessed impressive progress in developing quantum computing platforms, in particular based on superconducting circuits: Coherence times~\cite{PlaceNATCOMMUN2021,WangNPJQI2022} as well as gate fidelities~\cite{NegirneacPRL2021,SungPRX2021,KandalaPRL2021} have been substantially increased,  and multiqubit architectures have demonstrated quantum supremacy under specific conditions \cite{GoogleNATURE2019, GoogleNATURE2023, IBMNATURE2023}. This led to the first implementations of quantum algorithms for noisy intermediate-scale quantum (NISQ) devices~\cite{HavlicekNATURE2019,KandalaNATURE2017,RossmannekJPCL2023,WillschQIP2020,PerezNPJQI2020,FarrellPRD2024}. 

It has become clear that further progress requires a much better \emph{quantitative}  description of qubit operations in the presence of relevant noise sources. Indeed, with progressively increasing coherence times and fidelities, even subtle details of environmental effects, not seen in the previous generation of devices, now turn into the focus. This applies specifically to quantum correlations between individual qubits and thermal reservoirs and retardation effects in time induced by quantum fluctuations (non-Markovianity).
It was pointed out that a detailed understanding of these effects is crucial to fully exploit error correction~\cite{DevittRPP2013} and error mitigation~\cite{CaiRMP2023} because those techniques highly depend on elusive properties of noise~\cite{GulasciPRB2022,PapicARXIV2023}.
In addition, identification of the origin of noise-induced errors (bit-flip and phase errors) through the analysis of the qubit dynamics during sequences of gate operations may trigger optimized pulse shapes, protocols, and circuit designs. For this purpose, single-qubit devices may themselves function as ultrasensitive probes, for example, to monitor the emergence of quasiparticle noise in superconducting circuits \cite{RisteNATCOMMUN2013,CardaniNATCOMMUN2021,PanNATCOMMUN2022}.

An immediate consequence is that the development of the next generation of qubit devices with the required high fidelities has to go hand in hand with highly accurate numerical simulations.
Conventionally adopted methods, including the Redfield equation and Lindblad equation, provide only qualitative  results in this context (and in many cases not even this), and it seems that those treatments should not be used in order to contribute to the required improvements in the near future.
In fact, studies have already been conducted to go beyond the imposed Born--Markov approximation and to account for higher order quantum correlations.
Recent examples include studies of the population of steady states~\cite{TuorilaPRR2019}, leakage to higher-excited states during pulse applications~\cite{BabuNPJQI2021}, experimental protocols that can detect non-Markovianity~\cite{GulasciPRB2022,NakamuraPRB2024}, origins of noise~\cite{PapicARXIV2023}, accuracy of error correction codes~\cite{BabuPRR2023}, and two-spin systems that mimic a spin bath to account for both non-Markovian and non-Gaussian effects~\cite{RomeroQST2024,AgarwalARXIV2023}.
Most studies are limited to a \emph{single} pulse application or a \emph{single} free evolution (idle phase) though. It seems intuitive and has also been suggested in previous paper \cite{TuorilaPRR2019} that on the timescale of a single-gate pulse, higher-order reservoir-induced quantum effects are less relevant. However, this picture is expected to drastically change when entire sequences consisting of several subsequent gate pulses interleaved by idle phases are considered: Time-retardation effects may then correlate the qubit dynamics between different segments so that its dynamics at a certain time is affected by the entire past of the compound. 
\textcolor{black}{While memory effects of the reservoir after long periods of time have been investigated in previous studies~\cite{LorenzoPRA2011,LainePRL2012,FanchiniPRL2014,SiltanenPRA2021}, those effects for time-dependent system Hamiltonians including the switching-on and off of driving fields remain unclear.
Especially, quantitative predictions of the memory effects in the parameter domain in which the qubit systems are operated are desired for further improvement, as mentioned above.}

Studies in this direction and based on rigorous methods have not been conducted so far, mainly because of highly nontrivial conceptual and technical problems.
Conceptually, one has to accurately follow the quantum time evolution of an open quantum system in the presence of complex external driving over relatively long timescales. This requires nonperturbative techniques, which, and this is the technical challenge, are sufficiently efficient, reliable, and allow for versatile applicability.  

Here, we attack these issues with the aid of a recent extension of the hierarchical equations of motion (HEOM) method \cite{XuPRL2022}. This method maps the formally exact Feynman--Vernon path integral expression for the reduced density operator of the qubit system onto a nested hierarchy of equations of motions. With its recent extension (free-pole HEOM, i.e., FP-HEOM) it is now possible to simulate open quantum systems for almost arbitrary reservoir spectral densities and over the entire temperature range down to zero temperatures. In addition, the FP-HEOM is so efficient that it allows us to accurately monitor the long-time behavior as well. In the sequel, we exploit this technique to reveal a comprehensive and quantitative precise picture of the noisy quantum dynamics of nontrivial single-qubit operations, thus establishing the methodology as a standard tool to guide further developments also for multiqubit structures.  

More specifically, an extensive analysis is provided, which comprises
a broad class of thermal reservoirs relevant for superconducting qubits from reservoirs with Ohmic characteristics to those with deep sub-Ohmic behavior (relatively large portion of low-frequency modes). Three different gate operations with varying amplitudes and pulse durations are considered according to sequences depicted in Fig.~\ref{fig:pulseSequence}.
We reveal intricate correlations between the dynamics during pulse applications and idle phases. In this way, and in combination with varying rotation angles and rotation axes, bath-induced qubit errors are quantitatively investigated in terms of the fidelity. By ``numerically'' factorizing the total system into a system and a reservoir at certain times during a pulse sequence and by comparing the corresponding dynamics with the exact ones, we reveal interphase correlations caused by non-Markovianity. This opens ways to detect these subtle effects in actual experiments.


This paper is organized as follows. In Sec.~\ref{sec:model}, we introduce a model Hamiltonian for single-qubit dynamics and explain how rotation operators are expressed with a time-dependent Hamiltonian. 
An exact time-evolution method, HEOM, is also introduced.
In Sec.~\ref{sec:Parameters}, we discuss quantities that characterize the reservoir and relations between those quantities and noise models proposed in previous studies.
Sections~\ref{sec:resSequence} and \ref{sec:resCorrelation} are devoted to the numerical results:
In Sec.~\ref{sec:resSequence}, we study detrimental effects induced by non-Markovian dynamics of reservoirs in terms of the fidelity between an ideal state and numerically obtained one.
Dynamics of a single qubit subject to a sequence of gate operations are considered there.
In Sec.~\ref{sec:resCorrelation}, we focus on the non-Markovianity of the reservoir.
Correlations between a pulse-application phase and an idle phase and between two idle phases interleaved with an impulsive pulse are investigated.
We summarize the paper and draw conclusions in Sec.~\ref{sec:conclusion}.

\section{Model and methods} \label{sec:model}
\subsection{Open qubit dynamics and rotation operators}
In this paper, we consider a single qubit (two-level system) and its manipulation by external time-dependent pulses described by
\begin{align}
    & \hat{H}_S(\Omega, \phi; t)  \\
    = & \frac{\hbar \omega_q}{2} \hat{\sigma}_z
    + \frac{\hbar\Omega}{2} \left[\hat{\sigma}_x
    \cos\left(\omega_\mathrm{ex} t + \phi\right)
    + \hat{\sigma}_y \sin\left(\omega_\mathrm{ex} t + \phi\right)\right]\, ,
    \\
    \label{eq:H_S}
\end{align}
where $\hat{\sigma}_\alpha\, (\alpha \in \{x, y, z\})$ are the Pauli matrices, and $\omega_q$ is the qubit frequency. The second term corresponds to the external field that rotates the qubit with the amplitude $\Omega$, angular frequency $\omega_\mathrm{ex}$, and static phase $\phi$. Note that this Hamiltonian has also been derived for the pulse application on the basis of the input--output theory~\cite{BlaisRMP2021}.
\begin{figure}[h]
    \centering
    \includegraphics[width=0.9\linewidth]{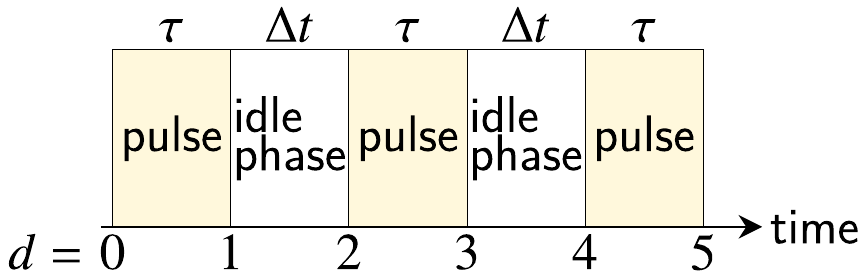}
    \caption{Schematic of the pulse sequence considered in this paper. The label $d$ is introduced to indicate the end of each phase ($d = 1, \ldots, 5$) as well as the initial time ($d = 0$).
    \label{fig:pulseSequence}}
\end{figure}

In order to set the stage for rotations in the presence of environmental degrees of freedom, let us first briefly recall the bare situation.
With the Bloch vector of a single qubit  
\begin{align}
    \ev*{\hat{\sigma}_\alpha(t)} 
    = \mathrm{tr}\{\hat{\sigma}_\alpha \hat{\rho}_S(t)\}\, ,
    \  \alpha \in \{x, y, z\}\, ,  \label{eq:Bloch}
\end{align}
the qubit's density operator can be written as $\hat{\rho}_S(t) = (\hat{1}+\sum_\alpha \ev*{\hat{\sigma}_\alpha(t)} \hat{\sigma}_\alpha)/2$ with $\hat{1}$ being the identity operator. Thus, a general rotation of the system on the Bloch sphere by an angle $\theta$ around an $\alpha$ axis is given by $\hat{R}_\alpha(\theta) = \exp[-i \theta \hat{\sigma}_\alpha/2]$. In particular, 
the time dependence in $\hat{H}_S(\Omega, \phi; t)$ can be \emph{exactly} gauged away via a unitary transformation $\hat{R}_z(-\omega_\mathrm{ex}t)$ to read
\begin{align}
    \tilde{H}_S(\Omega, \phi) 
    = \frac{\hbar(\omega_q-\omega_\mathrm{ex})}{2}\hat{\sigma}_z
    + \frac{\hbar\Omega}{2}\left(\hat{\sigma}_x \cos \phi + \hat{\sigma}_y \sin \phi\right)\, .
\end{align}
Accordingly, setting $\omega_q=\omega_\mathrm{ex}$ implies (cf.\ Appendix~\ref{sec:appRotatingFrame}, Figs.~\ref{fig:pulseSequence} and \ref{fig:Bloch}) that, in the rotating frame, rotations $\hat{R}_x(\theta)$ are generated by $\tilde{H}(\Omega, 0)$ and rotations $\hat{R}_y(\theta)$ by $\tilde{H}(\Omega, \pi/2)$.
Using the back-transformation 
$\hat{\rho}_S(t)=\hat{R}_z(\omega_q t)\,  \tilde{\rho}_S(t)\, \hat{R}_z(-\omega_q t)$ with the rotating-frame density operator $\tilde{\rho}_S(t)$, one verifies that in the laboratory frame rotation operations on the qubit correspond one-to-one to the time evolution generated by the Hamiltonian [Eq.~\eqref{eq:H_S}]  with a certain choice of parameter values $\Omega$ and $\phi$. The rotation angle $\theta$ is set by $\theta=\Omega\tau$ with the pulse duration $\tau$. In practice,  one fixes $\theta$ and $\Omega$ to adjust $\tau$ accordingly.


The common modeling of qubit systems interacting with reservoirs is formulated in the context of open quantum systems. It starts from a system+reservoir Hamiltonian $\hat{H}_\mathrm{tot}(\Omega, \phi; t)=\hat{H}_S(\Omega, \phi; t)- \hat{V} \hat{X}+ \hat{H}_R$, where for the sake of simplicity we assume a bilinear coupling between the qubit system with coupling operator $\hat{V}$ and a reservoir $\hat{H}_R$ with $\hat{X}$ \cite{Breuer2002,Weiss2012}.
\textcolor{black}{Note that this assumption is not a severe constraint, provided that systems subject to Gaussian noise with arbitrary intensity are exactly described with this model, which is true for most of the qubit systems.}
In the sequel, we consider bit-flip errors and adopt the same form for $\hat{V}$ as in a previous study~\cite{TuorilaPRR2019}, which is described as $\hat{V} = \hbar\hat{\sigma}_x$. 
Note that a different form, $\hat{V} = \hbar\hat{\sigma}_y$, is also proposed in previous studies~\cite{GulasciPRB2022,BlaisRMP2021}.

Now, during the time evolution of the total system in the presence of pulse sequences on the qubit part, the external field with a nonzero amplitude of $\Omega$ rotates the qubit (gate operation) and the time evolution without the external field ($\Omega=0$) between two of these pulse operations, see Fig.~\ref{fig:pulseSequence}, is referred to as the ``idle phase''. It must be taken into account, for example, considering that there must be synchronizations of qubits during a given multiqubit protocol. Of course, the state of the isolated qubit in the rotating frame does not change during idle phases. However, for a qubit interacting with reservoirs, decoherence sets in and correlations between qubits and reservoirs evolve such that they are expected to influence the next gate operation. Physically, these latter effects originate from the retarded feedback of the reservoir onto the qubit system, which always occurs at sufficiently low temperatures and induces time nonlocality in the qubit dynamics. Below, we will analyze these effects in more detail. In summary, we vary the amplitude $\Omega$ during a pulse sequence as 
\begin{align}
    \left\{
    \begin{aligned}
        \Omega \neq & 0 &&
        \mbox{(a fixed value during a pulse operation)} \\
        \Omega = & 0 && \mbox{(during idle phases)}
    \end{aligned} \right. \, ,
\end{align}
with the coupling between the system and reservoir always taken into account. For the sake of simplicity, we model the switching-on and off of the external field as a step function; improvements can be achieved by taking into account the rise time~\cite{TuorilaPRR2019,BabuNPJQI2021}.

In order to describe the open dynamics of the qubit during pulse applications of the length $\tau$, we have to take 
\begin{align}
   \hat{U}(\Omega, \phi; t, \tau)= {\mathcal T}_+\exp[-\frac{i}{\hbar} \int_{t}^{t+\tau} \hspace{-1em}dt' \hat{H}_\mathrm{tot}(\Omega, \phi; t')]
   \label{eq:timeEvo}
\end{align}
as the rotation operators instead of the bare system generator Eq.~\eqref{eq:H_S}.
Here, $\mathcal{T}_+$ is the positive time-ordering operator. The time evolution of the total density operator is then expressed as 
\begin{align}
    \hat{\rho}_\mathrm{tot}(\tau+t)
    = &\mathcal{U}_\mathrm{p}(\theta, \phi)
    \hat{\rho}_\mathrm{tot}(t)  \\
    = & \hat{U}(\Omega> 0, \phi; t, \tau)\,  \hat{\rho}_\mathrm{tot}(t)\, \hat{U}^\dagger(\Omega> 0, \phi; t, \tau)\\
    \label{eq:Upulse}
\end{align}
and the reduced density operator of the qubit follows by taking the partial trace over the environmental degrees of freedom, i.e., as $\hat{\rho}_S(t)=\mathrm{tr}_R\{\hat{\rho}_\mathrm{tot}(t)\}$.
Above, we have introduced the superoperator for the pulse application $\mathcal{U}_\mathrm{p}(\theta, \phi)$ with the relation $\theta = \Omega \tau$.
For the idle phase, we define 
\begin{align}
    \hat{\rho}_\mathrm{tot}(\Delta t + t) 
    = & \mathcal{U}_\mathrm{i}(\Delta t) 
    \hat{\rho}_\mathrm{tot}(t) \\
    = & \hat{U}(\Omega\!=\!0, \phi; t, \Delta t) \hat{\rho}_\mathrm{tot}(t)
    \hat{U}^\dagger(\Omega\!=\!0, \phi; t, \Delta t)\, .
    \label{eq:idle-pulse}
    \noeqref{eq:idle-pulse}
\end{align}
Because the pulse amplitude is zero, an arbitrary phase $\phi$ does not affect the time evolution of the system, and the Hamiltonian is manifestly time independent.

During the course of this analysis, we also consider impulsive pulses for which we take the limit $\Omega \to \infty$. In this limit, we can ignore the coupling term between the system and reservoir, and the pulse operation is expressed in the following form:
\begin{align}
    \hat{\rho}_\mathrm{tot}(t) 
    \leftarrow \, 
    & \mathcal{U}_\mathrm{imp}(\theta, \phi) \hat{\rho}_\mathrm{tot}(t)\\
    = & \hat{R}_z(\omega_qt) \hat{R}_\phi(\theta) \hat{R}_z(-\omega_qt) \\
    & \times \hat{\rho}_\mathrm{tot}(t) \hat{R}_z(\omega_qt)
    \hat{R}_\phi(-\theta) \hat{R}_z(-\omega_qt)\, .
    \label{eq:UpulseImp}
\end{align}
Here, the superoperator $\mathcal{U}_\mathrm{imp}(\theta, \phi)$ denotes the application of the impulsive pulse, and we have introduced the operator $\hat{R}_\phi(\theta) = \exp\left[-i \theta (\hat{\sigma}_x \cos \phi + \hat{\sigma}_y \sin \phi)/2\right]$. For more details of the derivation, see Appendix~\ref{sec:appRotatingFrame}.

\subsection{Exact Time evolution: Extended Hierarchical Equations Of Motion (FP-HEOM)}

Reservoirs with a macroscopic number of degrees of freedom are dominantly characterized by Gaussian fluctuations~\cite{FeynmanANNPHYS1963,MakriJPCB1999,SzankowskiJPHYS2017}, i.e., by autocorrelation functions $C(t)=\ev*{\hat{X}(t) \hat{X}(0)}_R$ given that $\ev*{\hat{X}(t)}_R=0$ and $\ev*{\bullet}_R=\mathrm{tr}\{\bullet \, \hat{\rho}_{R, eq}\}$ with the equilibrium density operator of the reservoir $\hat{\rho}_{R, eq} = e^{-\beta\hat{H}_R} / \mathrm{tr}\{e^{-\beta\hat{H}_R}\}$, and $\beta = 1/ k_\mathrm{B} T$. Equivalently, the noise properties of a respective reservoir follow from its spectral noise power
\begin{align}
    S_{\beta}(\omega) = \frac{1}{2\pi} \int_{-\infty}^{+\infty}dt C(t) e^{i\omega t}\, ,
    \label{Eq:fct}
\end{align}
where $S_{\beta}(\omega)$ and $S_{\beta}(-\omega)$ are related by the fluctuation-dissipation theorem that can be represented as 
\begin{align}
S_{\beta}(\omega) = \hbar [1+ n_\beta(\omega) ]\,  J(\omega)\, .
\label{eq:spectralnoise}
\end{align}
Here, $n_\beta(\omega)=1/[\exp(\beta\hbar \omega)-1]$ is the Bose distribution, and the spectral density $J(\omega)$ is an antisymmetric function with finite bandwidth characterized by a cutoff frequency $\omega_c$.
Note that this spectral density is directly proportional to the absorptive part of the dynamical susceptibility of the reservoir that can be extracted experimentally. Hence, the spectral noise power serves as the only ingredient required for describing the impact of environmental degrees of freedom on the qubit dynamics. Below, we will discuss in more detail the most relevant noise sources for superconducting qubits and their spectral densities.
We already note here though that this modeling is not only limited to reservoirs with underlying bosonic degrees of freedom but, effectively, may also apply, for example, to low-energy excitations of quasiparticles around Fermi surfaces.
\textcolor{black}{However, non-Gaussian noise, including quasiparticle tunneling induced by ionizing radiation~\cite{VepsalainenNATURE2020} cannot be described with this model:
This is out of scope of this study and is left for future work.}

Studying the open quantum dynamics according to the above pulse protocol is a highly nontrivial task since the combined time evolution is not separable. Standard procedures are then second-order perturbative approaches based on the Born--Markov approximation, including the Bloch--Redfield and the Lindblad equation, respectively. However, these approaches turn out to be insufficient in light of the growing accuracy, and in turn sensitivity, of actual qubit devices. For example, it was suggested that a more elaborate method beyond the Born--Markov approximation is needed when we consider dephasing dynamics with $1/f$ noise~\cite{IthierPRB2005}. In addition, it was reported that the Born approximation causes errors for simulations with multiple pulses~\cite{Tanimura2015,tanimuraJCP20} and that it provides inaccurate predictions for ground-state populations after a reset via equilibration~\cite{TuorilaPRR2019}. 

Hence, in order to conduct numerical simulations in a rigorous manner valid in all ranges of parameter space and applicable to a broad class of reservoirs, we adopt the hierarchical equations of motion (HEOM). Its derivation starts from the formally exact Feynman--Vernon path integral representation of the reduced density operator (RDO) of the system, where the impact of the reservoir is completely determined by the correlation $C(t)$.
The corresponding reduced quantum dynamics can exactly be mapped onto a nested hierarchy of equations of motion for auxiliary density operators (ADOs). As we have shown recently~\cite{XuPRL2022}, the key ingredient is the barycentric representation of $S_\beta(\omega)$, which provides, to any given accuracy, a representation of the form
\begin{align}
    C(t) = \sum_{k=1}^{K} d_k e^{-i\omega_k t-\gamma_k t} \qquad (t > 0)
    \label{eq:CF}
\end{align}
with a minimal number $K$ of effective reservoir modes. These are characterized by frequencies $\omega_k$, damping rates $\gamma_k>0$, and complex-valued amplitudes $d_k = d'_k + i d''_k$. Thus, the correlation $C(t)$ is described by a set of a \emph{moderate} number of damped harmonic modes even at zero temperature and also for structured reservoir densities. While the conventional HEOM was limited to higher temperatures and smooth reservoir spectral densities, the representation Eq.~\eqref{eq:CF} turns it into an extremely efficient simulation tool of general applicability. 

\textcolor{black}{Before we provide the explicit form of the equations of motion, we briefly discuss the relation between the correlation function and dynamics of the reduced systems. When the autocorrelation is proportional to the Dirac delta function, $C(t) \propto \delta(t)$, the spectral noise power is reduced to a constant function with respect to $\omega$, leading to white noise.
In this limit, the time derivative of the RDO depends only on the current state, and the process is ``memoryless'' in this sense.
Hence, the open quantum dynamics under the above condition are referred to as the ``Markovian'' in the field of quantum statistical physics.
By contrast, when the autocorrelation function is not a delta function, the dynamics of the RDO depend on the previous states as well, which results from the retarded feedback of the reservoir.
We refer to this process as the ``non-Markovian'' in this study.
}

Here, we display the structure of the new free-pole HEOM (FP-HEOM): See Appendix~\ref{sec:appHEOM} for more details. The dynamics of the ADOs follow from
\begin{align}
   \frac{\partial \hat{\rho}_{\vec{m},\vec{n}}(t)}{\partial t} =& -i\mathcal{L}_S\hat{\rho}_{\vec{m},\vec{n}}(t) -\sum_{k=1}^K (m_k z_k+n_k z_k^*) \hat{\rho}_{\vec{m},\vec{n}}(t) \\
    & -i\sum_{k=1}^K \mathcal{L}_k^{+} \hat{\rho}_{\vec{m},\vec{n}}(t) -i\sum_{k=1}^K \mathcal{L}_k^{-} \hat{\rho}_{\vec{m},\vec{n}}(t)
    \label{eq:HEOM}
\end{align}
with multi-index $(\vec{m},\vec{n}) \equiv \{m_1,\ldots,m_K,n_1,\ldots,n_K\}$ associated with forward and backward system path in the original path integral, complex-valued coefficients $z_k \equiv \gamma_k + i\omega_k$ according to Eq.~\eqref{eq:CF}, 
and raising and lowering superoperators $\mathcal{L}_k^+$ and $\mathcal{L}_k^-$, acting on the $k$th quasimode and involving $\hat{V}$.
\textcolor{black}{The superoperator $\mathcal{L}_S$ is the commutator of the system Hamiltonian, $\mathcal{L}_S \bullet = [\hat{H}_S(\Omega, \phi;t), \bullet]/\hbar$.}
It can be shown that, in fact, this equation is the Fock state representation in an extended Hilbert space including the qubit as well as the quasimodes \cite{Xu2023}. 

To obtain a closed set of equations for numerical calculations, Eq.~\eqref{eq:HEOM} is truncated by defining the depth of the hierarchy as $\mathcal{N} = \sum_{k=1}^{K} (m_k + n_k)$, and always set $\hat{\rho}_{\vec{m}, \vec{n}}(t) = 0$ for the ADOs with $\mathcal{N} > \mathcal{N}_{\max}$. 
We set $\mathcal{N}_{\max}$ to a sufficiently large integer to obtain converged results.
The physical RDO of the qubit system appears as $\hat{\rho}_S(t) = \mathrm{tr}_R\{\hat{\rho}_\mathrm{tot}(t)\} = \hat{\rho}_{\vec{0},\vec{0}}(t)$.

\section{Preliminaries: Spectral densities and parameters}\label{sec:Parameters}
In this section, we specify a class of spectral densities relevant for superconducting qubit platforms, particularly of the transmon type. We note that a substantial number of studies \cite{BylanderNP2011,CardaniNATCOMMUN2021,IthierPRB2005,PapicARXIV2023} have provided a quite accurate picture of relevant noise sources on a broad range of timescales from intrinsic qubit timescales of nanoseconds to macroscopic scales of hours for rare events. Here, we are interested in decoherence processes on frequency scales from GHz down to the range of MHz or kHz. On these scales three dominant noise sources have been identified:
\begin{enumerate}[label=(\roman*)]
    \setlength{\leftskip}{2em}
    \item Electromagnetic fields: Fluctuations of electromagnetic modes are typically modeled according to an Ohmic reservoir (Ohmic resistor) with $J_\mathrm{em}(\omega)\sim \omega$ up to some cutoff frequency $\omega_c$~\cite{Weiss2012,Barone1982,WendinARXIV2005}.
    \item Two-level fluctuators (TLFs): It has been suggested that transmon qubits are affected by $1/f^{\varepsilon}$ ($\varepsilon > 0$) noise in the low-frequency region $f \ll 1$~\cite{IthierPRB2005,BylanderNP2011}, and it was reported that two-level fluctuators (random telegraph noise) in the circuits work as a source of this noise~\cite{MachlupJAP1954,Weiss2012,PaladinoRMP2014,MullerRPP2019}. 
    \item Quasiparticles: Residual quasiparticles in superconductors have detrimental effects on the qubit performance; under some approximations the spectral noise power of quasiparticle noise has been shown to be of the form $S_{\beta, \mathrm{qp}}(\omega)\sim 1/\sqrt{\omega}$ at low temperatures~\cite{GlazmanSPPLN2021}.
\end{enumerate}
These major types of noise can be captured by the following class of spectral densities:
\begin{align}
    J_s(\omega) = \mathrm{sgn}(\omega)
    \frac{\kappa \omega_\mathrm{ph}^{1-s} |\omega|^{s}}
    {\left(1+(\omega/\omega_c)^2\right)^2}
    \label{eq:SD}
\end{align}
parameterized by a spectral exponent $s$.
Here,  $\kappa$ is the coupling rate between the system and reservoir, and $\omega_c$ a cutoff frequency.
The ``phononic reference frequency'' $\omega_\mathrm{ph}$ is usually introduced to fix the unit of $\kappa$ irrespective of the exponent $s$:
\textcolor{black}{For example, the quantity $\kappa$ corresponds to the viscosity, and $\omega_\mathrm{ph}$ determines the low-frequency behavior of $J(\omega)$ for damped systems.
As implied in Eq.~\eqref{eq:1/f-SD}, the scaled frequency $\omega / \omega_\mathrm{ph}$ plays a crucial role in thermodynamics~\cite{Breuer2002}.
Note that for the Debye model, the reference frequency $\omega_\mathrm{ph}$ corresponds to the Debye frequency $\omega_\mathrm{D}$~\cite{Breuer2002}.}
Here we put $\omega_\mathrm{ph}=\omega_q$ so that $J_s(\omega_q)$ takes the same value regardless of $s$. The sign function $\mathrm{sgn}(\omega)$ guarantees the property $J_s(-\omega) = -J_s(\omega)$.
For convenience and following a previous study~\cite{TuorilaPRR2019}, the cutoff function is chosen to be $1/(1+(\omega/\omega_c)^2)^2$, where the dependence of explicit results on this specific form is negligible as long as $\omega_c\gg \omega_q$ \cite{Weiss2012}.

The above class of spectral densities includes the Ohmic case ($s=1$) as well as sub- ($0 < s < 1$) and super-Ohmic ($s > 1$) baths, respectively. More specifically, in the low-frequency range the corresponding spectral noise power [Eq.~\eqref{eq:spectralnoise}] saturates to a finite value in the Ohmic case (i), i.e., $S_{\beta, s=1}(\omega=0) = \kappa k_\mathrm{B} T$, while in the sub-Ohmic case it scales according to
\begin{align}
    S_{\beta, s<1} (\omega\to 0) \simeq \kappa k_\mathrm{B} T (\omega_\mathrm{ph}/\omega)^{1-s}\, .
    \label{eq:1/f-SD}
\end{align}
With the relation $\varepsilon = 1-s$, this spectral noise power exhibits $1/f^\varepsilon$-like behavior (ii), and it captures quasiparticle noise (iii) for $s=1/2$. It is worth noting that for the TLF-noise (ii), the linear dependence on the temperature in Eq.~\eqref{eq:1/f-SD} corresponds to previous studies~\cite{EromsAPL2006,NugrohoAPL2013}, while a $T^2$ dependence has also been reported ~\cite{KenyonJAP2000,WellstoodAPL2004}.

In this study, we particularly investigate how the qubit dynamics changes with respect to the spectral exponent $s$, by considering values $s = 1, 1/2, 1/4, 1/8,$ and $1/14$.
Further, we set $\omega_q$ as the unit of frequency and fix parameter values to low temperatures $\beta \hbar \omega_q = 5$, high cutoff frequency $\omega_c/\omega_q = 50$, and weak coupling to the reservoir $2\pi \hbar \kappa = 0.04$.
For example, at a reservoir temperature of $T=30$~mK, this corresponds to $\omega_q\approx 2\pi \times 3.1$~GHz.
Pulse amplitudes $\Omega$ (pulse durations $\tau$ accordingly) and durations of the idle phase $\Delta t$ are tuned over a wide range of parameters.

\begin{figure*}
    \includegraphics[width=\linewidth]{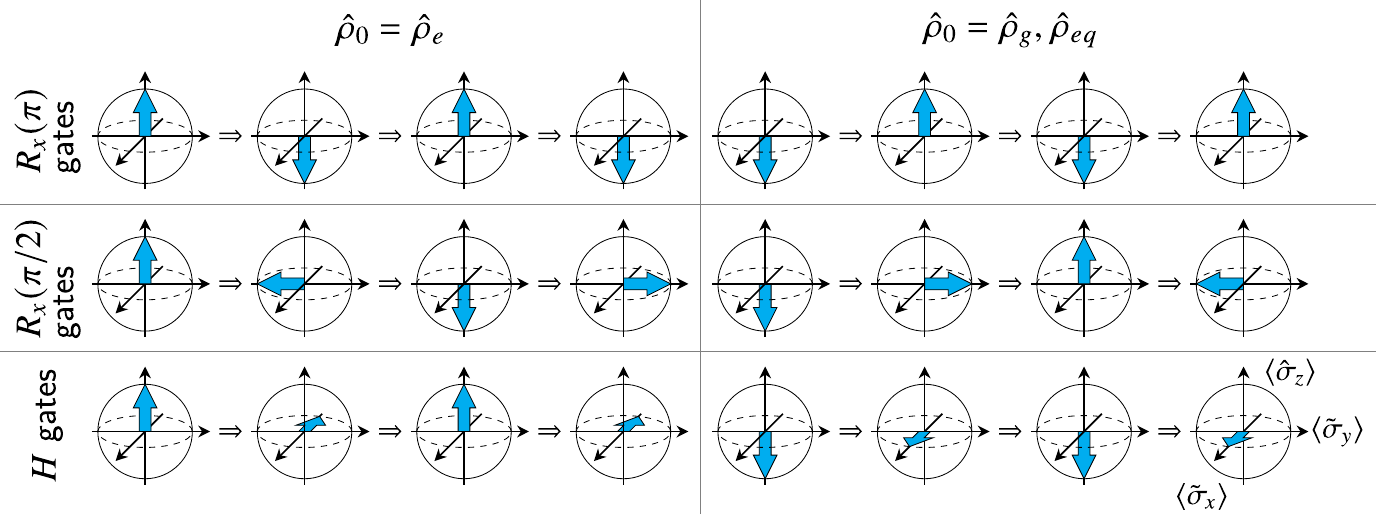}
    \caption{Schematics of time evolution of ideal Bloch vectors in the rotating frame. The right arrows ($\Rightarrow$) indicate the pulse applications. \textcolor{black}{Note that in the case $\hat{\rho}_{eq}$, the length of the Bloch vector is slightly smaller than $1$ [from $0.92$ to $0.97$, depending on the spectral exponent $s$ in Eq.~\eqref{eq:SD}; see the inset of Fig.~\ref{fig:relax} below]}, while the direction of the vector is same as the direction in the case $\hat{\rho}_g$.
    \label{fig:Bloch}}
\end{figure*}

\section{Sequences of gate operations} \label{sec:resSequence}
In this section, we analyze the performance of a single qubit subject to sequences of gate operations in the presence of noise sources according to Eq.~\eqref{eq:SD}. We emphasize that the numerical simulations based on the FP-HEOM [Eq.~\eqref{eq:HEOM}] provide highly accurate data including the full non-Markovianity, \textcolor{black}{that is, any higher-order system--reservoir correlations with infinitely long time memories beyond conventional perturbation theories.}  By sweeping parameters over a wide range of values we obtain a comprehensive picture of the qubit performance and the relevance of qubit--environmental correlations. 

More specifically, we consider pulse sequences that consist of three gate operations separated by two idle phases, see Fig.~\ref{fig:pulseSequence}. For the gate operations our focus lies on three types of operations, namely, (i) rotations with angle $\pi$ about the $x$ axis, denoted $R_x(\pi)$, (ii) rotations with angle $\pi/2$ about the $x$ axis, denoted $R_x(\pi/2)$, and (iii) a Hadamard gate, denoted $H$. The duration of the three pulses $\tau$ is set equal during the sequence as well as the time span $\Delta t$ for the two idle phases. Each sequence of gate operations is then described by a set of gate-specific superoperators $\mathcal{U}_\mathrm{p}(\theta, \phi)$ while during the idle phases the time evolution is generated by $\mathcal{U}_\mathrm{i}(\Delta t)$. 


Numerically, we practice the following procedure:
\textcolor{black}{For the first pulse application, the time evolution is calculated for a fixed value $\phi$ up to the pulse duration $\tau = \theta / \Omega$ by numerically integrating the FP-HEOM of Eq.~\eqref{eq:HEOM}.
We here symbolically represent the integration as $\hat{\rho}_{\vec{m}, \vec{n}}(\tau) \leftarrow \mathtt{U}(\Omega > 0, \phi; 0, \tau) \hat{\rho}_{\vec{m}, \vec{n}}(0)$.
The following time evolution for the first idle phase is calculated under the condition $\Omega = 0$ up to $\Delta t$, which is expressed as $\hat{\rho}_{\vec{m}, \vec{n}}(\tau+\Delta t) \leftarrow \mathtt{U}(\Omega = 0, \phi; \tau, \Delta t) \hat{\rho}_{\vec{m}, \vec{n}}(\tau)$.}
We repeat these calculations for the subsequent pulses/idle phases.
\textcolor{black}{In the situation of impulsive pulses, $\mathcal{U}_\mathrm{p}(\theta, \phi)$ is replaced by $\mathcal{U}_\mathrm{imp}(\theta, \phi)$, and this superoperator is applied to all RDOs and ADOs.
The latter is expressed as $\hat{\rho}_{\vec{m}, \vec{n}}(t) \leftarrow \mathcal{U}_\mathrm{imp}(\theta, \phi) \hat{\rho}_{\vec{m}, \vec{n}}(t)$, which corresponds to the replacement of $\hat{\rho}_\mathrm{tot}(t)$ in Eq.~\eqref{eq:UpulseImp} with $\hat{\rho}_{\vec{m}, \vec{n}}(t)$}.
Note that we treat the open quantum dynamics over the full sequence such that the RDO and ADOs obtained at the end of a previous phase are used as the initial states for the subsequent ones.

For the initial states prior to the first gate operation, we consider three initial states:
(a) the qubit is in the excited state, $\hat{\rho}_{\vec{0}, \vec{0}}(0) = \ketbra{1}{1}$ and $\hat{\rho}_{\vec{m}\neq\vec{0}, \vec{n}\neq\vec{0}}(0) = 0$, (b) the qubit is in the ground state, $\hat{\rho}_{\vec{0}, \vec{0}}(0) = \ketbra{0}{0}$ and $\hat{\rho}_{\vec{m}\neq\vec{0}, \vec{n}\neq\vec{0}}(0) = 0$, and (c) the qubit resides in the equilibrium state of the \emph{total} Hamiltonian.
Here, we have introduced the ket-vector of the ground (excited) state of the system as $\ket{0}$ ($\ket{1}$).
The initial states (a) and (b) correspond to factorized states $\ketbra{1}{1}\otimes\hat{\rho}_{R, eq}$ and $\ketbra{0}{0}\otimes\hat{\rho}_{R, eq}$, respectively.
For case (c), before the pulse applications, the relaxation dynamics of the qubit without the external field ($\Omega = 0$) starting from $\ketbra{1}{1}\otimes\hat{\rho}_{R, eq}$ is evaluated until a steady state is reached.
Since the system reaches the same steady state irrespective of the initial state, we identify this steady state with the correlated thermal equilibrium state of the total compound, i.e., $\hat{\rho}^\mathrm{c}_{S, eq}=\mathrm{tr}_R \{e^{-\beta\hat{H}_\mathrm{tot}(\Omega=0, \phi;t)}\}/\mathrm{tr}\{e^{-\beta\hat{H}_\mathrm{tot}(\Omega=0, \phi;t)}\}$~\cite{Tanimura2014}.
For more details of the preliminary equilibration process, see Appendix~\ref{sec:appRelax}. In short, we denote these three types of initial states as (a) $\hat{\rho}_0 = \hat{\rho}_e$, (b) $\hat{\rho}_g$, and (c) $\hat{\rho}_{eq}$, respectively.
\textcolor{black}{Note that the initial state $\hat{\rho}_{eq}$ is a mixed state with $\ketbra{0}{0}$ and $\ketbra{1}{1}$ and is close to $\ketbra{0}{0}$ ($0.92 \lesssim \mel{0}{\hat{\rho}^\mathrm{c}_{S, eq}}{0} \lesssim 0.97$, see the inset of Fig.~\ref{fig:relax} below) because of the low temperature we consider in this study:
Differences originating from small variations in the initial states are illustrated through the comparison of cases (b) and (c) in the following results.}

From the perspective of experimental implementations, we assume
that the qubit prior to the gate sequence is equilibrated with respect to the total Hamiltonian. In the domain, where superconducting qubits are operated and at weak couplings to the environment, this implies basically only weak qubit--reservoir correlations in thermal equilibrium. An initialization pulse can then be assumed to prepare the compound into either 
(a) $\ketbra{1}{1} \otimes \hat{\rho}_{R, eq}$ or (b) $\ketbra{0}{0} \otimes \hat{\rho}_{R, eq}$ with equilibrium correlations between the qubit and reservoir basically destroyed.
For case (c), we simply use the total equilibrium state as the initial state, and no initialization pulse is required; the qubit resides almost exclusively in its ground state correlated with the reservoir (when projecting $\hat{\rho}^\mathrm{c}_{S, eq}$ onto $\ketbra{0}{0}$).


To illustrate the described protocol, we show in Fig.~\ref{fig:Bloch} a cartoon displaying the status of the qubit's Bloch vector in the rotating frame after the application of the respective gate operation starting from a specific initial state.
The dynamics during the idle phases, which appear between the second and the third and the third and the fourth snapshot, is not shown, since the system in the rotating frame ideally remains in a certain state during an idle phase.

The Bloch vector in the rotating frame is defined as
\begin{align}
    \ev*{\tilde{\sigma}_\alpha (t)} = \mathrm{tr}\{\hat{\sigma}_\alpha \hat{R}_z(-\omega_q t)\hat{\rho}_S(t)\hat{R}_z(\omega_q t)\}\,  
    \label{eq:BlochRotFrame}
\end{align}
with the reduced density operator $\hat{\rho}_S(t)$.


In order to quantify the (detrimental) impact of reservoirs onto the performance of the qubit under the gate operations, we introduce the fidelity
\begin{align}
    F(t) = \left(\mathrm{tr}\left\{\sqrt{\sqrt{\hat{\rho}_\mathrm{iso}(t)}
    \hat{\rho}_S(t) \sqrt{\hat{\rho}_\mathrm{iso}(t)}}
    \right\}\right)^2\, .
\end{align}
Here the density operator for the \emph{isolated} qubit system is introduced as $\hat{\rho}_\mathrm{iso}(t)$. For its evaluation identical pulse sequences compared to the dissipative case are considered with the only difference that we set $\hat{V}=0$ with initial states $\hat{\rho}_\mathrm{iso}(0) = \ketbra{1}{1}$, $\ketbra{0}{0}$ and $\hat{\rho}^\mathrm{c}_{S, eq}$ respectively. 

\begin{figure*}
    \centering
    \includegraphics[width=0.88\linewidth]{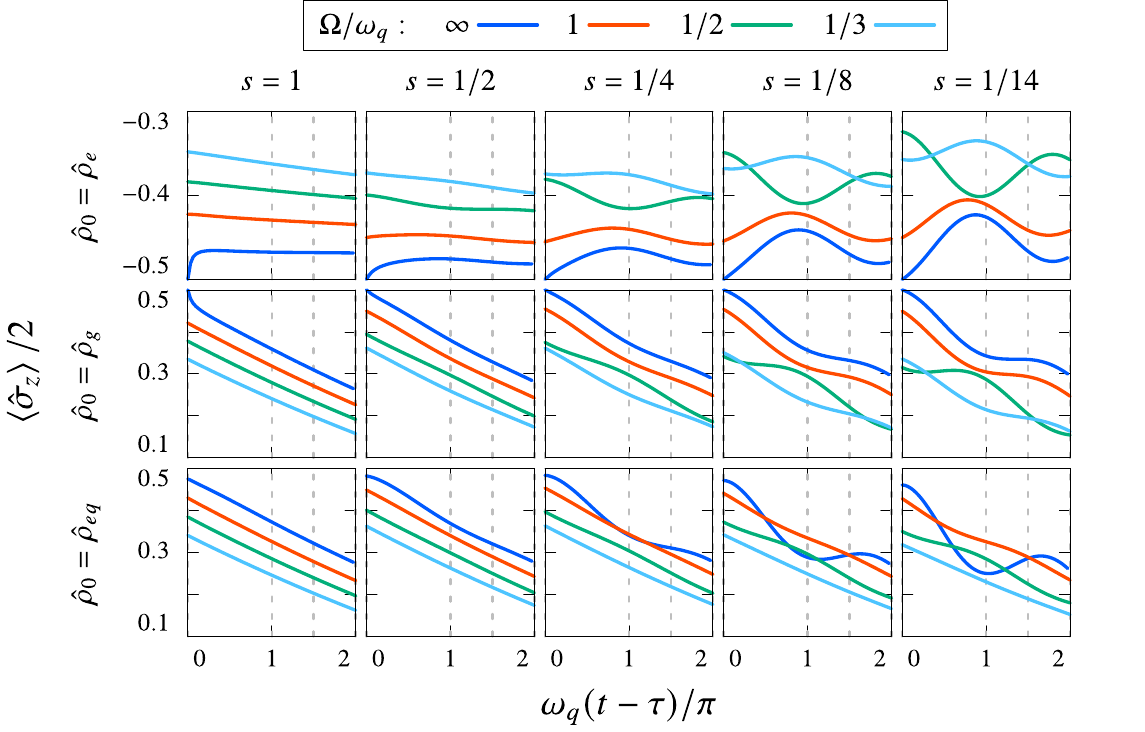}
    \caption{Dynamics of the expectation value $\ev*{\hat{\sigma}_z(t)}/2$ during the first idle phase with various initial states $\hat{\rho}_0$ and spectral exponents $s$. The sequence of $R_x(\pi)$ gates is considered. The gray-vertical-dashed lines indicate the time duration $\Delta t \omega_q /\pi = 0, 1, 3/2, 2$, respectively, which corresponds to the time $d=2$ in Fig.~\ref{fig:pulseSequence}. 
    \label{fig:idle_z}}
\end{figure*}

\subsection{$R_x(\pi)$ gates} \label{sec:resPI}

Expressed in terms of the superoperator introduced in Eqs.~\eqref{eq:Upulse}--\eqref{eq:UpulseImp}, the sequence of three $R_x(\pi)$ gates is described by the evolution
\begin{align}
   \mathcal{U}_\mathrm{p}(\pi, 0)\mathcal{U}_\mathrm{i}(\Delta t)
    \mathcal{U}_\mathrm{p}(\pi, 0)\mathcal{U}_\mathrm{i}(\Delta t)
    \mathcal{U}_\mathrm{p}(\pi, 0)\, .
    \label{eq:seq_PI}
\end{align}

We start by discussing typical dynamical features that reveal interesting physics and have direct impact on the fidelities to be analyzed below.
By way of example, we depict in Fig.~\ref{fig:idle_z} snapshots of the qubit dynamics during the first idle phase (segment $1 < d \leq 2$ in Fig.~\ref{fig:pulseSequence}) for various driving amplitudes $\Omega$, i.e., pulse durations $\tau=\pi/\Omega$, and spectral exponents $s$.
Clearly, for $\tau=0$, the qubit starts after the impulsive $\pi$-pulse in the ideally rotated state with $\ev*{\hat{\sigma}_z}/2=\pm 0.5$ (for ground/excited state initial preparation).
It then tends to relax monotonously for reservoirs with $s>1/2$ with the initial states $\hat{\rho}_0 = \hat{\rho}_g$ and $\hat{\rho}_{eq}$, while for smaller spectral exponents (towards $1/f$ noise) an oscillatory behavior sets in as a result of the stronger portion of low-frequency modes, which induce a sluggish dynamics and strongly retarded feedback of the reservoir.
For finite duration of the first gate pulse (finite $\Omega$, $\tau>0$), the qubit starts progressively further away from its ideal value since relaxation happens to occur already during the gate pulse and pursues in the subsequent idle phase.
In relative terms, this process is more pronounced when starting initially from an excited state compared to a ground-state preparation.
Interestingly, deeper into the sub-Ohmic domain, $s \leq 1/8$, and with increasing duration $\Delta t$ of the idle phase [larger $\omega_q (t-\tau) / \pi$ in Fig.~\ref{fig:idle_z}], the qubit dynamics for different $\tau$ interchange:
Less ideal $\ev*{\hat{\sigma}_z}$ values at the beginning of the idle phase [e.g., a green curve in the upper-right panel in Fig.~\ref{fig:idle_z} ($\hat{\rho}_0 = \hat{\rho}_e, s = 1/14$ and $\Omega/\omega_q = 1/2$)] are overcompensated by an oscillatory reservoir-induced dynamics such as to exceed those with more ideal starting values [a cyan curve in the same panel ($\Omega / \omega_q = 1/3$)].
This ``switching'' may lead to a somewhat counter-intuitive behavior of respective fidelities as we will see now.

\begin{figure*}
    \includegraphics[width=\linewidth]{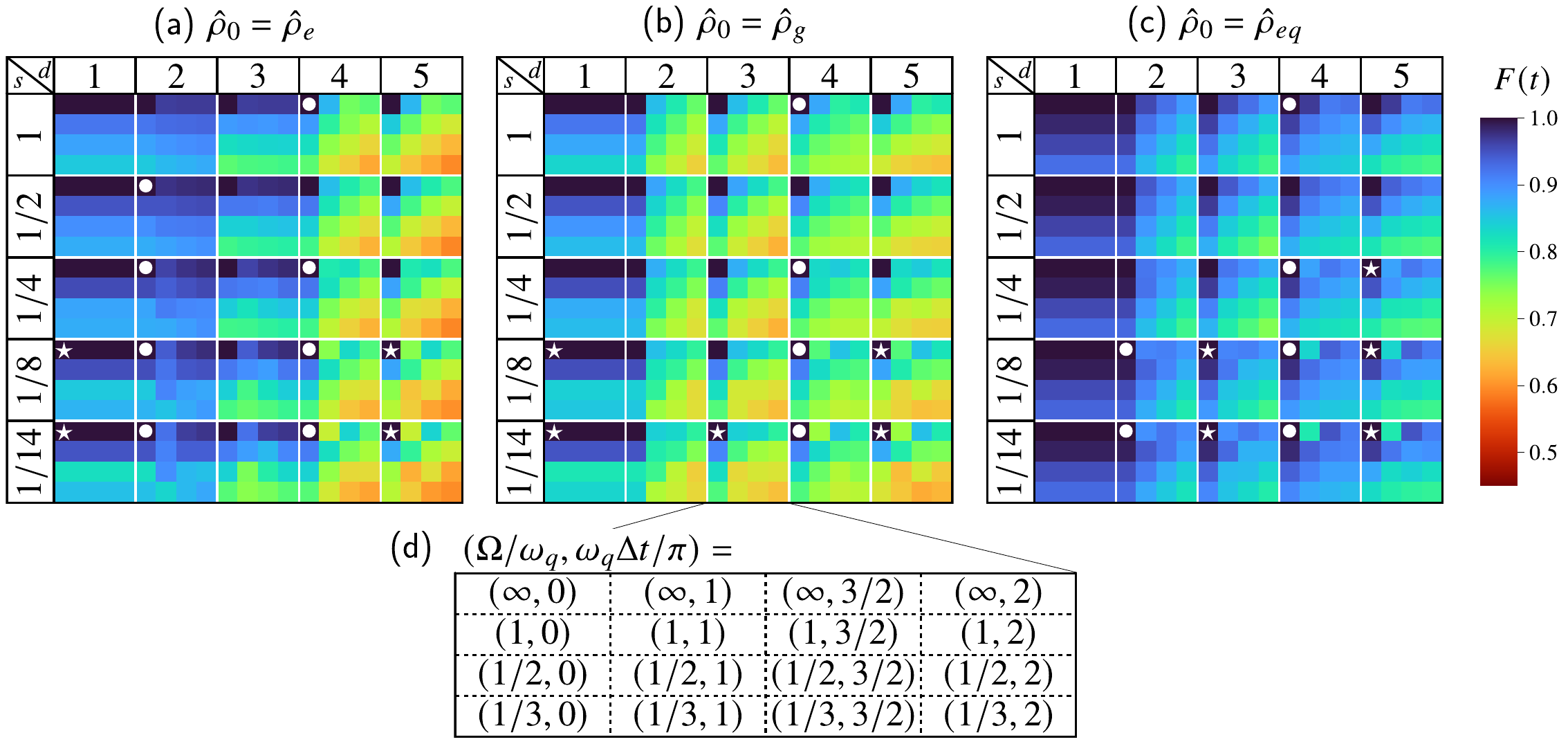}
    \caption{[(a)--(c)] Heatmaps of the fidelity of the $R_x(\pi)$ gates measured at the end of each phase, $d$, with various spectral exponents, $s$. The initial states are given by (a) $\hat{\rho}_e$, (b) $\hat{\rho}_g$, and (c) $\hat{\rho}_{eq}$, respectively. The whole heatmap is divided into $5 \times 5$ supercells, and a supercell consists of $4 \times 4$ cells. For each supercell, the values of $\Omega$ and $\Delta t$ are varied, while $d$ and $s$ are fixed. The circle and star symbols in the upper-left corner of the supercell indicate the violation of the expected order: see the main text for the definition of the expected order. (d) Legend for the supercell. 
    \label{fig:HMap_PI}}
\end{figure*}

Figures~\ref{fig:HMap_PI}(a)--\ref{fig:HMap_PI}(c) display heatmaps of fidelities for the three respective initial states. For each initial preparation, the fidelity at the end of each of the five segments ($d=1, \ldots 5$) in Fig.~\ref{fig:pulseSequence} is depicted for various values of the spectral exponent $s$, from an Ohmic noise source ($s=1$) to a reservoir with deep sub-Ohmic fluctuations ($1/f$ noise). Each pair $(d, s)$ defines a supercell consisting of $4\times 4$ cells, for which the driving amplitude $\Omega$ and the duration of the idle phase $\Delta t$ are varied, see Fig.~\ref{fig:HMap_PI}(d). 

Before we come to the details, we summarize the \emph{overall picture}: The tendency towards lower fidelities can be seen (i) for weaker drive amplitudes (i.e., longer gate pulse durations), (ii) for longer idle times (with some exceptions, see below), and (iii) when starting from the excited state.
The overall dependence on the spectral exponent $s$ (type of noise) is weak for the $R_x(\pi)$ gate.
In general, fidelities in case (c) are higher than those in case (b), although the initial states $\hat{\rho}_g$ and $\hat{\rho}_{eq}$ are close to each other, and the dynamics are similar (the middle and bottom panels of Fig.~\ref{fig:idle_z}).
This is mainly caused by the difference of the reference density operator $\hat{\rho}_\mathrm{iso}(t)$;
$\hat{\rho}_\mathrm{iso}(0) = \ketbra{0}{0}$ in case (b) and $\hat{\rho}_\mathrm{iso}(0) = \hat{\rho}^\mathrm{c}_{S, eq}$ in case (c).
However, this picture is blurred as some counter-intuitive features appear.

\begin{figure}
    \centering
    \includegraphics[width=\linewidth]{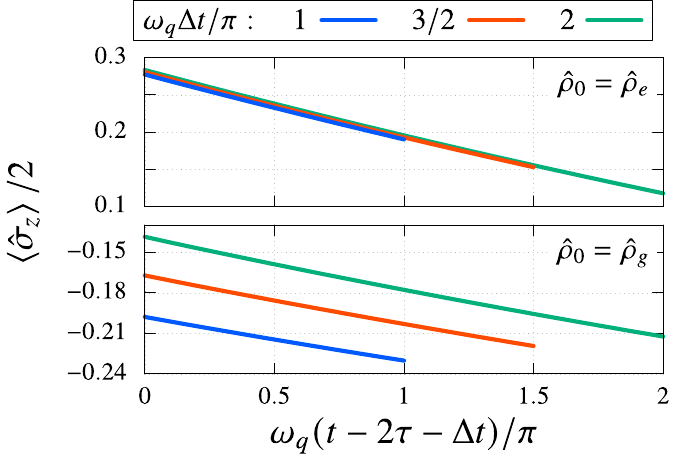}
    \caption{Dynamics of the expectation value $\ev*{\hat{\sigma}_z(t)}/2$ during the second idle phase for the Ohmic reservoir ($s=1$) with the pulse duration $\Omega / \omega_q = 1/3$. The initial and terminal point of each curve correspond to the time $d = 3$ and $4$, respectively.
    \label{fig:2ndIdle}}
\end{figure}

This brings us to a more \emph{detailed discussion}: 
We first mention that at low temperatures, relaxation occurs predominantly from an excited state toward the ground state.
Hence, the expectation is that whenever the qubit, after a gate pulse, is ideally positioned in the excited state, the fidelity at the end of an idle phase is smaller compared to the situation when it is supposed to be in the ground state.
This is confirmed by comparing columns $d=2$ in Figs.~\ref{fig:HMap_PI}(b) and \ref{fig:HMap_PI}(a):
\textcolor{black}{For case (b), the system is prepared into the excited state after the first pulse, and the more significant decrease of the fidelity is observed during the first idle phase ($1 < d \leq 2$) compared to case (a), where the ideal state after the first pulse is the ground state.}
One also observes that with increasing duration of the idle phase for $d=2$ in case (a) the fidelity generally increases while it decreases with growing $\Delta t$ for $d=2$ in case (b).
Note that the former tendency is opposed to the general tendency (ii), i.e., this is the exceptional case mentioned above.

To analyze the tendency at $d = 4$, we display typical dynamics during the second idle phase in Fig.~\ref{fig:2ndIdle}.
Comparing the fidelities at $d = 3$ and $4$ for each gate sequence, which correspond to the initial and terminal point of each curve in Fig.~\ref{fig:2ndIdle}, respectively, one expects qualitatively an opposite behavior from cases at $d=2$:
loss of the fidelity in case (a) (because the gate has positioned the qubit close to its excited state) and partial recovery of the fidelity in case (b) (when the second gate has positioned it back into the ground state). This is indeed the case and leads to relatively larger fidelities at the end of the sequence ($d=5$) in case (b) compared to case (a).
In terms of the duration $\Delta t$, the fidelity at $d = 4$ is expected to shrink in both cases as $\Delta t$ grows:
The relaxation behavior directly results in this expectation in case (a).
Although the partial recovery occurs in case (b), the initial difference in Fig.~\ref{fig:2ndIdle} with respect to $\Delta t$ is too large to be compensated.
In case (c), the same expectation as in case (b) holds.

However, there are deviations from this behavior.
They are indicated by the circle symbols in Figs.~\ref{fig:HMap_PI}(a)--\ref{fig:HMap_PI}(c) and are because of the following two factors:
(i) Instantaneous gate pulses ($\tau=\pi/\Omega=0$):
In this situation, drastic changes of the qubit--reservoir correlations emerge during the idle phase and nonmonotonous behavior in the qubit populations is observed (blue curves in Fig.~\ref{fig:idle_z}).
Note in passing that the fidelities for $\tau = \Delta t = 0$ are always $1$ because no relaxation and decoherence occur in this case.
(ii) Long-range qubit--reservoir correlations (memory effects): Since the duration of idle phases is much shorter than equilibration times of the qubit, nonequilibrium dynamics appear throughout the complete sequence. This implies that retardation effects induce \emph{correlations} between the dynamics in subsequent idle phases ($1 < d \leq 2$ and $3 < d \leq 4$) as well as between idle phases and gate segments. These memory effects influence the fidelities as well and, as detailed inspection reveals, lead in some cases to deviations from the general picture described above. 
For example, in the deep sub-Ohmic regime ($s=1/8, 1/14$), the qubit and reservoir can coherently interact with each other multiple times (for more details, see Appendix~\ref{sec:appCF}).
This non-Markovian effect induces oscillations in the populations (Fig.~\ref{fig:idle_z}) and a nonmonotonous trend in the fidelities when sweeping pulse and idle-phase parameters.
We emphasize that within the Born--Markov approximation the reservoir is treated such that it were always in the bare equilibrium state $\hat{\rho}_{R, eq}$ and nonmonotonous phenomena (i) and (ii) are not predicted within the framework of Bloch--Redfield and Lindblad simulations.

There is another very interesting observation that we stress here. 
One expects that for longer gate pulses the fidelity deteriorates because of a longer interaction time between the system and reservoir.
We have confirmed this tendency within the frame of Lindblad equations (results are not shown).
This implies that in Figs.~\ref{fig:HMap_PI}(a)--\ref{fig:HMap_PI}(c) the fidelity aligns in descending order from top to bottom at $d = 1, 3$, and $5$. The star symbols indicate the violation of this expectation. 
The reason for this deviation is the following:
Both, the angle between the experimentally obtained and the ideal Bloch vectors, as well as the length of the Bloch vector contribute to the fidelity.
\textcolor{black}{As depicted in Appendix~\ref{sec:appRamsey}, the bare qubit frequency $\omega_q$ (no reservoir) differs from the effective qubit frequency obtained from Ramsey experiments in the presence of a reservoir.}
If the frequency of the external pulse is misaligned with the effective qubit frequency, the rotation axis changes from the desired one, and the fidelity deteriorates at the end of the pulse application, $\tau = \theta / \Omega$. Furthermore, the effective qubit frequency varies in time so that the pattern of the fidelity may not be intuitive.

\begin{figure*}
    \includegraphics[width=\linewidth]{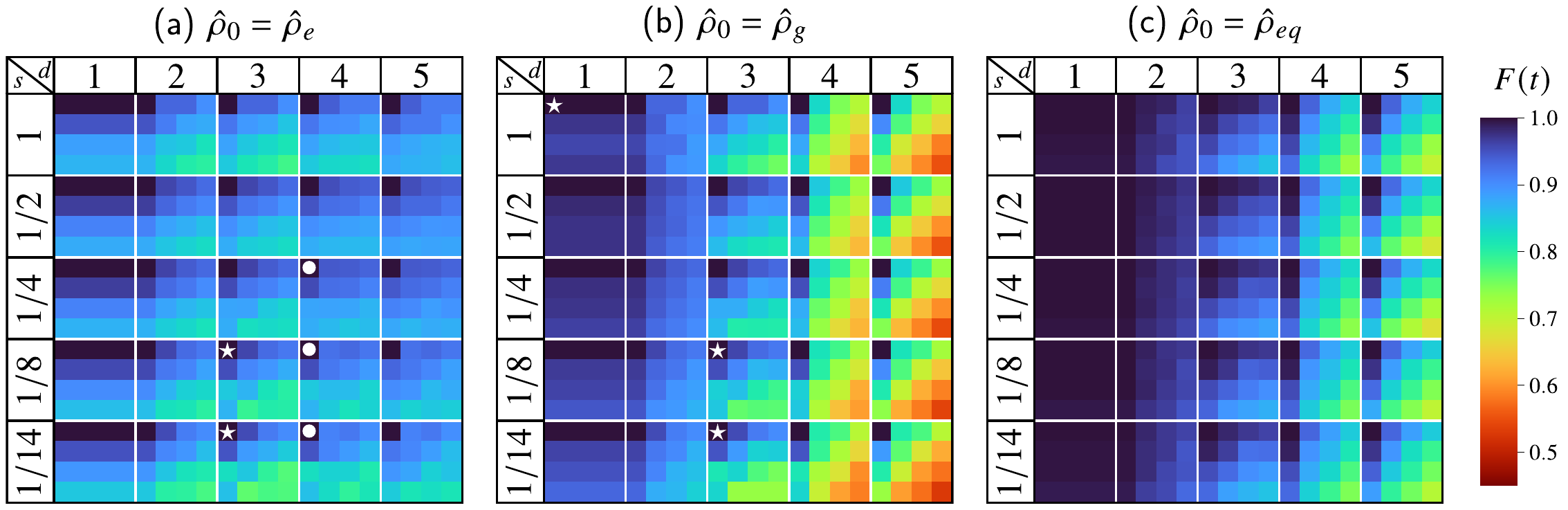}
    \caption{Heatmaps of the fidelity of the $R_x(\pi/2)$ gates measured at the end of each phase, $d$, with various spectral exponents, $s$. The initial states are given by (a) $\hat{\rho}_e$, (b) $\hat{\rho}_g$, and (c) $\hat{\rho}_{eq}$, respectively. The parameter values of each cell are the same as Fig.~\ref{fig:HMap_PI}. The circle and star symbols in the upper-left corner of the supercell indicate the violation of the expected order, in the same way as Fig.~\ref{fig:HMap_PI}.
    \label{fig:HMap_HPI}}
\end{figure*}

\subsection{$R_x(\pi/2)$ gates} \label{sec:resHPI}
Let us now turn to the $R_x(\pi/2)$ gate, which follows from the following sequences [cf.\ Eqs.~\eqref{eq:Upulse}--\eqref{eq:UpulseImp}]:
\begin{align}
    \mathcal{U}_\mathrm{p}\left(\frac{\pi}{2}, 0\right)
    \mathcal{U}_\mathrm{i}(\Delta t)
    \mathcal{U}_\mathrm{p}\left(\frac{\pi}{2}, 0\right)
    \mathcal{U}_\mathrm{i}(\Delta t)
    \mathcal{U}_\mathrm{p}\left(\frac{\pi}{2}, 0\right)\, . 
    \label{eq:seq_HPI}
    \noeqref{eq:seq_HPI}
\end{align}
The qubit dynamics during the first idle phase can be seen in Fig.~\ref{fig:idle_y} in Appendix~\ref{sec:appHPI}.
They reveal again non-Markovian behavior depending on the spectral exponents.
Corresponding fidelities are seen in Fig.~\ref{fig:HMap_HPI} with the same structuring and the same value of the parameters as above for the $\pi$ gate. However, the pulse duration is only half of that for the $\pi$ gate, of course, given by $\tau = \pi / (2\Omega)$. 

The final fidelity is the largest for the initialization (a). As discussed in Sec.~\ref{sec:resPI}, this is because at the beginning of the second idle phase, the desired state of the qubit is the ground state $\ketbra{0}{0}$ (Fig.~\ref{fig:Bloch}), and the relaxation process constructively supports this state.
The cause of the difference of the fidelity between cases (b) and (c) is mainly the difference of the reference state, which is the same as in the case of the $R_x(\pi)$ sequence.
With a fixed $d$, $\Omega$, and $\Delta t$, the fidelity in general tends to take a maximum value for medium sub-Ohmic reservoirs with $s = 1/2$ and $1/4$ while minimum values appear for exponents $s = 1$ (Ohmic) and $1/14$ (deep sub-Ohmic), i.e., reservoirs with intermediate exponents have the least detrimental impact on fidelities for $R_x(\pi/2)$ gates.

Comparing the supercells of initial preparations 
$\hat{\rho}_e$ and $\hat{\rho}_g$ in column $d = 1$, we find that the fidelity of the latter exceeds that of the former. As depicted in Fig.~\ref{fig:Bloch}, the first pulse application corresponds in the rotating frame to the rotation of the Bloch vector from $(0,0,1)$ to $(0,-1,0)$ in case (a), and from $(0,0,-1)$ to $(0,1,0)$ in case (b). The positive $\ev*{\hat{\sigma}_z(t)}$ (excited state) is converted to a superposition (coherence) via the $\pi/2$-pulse in case (a), while the negative $\ev*{\hat{\sigma}_z(t)}$ (ground state) contributes in case (b):
In any case, larger absolute values, $|\ev*{\hat{\sigma}_z(t)}|$, result in larger fidelities after the pulse application.
Because decay of $|\ev*{\hat{\sigma}_z(t)}|$ is more pronounced in the excited state, a superposition with a lower fidelity is created in case (a).
This in turn suggests creation of superposition states with higher fidelities starting initially from a ground state.

Similar to the $R_x(\pi)$ gates, we expect the following tendency of the fidelity in each supercell:
In descending order from top to bottom in $d = 1, 3$ and $5$, and from left to right in $d = 2$ and $4$, respectively.
In the case of the $R_x(\pi/2)$ gates, however, we cannot observe the violation of this expected order in $d = 2$ [as is the case for the $R_x(\pi)$ gate].
Namely, during the first idle phase, the $\ev*{\tilde{\sigma}_y}$ element of the Bloch vector mainly contributes to the fidelity.
Because the reservoir affects the fidelity during this phase through the decoherence process rather than the population-relaxation process, the tendency of the order is different compared to the $R_x(\pi)$ gates.
As depicted in Appendix~\ref{sec:appHPI}, the peculiar behavior for $\Omega = \infty$ or small $s$ found in the $R_x(\pi)$ sequences is not observed here, or rather, the expected order is obtained.
In $d = 4$, the fidelity is again determined mainly by $\ev*{\hat{\sigma}_z}$, and the oscillatory pattern of the populations again contributes to the development of the fidelities.
Similar to the $R_x(\pi)$ gates (Fig.~\ref{fig:idle_z} and Sec.~\ref{sec:resPI}), the intrasegment oscillations change the order for lower $s$ when the qubit state is close to the ground state, which is found in case (a) for reservoirs with $s = 1/4, 1/8$ and $1/14$. 

Overall, a violation of the expected order during the pulse application ($d = 1$, $3$, and $5$) is observed only in a smaller number of supercells compared to the case of $R_x(\pi)$ gates. The differences between initial and final states (Fig.~\ref{fig:Bloch}), and the length of the pulse duration may be responsible for this different behavior.

\begin{figure*}
    \includegraphics[width=\linewidth]{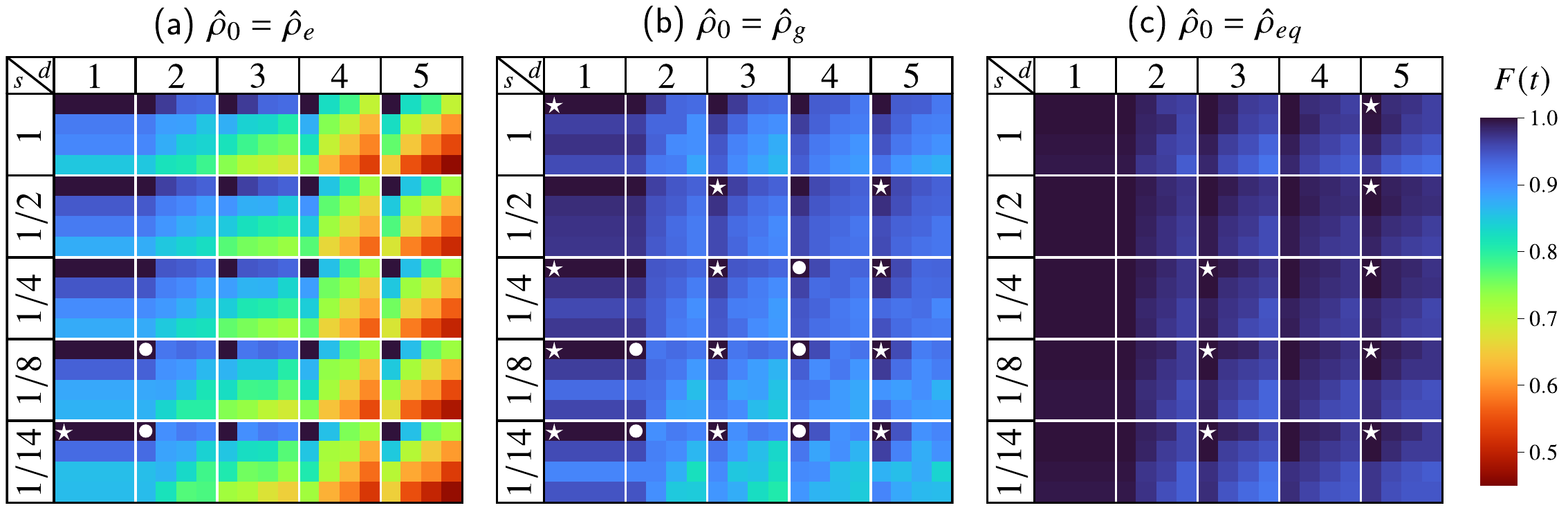}
    \caption{Heatmaps of the fidelity of the Hadamard ($H$) gates measured at the end of each phase $d$, with various spectral exponents $s$. The initial states are given by (a) $\hat{\rho}_e$, (b) $\hat{\rho}_g$, and (c) $\hat{\rho}_{eq}$, respectively. The parameter values of each cell are the same as Fig.~\ref{fig:HMap_PI}. The circle and star symbols in the upper-left corner of the supercell indicate the violation of the expected order, in the same way as Fig.~\ref{fig:HMap_PI}.
    \label{fig:HMap_H}}
\end{figure*}

\subsection{Hadamard ($H$) gates}
The third gate sequence that we analyze here consists of three Hadamard gates according to Eqs.~\eqref{eq:Upulse}--\eqref{eq:UpulseImp}
\begin{align}
    \mathcal{U}_\mathrm{p}\left(\frac{\pi}{2}, -\frac{\pi}{2}\right)
    \mathcal{U}_\mathrm{i}(\Delta t)
    \mathcal{U}_\mathrm{p}\left(\frac{\pi}{2}, \frac{\pi}{2}\right)
    \mathcal{U}_\mathrm{i}(\Delta t)
    \mathcal{U}_\mathrm{p}\left(\frac{\pi}{2}, -\frac{\pi}{2}\right)\, .
    \label{eq:seq_H}
\end{align}
Note that the virtual $Z$ gate~\cite{McKayPRA2017} is considered here.
In Fig.~\ref{fig:HMap_H}, we depict corresponding heatmaps of the fidelity, where the parameter values $s$, $d$, $\Omega$, and $\Delta t$ for each cell are the same as in Fig.~\ref{fig:HMap_PI} while the pulse duration is the same as the one in Fig.~\ref{fig:HMap_HPI}.

Overall, the fidelity is maximum in the case of an equilibrium initial preparation $\hat{\rho}_{eq}$, while it takes minimum values for the qubit being initially in an excited state $\hat{\rho}_{e}$:
The value $F(t) = 0.467$ for Ohmic reservoirs ($s=1$) in $d = 5$ with $\Omega /\omega_q = 1/3$ and $\omega_q \Delta t / \pi= 2$ is the worst for all the cells in Figs.~\ref{fig:HMap_PI}, \ref{fig:HMap_HPI}, and \ref{fig:HMap_H}.
This is mainly because of two factors, namely, the first pulse application ($0 < d \leq 1$) and the second idle phase ($3 < d \leq 4$).
As discussed in Sec.~\ref{sec:resHPI}, the rotation with an angle $\pi/2$ starting from the excited state is most subject to noise.
This tendency was found to be independent of the rotation axis.
In addition, at the beginning of the second idle phase the qubit ideally starts again from an excited state when it was prepared there before the first gate pulse. Since the relaxation process causes more detrimental effects on the excited state than on the ground state, as discussed in Sec~\ref{sec:resPI}, 
these two contributions add up to reduce the fidelity substantially.

The reason for the better performance starting from an initial state $\hat{\rho}_{eq}$ compared to that for $\hat{\rho}_g$ is the same as above for the $R_x(\pi)$ and $R_x(\pi/2)$ gates. In terms of the reservoir exponent $s$, it is true also for $H$ gates that the fidelity for $s = 1/2$ and $1/4$ is maximum while that with $s = 1$ and $1/14$ is minimum for fixed $d$, $\Omega$, and $\Delta t$. In fact, this tendency is here even more significant than in the case of $R_x(\pi/2)$ gates.

It is worth noting that at $d = 2$, a violation of the expected order for the fidelity is observed in the deep sub-Ohmic domain $s = 1/8$ and $1/14$ for initial states $\hat{\rho}_e$ and $\hat{\rho}_g$.
Here, the expected order is defined in the same way as in Sec.~\ref{sec:resHPI}.
As depicted in Appendix~\ref{sec:appH}, a significant decoherence that is not observed in the $R_x(\pi/2)$ gates contributes to this violation.
Namely, the different rotation axis leads to significantly different behavior during the first idle phase.
During the second idle phase starting with $\hat{\rho}_g$, the qubit is close to the ground state. This situation is similar to that in Fig.~\ref{fig:HMap_HPI}(a) at $d=4$:
Again the strongly non-Markovian behavior corresponding to oscillatory qubit dynamics for $s \leq 1/4$ induces a violation of the expected order, see Fig.~\ref{fig:HMap_H}(b) [cf.\ with Fig.~\ref{fig:HMap_HPI}(a)].

As for the pulse-application phase, a violation is observed in different cells in Fig.~\ref{fig:HMap_H} compared to Fig.~\ref{fig:HMap_HPI}.
The static phase of the external field $\phi$ induces differences in the appearance, as gate operations $\hat{R}_x(\pi/2)$ are used in Fig.~\ref{fig:HMap_HPI} while $\hat{R}_y(\pm\pi/2)$ in Fig.~\ref{fig:HMap_H}.

\subsection*{\textcolor{black}{Suggestion for experiments}}
As discussed above, the shortest operation time does not always result in the best performance in terms of the fidelity.
From our results, we suggest the following for experiments:
Monitoring the gate fidelity with respect to the pulse duration might be beneficial to find optimal gate times.
For multiqubit systems, extension of idling times for qubit synchronizations might lead to improvement of the performance.

The spectral exponent $s$ as well as the duration of the pulse application and idle phase ($\tau$ and $\Delta t$, respectively) plays a crucial role for the final fidelity.
\textcolor{black}{Although the spectral exponent is intrinsic to each material and circuit and cannot be controlled in general, methods to engineer the spectral density have been proposed in previous studies, especially for trapped-ion arrays~\cite{BiercukNATURE2009}.
Engineering of the reservoir is also found in a recent study of transmon qubits to observe transition from non-Markovian to Markovian behavior~\cite{GaikwadPRL2024}:
Experiments with varying the exponent $s$ might be feasible in the future.}

\begin{figure}
    \centering
    \includegraphics[width=0.93\linewidth]{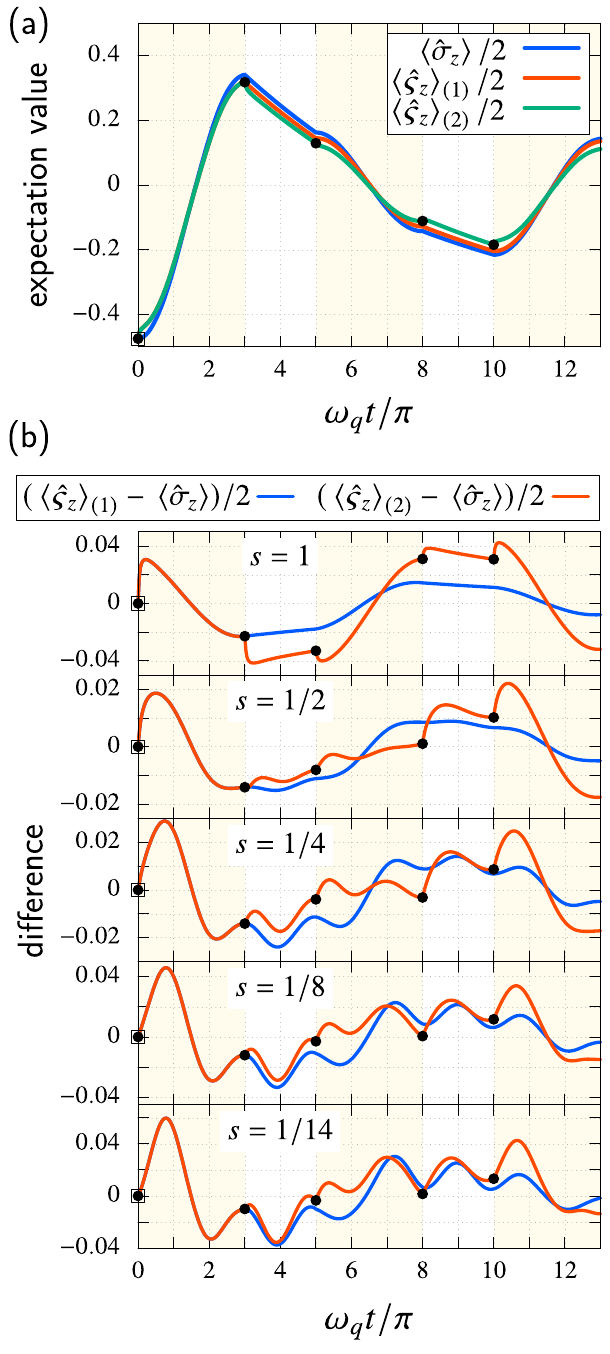}
    \caption{(a) Whole dynamics of the expectation value $\ev*{\hat{\sigma}_z(t)}/2$, $\ev*{\hat{\varsigma}_z(t)}_{(1)}/2$, and $\ev*{\hat{\varsigma}_z(t)}_{(2)}/2$ with the $R_x(\pi)$ sequence. The initial state is given by $\hat{\rho}_0 = \hat{\rho}_{eq}$, and $\Omega/\omega_q$ and $\omega_q \Delta t/\pi$ are set to $1/3$ and $2$, respectively. The shaded areas indicate the pulse-application phase, which corresponds to the schematic of Fig.~\ref{fig:pulseSequence}. We applied the projection operator $\mathcal{P}[\hat{\rho}_\mathrm{tot}(t)] = \mathrm{tr}_R \{\hat{\rho}_\mathrm{tot}(t)\} \otimes \hat{\rho}_{R, eq}$ at times indicated by the empty squares [at $t = 0$, $\ev*{\hat{\varsigma}_z(t)}_{(1)}$] and filled circles [$\ev*{\hat{\varsigma}_z(t)}_{(2)}$]. The dynamics without the projection operators $\ev*{\hat{\sigma}_z(t)}$ are also depicted as a reference. As a representative, the case for the Ohmic reservoir ($s = 1$) is depicted. (b) Time trace of the difference of the expectation value between the dynamics with the projection operators and exact dynamics.
    \label{fig:compare}}
\end{figure}

\section{Qubit--reservoir correlations} \label{sec:resCorrelation}
In this section we discuss in more detail means to monitor directly feedback effects from the reservoir onto the qubit dynamics (non-Markovianity) and demonstrate their significance. 

\subsection{Interphase correlations} \label{sec:interPhase}
Here, we study the limitation of the Born--Markov approximation through a deeper investigation of system--reservoir correlations.
Strictly, the dynamics of the qubit at a time $t$ are affected by its properties at previous times because of the finite-time (finite-frequency) retardation of the reservoir (non-Markovianity). In contrast, the Born--Markov approximation assumes an instantaneous interaction and can thus not describe corresponding correlations.
In our case, the qubit dynamics during a certain phase of a specific gate sequence are correlated with its dynamics during previous phases. We refer to these correlations as ``interphase correlations''.

In addition, when we consider a thermal initial state $\hat{\rho}_0 = \hat{\rho}_{eq}$, static system--reservoir correlations at time $t = 0$ emerge and also affect the future dynamics of the qubit.

To study these correlations, we conduct numerical simulations in which we ``decouple'' the system and reservoir by means of the projection operator $\mathcal{P}[\hat{\rho}_\mathrm{tot}(t)] = \mathrm{tr}_R\{\hat{\rho}_\mathrm{tot}(t)\}\otimes\hat{\rho}_{R, eq}$ at the end of each phase $d$ as well as at time $t=0$, and compare results of these simulations with those of the full dynamics, i.e., without projection operators. Note that the above projection operator $\mathcal{P}[\bullet]$ is the starting point to derive the Nakajima--Zwanzig equation~\cite{Alicki1987,Rivas2012}.
The same is true for the time-convolutionless (TCL) master equation ~\cite{Breuer2002}, which, as was pointed out in recent work, cannot describe interphase correlations~\cite{Tanimura2015,tanimuraJCP20}.
Considering that in the Born approximation one always assumes a factorization $\hat{\rho}_\mathrm{tot}(t) \simeq \hat{\rho}_S(t) \otimes \hat{\rho}_{R, eq}$, the introduced projection operator provides direct insight into the limitations of this approximate treatment. Within the HEOM, the application of this projection operator corresponds to the reset of the ADOs to $0$ at the time $t$, which is expressed by $\hat{\rho}_{\vec{m}\neq\vec{0}, \vec{n}\neq\vec{0}}(t) \to 0$.

In order to analyze these correlations in more detail, we not only consider the full dynamical expectation value $\ev*{\hat{\sigma}_z(t)}/2$ but also introduce $\ev*{\hat{\varsigma}_z(t)}_{(\alpha)}/2$ ($\alpha=1, 2$).
These describe $z$ elements of the Bloch vector during a pulse sequence with the following decoupling scheme: ($\alpha=1$) only one projection operator is applied at time $t=0$, and ($\alpha=2$) this operator is applied initially and at the end of each phase.
Corresponding time-dependent data are depicted in Fig.~\ref{fig:compare}(a) for an Ohmic reservoir as a representative.
As the initial state the equilibrium state, $\hat{\rho}_0 = \hat{\rho}_{eq}$, is chosen, and the parameter values are $\Omega/\omega_q = 1/3$ and $\omega_q \Delta t / \pi = 2$. 
\textcolor{black}{The shaded areas correspond to the pulse-application phase, in which the qubit ideally rotates from $\ket{0}$ ($\ev*{\hat{\sigma}_z}/2 = -0.5$) to $\ket{1}$ ($\ev*{\hat{\sigma}_z}/2 = 0.5$) for the first and third pulse (left and right shaded area), while $\ket{1}$ to $\ket{0}$ for the second pulse (middle shaded area).}

To stress contributions of the interphase correlations, differences between expectation values with projection operators applied and the exact ones are shown in Fig.~\ref{fig:compare}(b) for various spectral exponents $s$.
In the Ohmic case ($s=1$), pronounced step-like deviations are observed right after the application of the projection operator, which tends to be smoother in the moderate sub-Ohmic domain.
Namely, for $s=1$, owing to the high-frequency modes of the reservoir, a fast reconfiguration from $\hat{\rho}_{R, eq}$ towards the correlated equilibrium state occurs.
As the exponent $s$ becomes smaller, the intensity of the spectral noise power $S_\beta(\omega)$ in the high-frequency region gradually decreases.
This results in a slower reconfiguration process.
However, deviations increase again in the deep sub-Ohmic domain ($s<1/4$), and we attribute this increase to the strongly growing portion of low-frequency modes.

More specifically, without projection onto the bare equilibrium state of the reservoir $\hat{\rho}_{R,eq}$ and with the parameter values chosen here, we observed a monotonic decay of the Bloch vector irrespective of the exponent $s$ during the idle phases (cf.\ Fig.~\ref{fig:idle_asym} in Appendix~\ref{sec:appAsym}). The oscillatory behavior seen in Fig.~\ref{fig:compare}(b) for the dynamics with projection must be thus attributed to the instantaneous change of the reservoir to $\hat{\rho}_{R,eq}$ each time in which the projection operator is applied. The destruction of qubit--reservoir correlations induces for lower spectral exponents $s=1/8$ and $1/14$ a sluggish oscillatory response to reestablish them. Because the enhancement of this oscillatory pattern is accompanied by a quantitative increase of deviations, we conclude that these correlate with each other.


For the dynamics of $\ev*{\hat{\varsigma}_z(t)}_{(1)}$, all interphase correlations are taken into account, while the static initial system--reservoir correlations are not. Notably, as seen in Fig.~\ref{fig:compare}(b), deviations to the exact dynamics can be observed even during the \emph{second} idle phase. This clearly shows the impact of static qubit--reservoir correlations even in the long-time regime (here time span $8 \leq \omega_q t/\pi \leq 10$). 

The general conclusion we draw from Fig.~\ref{fig:compare}(b) is that the static system--reservoir correlation at the initial time as well as the interphase correlations significantly contributes to the qubit dynamics and directly affect quantitatively predictions for gate performances. For the precise study of the qubit dynamics during pulse sequences, methods that go beyond the Born approximation must be applied.

\begin{figure}
    \centering
    \includegraphics[width=0.94\linewidth]{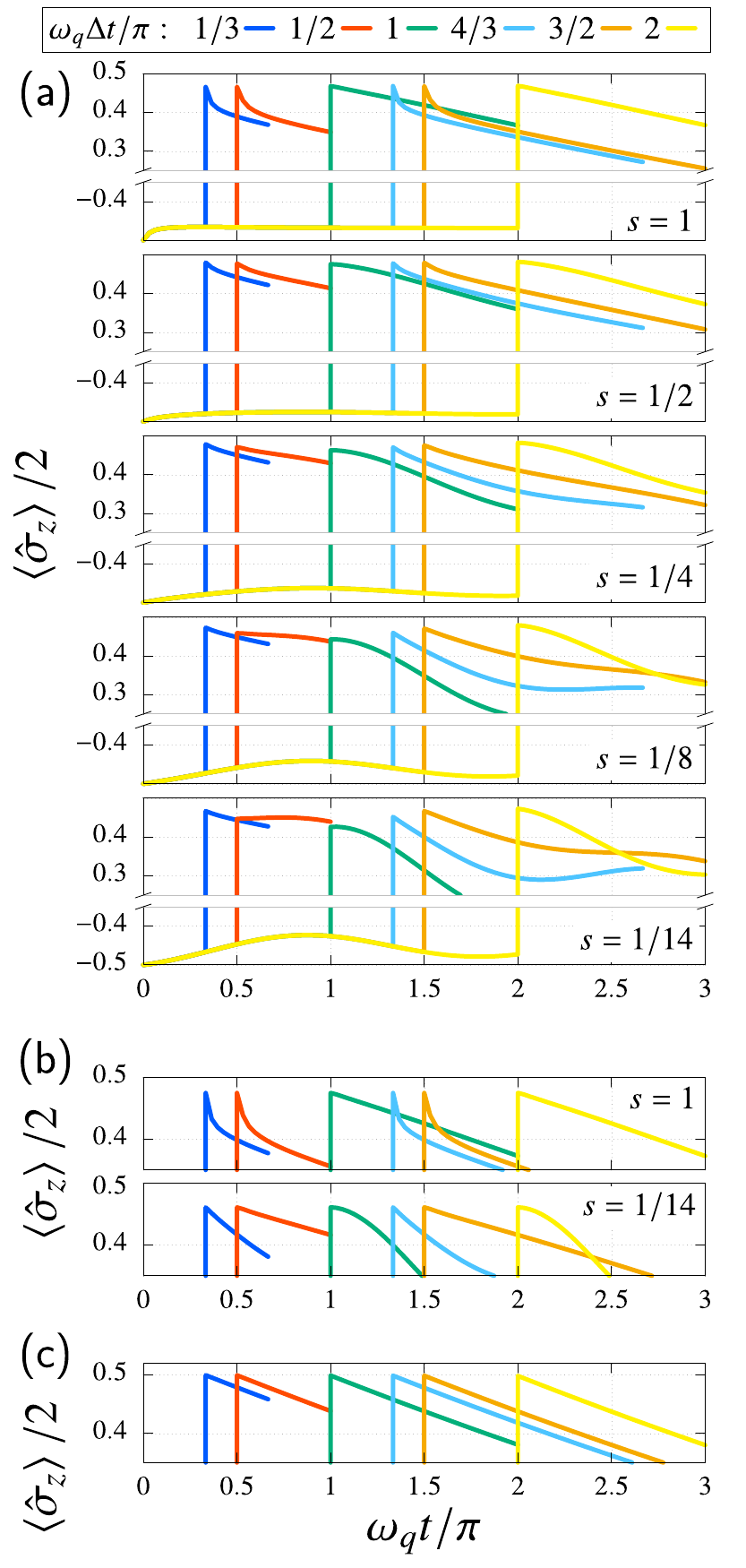}
    \caption{(a) Dynamics of the expectation value $\ev*{\hat{\sigma}_z(t)}/2$ during the sequence of three $R_x(\pi)$ gates, with various $\Delta t$ and a fixed $\Omega$ ($\Omega / \omega_q = \infty$). The vertical lines at the beginning and end of the sequence, which corresponds to the first and third pulse, are omitted. The initial states are the excited state, $\hat{\rho}_0 = \hat{\rho}_e$. (b) Dynamics of $\ev*{\hat{\sigma}_z(t)}/2$ with the two idle phases interleaved with one impulsive pulse [Eq.~\eqref{eq:PeriodReference}]. The initial states are the equilibrium state, $\hat{\rho}_S(0) = \hat{\rho}_{eq}$. As representatives, the dynamics with $s = 1$ and $1/14$ are displayed. The dynamics of the first idle phase are not depicted, because the expectation value $\ev*{\hat{\sigma}_z(t)}$ takes a same value during this phase, as inferred by the equilibrium initial states. (c) Same dynamics as case (a), but the dynamics are calculated with the Lindblad equation [Eq.~\eqref{eq:Lindblad}]. The dynamics during the first idle phase are not depicted. The results are independent of the spectral exponent $s$.
    \label{fig:periodic}}
\end{figure}

\subsection{Periodic behavior for impulsive-pulse sequences} \label{sec:periodic}
Here, we focus on the dynamics with the $R_x(\pi)$ sequence with the impulsive pulses, i.e., $\Omega = \infty$ and zero pulse-duration $\tau = \pi / \Omega = 0$. Figure~\ref{fig:periodic}(a) displays the dynamics of the expectation value $\ev*{\hat{\sigma}_z(t)}/2$ with $\hat{\rho}_e$ chosen as the initial state. The vertical lines correspond to the second pulse application while, for the sake of clarity, the vertical lines corresponding to the first and third pulse application are omitted (see Fig.~\ref{fig:pulseSequence} with segments $1 < d \leq 4$ with two idle phases).
The duration of the idle phase after the impulsive gates $\Delta t$ is varied from $\omega_q\Delta t/\pi=1/3$ to $\omega_q\Delta t/\pi=2$. 

When one compares the dynamics of $\Delta t/\pi$ with the $\pi$-shifted data (for example, $\omega_q \Delta t/\pi=1/2$ and $\omega_q \Delta t/\pi=3/2$), one observes a clear periodicity in the short-time region right after the beginning of the second idle phase in the cases $s = 1$ and $1/2$. Notably, this periodicity is not seen anymore for small spectral exponents (see e.g.,\ $s=1/14$ for $\omega_q \Delta t/\pi=1/2$ versus $\omega_q \Delta t/\pi=3/2$).

To better understand this behavior, we consider the dynamics of the qubit according to a reduced sequence of the form
\begin{align}
    \mathcal{U}_\mathrm{i}(\Delta t) \mathcal{U}_\mathrm{imp}(\pi, 0)
    \mathcal{U}_\mathrm{i}(\Delta t),
    \label{eq:PeriodReference}
\end{align}
taking now $\hat{\rho}_0 = \hat{\rho}_{eq}$ as the initial state (rather than the excited one as above).
The expectation value $\ev*{\hat{\sigma}_z(t)}/2$ does not change during the first idle phase because of the equilibrium state;
the right $\mathcal{U}_\mathrm{i}(\Delta t)$ in Eq.~\eqref{eq:PeriodReference} is introduced only to shift the time of the pulse application, and we focus on the dynamics during the second idle phase [left $\mathcal{U}_\mathrm{i}(\Delta t)$ in Eq.~\eqref{eq:PeriodReference}].
The extreme cases of Ohmic ($s = 1$) and deep sub-Ohmic ($s = 1/14$) reservoirs are shown in Fig.~\ref{fig:periodic}(b).
Now, the $\pi$ periodicity is observed also in the case for $s = 1/14$.
The reason for this different behavior can be traced back to the time dependence of the density operators:
In the equilibrium state, the \emph{off-diagonal} elements of the reduced density operator (RDO) are negligibly small so that the application of an impulsive $\pi$-pulse $\mathcal{U}_\mathrm{imp}(\pi, 0)$ to the RDO is a time-independent transformation.
By contrast, because the off-diagonal elements of the ADOs are not necessarily zero, the unitary transformation of the ADOs in Eq.~\eqref{eq:UpulseImp} is time dependent owing to the term $\hat{R}_z(\pm\omega_q t)$.
Hence, we conclude that since off-diagonal elements of all the ADOs are almost invariant in time when the total system is in its equilibrium state, for the $\pi$ rotations considered here, the qubit exhibits a $\pi$-periodical behavior for all spectral exponents.
This is exemplified in Fig.~\ref{fig:periodic}(b).
Initialized in the excited state, however, this only applies to $s=1$ and $1/2$ in Fig.~\ref{fig:periodic}(a), when the total state stays very close to the equilibrium state during the first idle phase.
By contrast, when the reservoir becomes more sub-Ohmic, during the first idle phase the qubit stays far from full equilibrium and oscillatory behavior emerges with no $\pi$ periodicity.

What happens when one simulates this situation within a Born--Markov treatment? This is seen in
Fig.~\ref{fig:periodic}(c): For the application of the impulsive pulse, the unitary transformation, Eq.~\eqref{eq:UpulseImp}, is applied to the RDO, so that during the idle phases the dynamics follow from the Lindblad equation
\begin{align}
    & \frac{\partial}{\partial t} \hat{\rho}_S(t) \\
    = & -\frac{i}{\hbar} \left[\hat{H}_S(\Omega=0, \phi;t), \hat{\rho}_S(t) \right] \\
    & + 2\pi S_\beta(\omega_q) \left(\hat{\sigma}_{-} \hat{\rho}_S(t)
    \hat{\sigma}_{+} - \frac{1}{2} \left\{\hat{\sigma}_{+}
    \hat{\sigma}_{-}, \hat{\rho}_S(t)\right\} \right) \\
    & + 2\pi S_\beta(-\omega_q) \left(\hat{\sigma}_{+} \hat{\rho}_S(t)
    \hat{\sigma}_{-} - \frac{1}{2} \left\{\hat{\sigma}_{-}
    \hat{\sigma}_{+}, \hat{\rho}_S(t)\right\} \right)\, . \\
    \label{eq:Lindblad}
\end{align}
Here, $[\bullet, \bullet]$ and $\{\bullet, \bullet\}$ denote the commutator and anticommutator, respectively.
The parameter values for $S_\beta(\omega_q)$ are identical to those of the HEOM calculations and independent of $s$. Note that we have ignored the Lamb shift here because it does not contribute to the dynamics of the diagonal elements. We do not depict the dynamics during the first idle phase in Fig.~\ref{fig:periodic}(c) because the expectation value $\ev*{\hat{\sigma}_z(t)}/2$ in the equilibrium state of the bare system $\hat{\rho}_{S, eq}$ is very close to $-0.5$ and significant changes are not observed during this phase. In terms of the numerical calculation, the ADOs are always zero within the framework of the Lindblad equation, and therefore the time-dependent properties of the pulse application cannot be described. For this reason, the dynamics of the RDO during the second idle phase exhibits almost the \emph{same linear relaxation} behavior irrespective of $\Delta t$.

We further mention that in the exact treatment with the HEOM, the total equilibrium state is sensitive to the direction of the rotation axis of the qubit because of the system--reservoir coupling. Namely, 
contributions of the terms $\hat{R}_z(\pm\omega_q t)$ in Eq.~\eqref{eq:UpulseImp} can be expressed through the initial phase of the external field $\phi$, and the change of the time duration $\Delta t$ corresponds to the change of the direction of the rotation axis. This sensitivity is absent in the framework of the Born--Markov approximation approach and can thus not been predicted by this treatment. It is a clear signature of qubit--reservoir correlations.

We note in passing that in a previous study different dynamics of the RDO caused by different rotation axes were numerically predicted~\cite{NakamuraPRB2024}.
The cause of these differences is the same as the one for the $\pi$ periodicity in this study.
For the dependence of the dynamics on $\Delta t$ with a finite amplitude $\Omega$, see Appendix~\ref{sec:appAsym}.

\subsection*{\textcolor{black}{Feasibility of experiments}}
There are growing activities in observing non-Markovian effects in specific set-ups, for example, we refer to Ref.~\onlinecite{GaikwadPRL2024} for a most recent development.
In previous studies, experimental protocols to observe signatures of the non-Markovianity were proposed~\cite{GulasciPRB2022,NakamuraPRB2024}.
Here, we point out that the above protocols discussed in Secs.~\ref{sec:interPhase} and \ref{sec:periodic} may provide  additional means:
The interphase correlations can be detected if we can reset the reservoir to the decoupled equilibrium state $\hat{\rho}_{R, eq}$.
This might be achievable by application of external fields, which is similar to the pulse application to the system qubits.
The protocol for the periodic behavior is much easier:
When we can prepare approximately impulsive pulses with a negligible width, then we simply vary the duration of the idle phase to obtain corresponding results.

\section{Summary and Conclusions}\label{sec:conclusion}
In this paper, high-precision quantitative predictions are provided for various sequences of single-qubit gate operations for a broad class of thermal environments. Reservoirs with spectral densities of the form $J_s(\omega)\propto \omega^s$ are considered, i.e., from the Ohmic ($s=1$) to the deep sub-Ohmic ($s\ll 1$) domain, thus covering prominent noise sources for superconducting qubits such as electromagnetic fluctuations, two-level fluctuators, and quasiparticle noise. As representative applications, gate sequences are chosen to consist of three pulses ($\pi$, $\pi/2$, and $H$ gate) of varying amplitudes separated by two idle phases of varying lengths. In this way, we are able to unfold a detailed and comprehensive picture of the dynamics and performance of major gate sequences for realistic superconducting circuit implementations in domains, where perturbative treatments (Lindblad, Redfield) fail.
Our paper clearly demonstrates the necessity to invoke highly advanced simulations techniques such as the HEOM to provide a detailed understanding of the intricate qubit--reservoir dynamics and to deliver quantitative predictions for complex gate operations matching the growing accuracy achieved experimentally.

The main results can be summarized as follows:
\begin{enumerate}
    \setlength{\leftskip}{1.2em}
    \item In the temperature domain, where superconducting qubits are operated, retardation effects of the reservoir induce long-range correlations during gate sequences, particularly between subsequent idle phases. This impact grows for reservoirs with more prominent sub-Ohmic characteristics (low to moderate frequency noise compared to qubit transition frequencies). 
    \item By varying parameters of gate sequences (amplitude/pulse duration, duration of idle phase) and depending on spectral exponents of reservoirs, we found a nonmonotonous pattern for gate fidelities for all three initial preparations (ground, excited, and thermal state). In contrast to simple expectations, the recovery of fidelities in subsequent idle phases originates from non-Markovian dynamics of $\hat{\rho}_S(t)$. By choosing proper parameters in each of these cases, our simulations lay the foundation for optimizing gate performances.
    \item In most cases, we observed that fidelities for gate sequences starting from the qubit's ground state or thermal state exceed those starting from the excited state. 
    \item Fidelities after the final pulse, decisive for the overall gate performance, strongly depend on the loss or recovery of fidelities during all preceding idle phases.
    \item The rotation axis of qubit gate operations on the Bloch sphere relative to the qubit--reservoir coupling has substantial influence on gate performances, as explicitly demonstrated for $R_x(\pi/2)$ and $H$ gates. 
    \item Long-range qubit--reservoir correlations were shown to induce interphase correlations during gate sequences depending on the relative portion of low-frequency modes in the reservoirs. Monitoring the qubit's population dynamics upon application of impulse $R_x(\pi)$ gates interleaved by idle phases of varying lengths allows us to reveal directly the significance of non-Markovian feedback in actual circuits. The latter appears to be imprinted in periodicities that are predicted to occur when comparing the qubit dynamics for idle phases $\omega_q \Delta t$ that differ by multiples of $\pi$.
     \item The developed and applied rigorous numerical simulation technique is highly efficient and very versatile so that it can be used in the laboratory to directly guide optimized designs of circuitries and gate pulse shapes.
     For example, the results reported here for a single run (a single cell in Fig.~\ref{fig:HMap_PI}) were obtained on a personal computer (Intel Core i9 CPU with $10$ cores) within a few seconds (Ohmic case) up to a few hours (very deep sub-Ohmic case). This can be further improved by implementing matrix product state (MPS) techniques within the FP-HEOM \cite{XuPRL2022}.
 \end{enumerate}
 
 In this paper, we restricted ourselves to the simulations of single-qubit gates. The pulse shape was also restricted to a rotating external field, with the ideal switching given by a step function. This can easily be extended by taking into account derivative removal adiabatic gates (DRAG)~\cite{MotzoiPRL2009} with a finite rise time. Leakage effects of pulses to the second and higher qubit excited states have also not been considered.
 \textcolor{black}{Studies of multiqubit dissipative systems must be conducted in the same line as this study.}
 However, because of its unique efficiency combined with its versatile applicability the presented numerical approach allows us to investigate these topics as well. Future extensions include two-qubit gate operations, circuitries with more complex impedances, and multiqubit correlations.

\section*{Acknowledgement}
The authors would like to thank J.~T.~Stockburger and M.~Xu for fruitful discussions and numerical assistance. This work was supported by the BMBF through QSolid and the Cluster4Future QSens (Project QComp) and the DFG through Grant No. AN336/17-1 (FOR2724). Support by the state of Baden-W{\"u}rttemberg through bwHPC
and the German Research Foundation (DFG) through grant No. INST 40/575-1 FUGG (JUSTUS 2 cluster) is also acknowledged.

\appendix
\section{Rotation operators and time evolution} \label{sec:appRotatingFrame}
In this appendix, we discuss in detail the rotation operators and Hamiltonian.
Following the main text, we move to the rotating frame with the rotation axis and angular frequency given by $z$ and $\omega_\mathrm{ex}$, respectively.
The system Hamiltonian is transformed as 
\begin{align}
    & \tilde{H}_S(\Omega, \phi) \\ 
    = & \hat{R}_z(-\omega_\mathrm{ex}t) \hat{H}_S(\Omega, \phi;t)
    \hat{R}_z(\omega_\mathrm{ex}t)
    + i \hbar \dot{\hat{R}}_z(-\omega_\mathrm{ex}t)
    \hat{R}_z(\omega_\mathrm{ex}t)  \\
    = & \frac{\hbar(\omega_q-\omega_\mathrm{ex})}{2}\hat{\sigma}_z
    + \frac{\hbar\Omega}{2}\left(\hat{\sigma}_x \cos \phi + \hat{\sigma}_y \sin \phi\right)\, .
\end{align}
Note that the system Hamiltonian in the rotating frame is time independent, and we omit the argument $t$ for $\tilde{H}_S$. When we choose $\omega_q$ for the frequency $\omega_\mathrm{ex}$, as discussed in the main text, the first term of this equation vanishes. Under this condition, we can easily confirm the relations $\tilde{H}_S(\Omega, 0) = \hbar\Omega \hat{\sigma}_x/2$ and $\tilde{H}_S(\Omega, \pi/2) = \hbar\Omega \hat{\sigma}_y/2$. This indicates that we can express the rotation operator with the time-evolution operator as
\begin{align}
    \hat{R}_x(\theta) & = \exp\left[-\frac{i}{\hbar} \tilde{H}_S(\Omega, 0) \tau \right]\, , \\
    \hat{R}_y(\theta) & = \exp\left[-\frac{i}{\hbar} \tilde{H}_S\left(\Omega, \frac{\pi}{2}\right) \tau \right]\, ,
\end{align}
where the frequency of the external field $\omega_\mathrm{ex}$ is set to $\omega_q$. Note that the pulse duration $\tau$ is determined with the condition $\tau = \theta / \Omega$.
Accordingly, we can obtain the rotation operators with the negative angle with $\phi = \pi$ for the $x$-axis rotation and $\phi = -\pi/2$ for the $y$-axis rotation.
A rotation operator about an arbitrary axis in the $x$--$y$ plane is expressed as $\hat{R}_\phi(\theta) = \exp[-i \theta (\hat{\sigma}_x \cos\phi + \hat{\sigma}_y \sin\phi)/2]$ and corresponds to the time-evolution operator $\exp[-i \tilde{H}_S(\Omega, \phi) \tau / \hbar]$.

By using the relation
\begin{align}
    & \exp\left[-\frac{i}{\hbar} \tilde{H}_S(\Omega, \phi) \tau \right] \\
    = & \hat{R}_z(-\omega_q [t+\tau])
    \mathcal{T}_+\exp\left[-\frac{i}{\hbar} \int_{t}^{t+\tau} \hspace{-1em}dt' \hat{H}_S(\Omega, \phi;t') \right] \\
    & \times \hat{R}_z(\omega_q t)\, ,
\end{align}
which transforms the time-evolution operator in the rotating frame to that in the laboratory frame, the rotation operator in the rotating frame is expressed as
\begin{align}
    \tilde{\rho}_S(t+\tau) = & \hat{R}_\phi(\theta) \tilde{\rho}_S(t) \hat{R}_\phi(-\theta) \\
    = & \exp\left[-\frac{i}{\hbar} \tilde{H}_S(\Omega, \phi) \tau \right] \tilde{\rho}_S(t)
    \exp\left[\frac{i}{\hbar} \tilde{H}_S(\Omega, \phi) \tau \right] \\
    = & \hat{R}_z(-\omega_q[t+\tau]) \mathcal{T}_+\exp[-\frac{i}{\hbar}\int_{t}^{t+\tau}\hspace{-1.5em}dt'
    \hat{H}_S(\Omega, \phi;t')] \\
    & \times \hat{R}_z(\omega_qt)\tilde{\rho}_S(t) \hat{R}_z(-\omega_qt) \\
    & \times \mathcal{T}_-\exp\left[\frac{i}{\hbar} \int_{t}^{t+\tau} \hspace{-1.5em}dt'
    \hat{H}_S(\Omega, \phi;t') \!\right] \!\!\hat{R}_z(\omega_q[t+\tau])\, ,
    \label{eq:ApplRotationOpe}
\end{align}
where the operators $\mathcal{T}_\pm$ are the positive and negative time-ordering operator.
Using the relation $\hat{\rho}_S(t) = \hat{R}_z(\omega_qt)\tilde{\rho}_S(t)\hat{R}_z(-\omega_qt)$, we obtain the rotation operator corresponding to the time evolution with the Hamiltonian in Eq.~\eqref{eq:H_S} in the laboratory frame.
We utilize the laboratory frame to introduce the reservoir operators.
Accordingly, the system Hamiltonian in Eq.~\eqref{eq:ApplRotationOpe} is replaced with $\hat{H}_\mathrm{tot}(\Omega, \phi;t) = \hat{H}_S(\Omega, \phi;t) - \hat{V}\hat{X} + \hat{H}_R$, and the time-evolution operator is expressed by Eq.~\eqref{eq:timeEvo}.
The density operator is also replaced with $\hat{\rho}_\mathrm{tot}(t)$.
There is no reason to assume that the reservoir rotates about the $z$ axis at the angular frequency $\omega_q$, and it is plausible that the qubit system couples with the reservoir in this form. 

Next, we consider the application of impulsive pulses.
As mentioned in the main text, we can ignore the system--reservoir coupling term when we consider the impulsive pulses. Equation~\eqref{eq:ApplRotationOpe} is then rewritten as
\begin{align}
    \tilde{\rho}_\mathrm{tot}(t+\tau)
    = & \exp \left[-\frac{i}{\hbar}
    \left(\tilde{H}_S(\Omega, \phi)
    + \hat{H}_R \right) \tau\right] \tilde{\rho}_\mathrm{tot}(t) \\
    & \times \exp\left[\frac{i}{\hbar}
    \left(\tilde{H}_S(\Omega, \phi)
    + \hat{H}_R \right) \tau\right] \\
    = & e^{-i\hat{H}_R \tau /\hbar} \hat{R}_\phi(\theta) \tilde{\rho}_\mathrm{tot}(t) 
    \hat{R}_\phi(-\theta) e^{i\hat{H}_R \tau/\hbar}\, .
\end{align}
Note that the transformation with $\hat{R}_z(-\omega_qt)$ does not change the reservoir Hamiltonian.
For the impulsive pulse, we take the limits $\Omega \to \infty$ and $\tau \to 0$, keeping $\theta$ a finite fixed value. With this operation, we obtain the equation for the application of the impulsive pulse in the rotating frame as follows:
\begin{align}
    \tilde{\rho}_\mathrm{tot} (t) \leftarrow & \hat{R}_\phi(\theta) \tilde{\rho}_\mathrm{tot}(t) \hat{R}_\phi(-\theta)\, .
\end{align}
When we go back to the laboratory frame from the rotating frame, we obtain Eq.~\eqref{eq:UpulseImp}.

\section{Details of the HEOM} \label{sec:appHEOM}
In this appendix, we illustrate the detailed derivation of the HEOM.
The total Hamiltonian consisting of the system and reservoir is described with
\begin{align}
   & \hat{H}_\mathrm{tot}(\Omega, \phi;t)\\
   = & \hat{H}_S(\Omega, \phi;t)  - \hat{V}\hat{X} + \hat{H}_R \\
   = &\hat{H}_S(\Omega, \phi;t)
   -\hat{V}\sum_j c_j \hat{x}_j
   + \sum_j \left(\frac{\hat{p}_j^2}{2m_j} 
   + \frac{1}{2} m_j \omega_j^2 \hat{x}_j^2 \right)\, .
   \label{eq:H_CL}
\end{align}
The reservoir is represented by an infinite number of harmonic oscillators, and $\hat{p}_j, \hat{x}_j, m_j$, and $\omega_j$ are the momentum, position, mass, and angular frequency of the $j$th bath, respectively.
The coupling strength between the system and $j$th bath is given by $c_j$, which defines the spectral density as
\begin{align}
    J(\omega) = \sum_{j} \frac{c^2_j}{2m_j\omega_j} \delta(\omega-\omega_j)\, .
\end{align}
The system part of the coupling $\hat{V}$ is set to $\hbar \hat{\sigma}_x$ as discussed in the main text.

To obtain the equation for the open quantum dynamics without any approximations, we exploit the Feynman--Vernon path integral representation.
The reduced density operator (RDO) of the system $\hat{\rho}_S(t) = \mathrm{tr}_{R}\{e^{-i\hat{H}_\mathrm{tot}t/\hbar} \hat{\rho}_\mathrm{tot}(0) e^{i\hat{H}_\mathrm{tot}t/\hbar}\}$ is expressed as
\begin{align}
    & \mel{\mu}{\hat{\rho}_S(t)}{\mu'}  \\
    = & \int \frac{d^2 \mu_i d^2 \mu'_i}{N(\mu_i)N(\mu'_i)}
    \int_{\mu(0)=\mu_i}^{\mu(t)=\mu}\hspace{-5ex}\mathcal{D}[\mu(\cdot)]
    \int_{\mu'(0)=\mu'_i}^{\mu'(t)=\mu'}
    \hspace{-5ex}\mathcal{D}[\mu'(\cdot)]
     \\
    & \times \exp\left[\frac{i}{\hbar}\int_{0}^{t} dt' L_{S}(\mu;t')
    -\frac{i}{\hbar}\int_{0}^{t} dt' L_{S}(\mu';t')\right]  \\
    & \times \mel{\mu_i}{\hat{\rho}_S(0)}{\mu'_i}
    \mathscr{F}[\mu, \mu'; t]\, .
\end{align}
Here, we consider the spin-coherent state $\ket{\mu}$~\cite{SpinCoherent}, and $L_S(\mu; t)$ is the Lagrangian of the system. The functional $\mathscr{F}[\mu, \mu';t]$ is the influence functional, which is given by
\begin{align}
    & \mathscr{F}[\mu, \mu';t]  \\
    = & \exp\biggl[-\frac{1}{\hbar^2}
    \int_{0}^{t}dt'\int_{0}^{t'} dt'' V^\times(\mu, \mu'; t')  \\
    & \times \{C(t'\!-\!t'')V(\mu;t'')
    -C^{*}(t'\!-\!t'')V(\mu';t'')\} \biggr]
    \label{eq:IF1} \\
    = & \exp\biggl[-\frac{1}{\hbar^2}
    \int_{0}^{t}dt'\int_{0}^{t'} dt'' V^\times(\mu, \mu'; t')  \\
    & \times \{C'(t'\!-\!t'')V^\times(\mu, \mu';t'')
    +iC''(t'\!-\!t'')V^\circ(\mu, \mu';t'') \biggr]\, .
    \label{eq:IF2}
\end{align}
The quantity $1/N(\mu) = 2/\{\pi(1+|\mu|^2)^2\}$ is the normalization factor for the spin-coherent states.
The function $V(\mu;t)$ is the path-integral representation of the operator $\hat{V}$, and $V^\times(\mu, \mu';t) = V(\mu;t) - V(\mu';t)$ and $V^\circ(\mu, \mu';t) = V(\mu;t) + V(\mu';t)$ are the corresponding commutator and anticommutator, respectively.
The Lagrangian of the system $L_S(\mu; t)$ is also defined in the path-integral representation.
For more details of the path integral in the spin-coherent representation, we refer the readers to Ref.~\onlinecite{Nakamura18PRA}.
The function $C^*(t)$ indicates the complex conjugate of $C(t)$, and we have utilized the relation $C^*(t) = C(-t)$ to obtain the expression of Eq.~\eqref{eq:IF2}.
The real and imaginary part of the two-time correlation function is defined as $C(t) = C'(t) + iC''(t)$. Here, we assume the factorized initial state $\hat{\rho}_\mathrm{tot}(0) = \hat{\rho}_S(0)\otimes\hat{\rho}_{R, eq}$.

We express the two-time correlation function $C(t)$ with the complex-valued exponential functions [Eq.~\eqref{eq:CF}]. The original free-pole HEOM (FP-HEOM)~\cite{XuPRL2022} is based on the representation of Eq.~\eqref{eq:IF1}, and its form is expressed in Eq.~\eqref{eq:HEOM}, where the auxiliary density operators (ADOs) are not Hermitian operators. In this paper, we derive the HEOM with Hermitian ADOs to reduce the computational costs.
To achieve this goal, we utilize the representation of Eq.~\eqref{eq:IF2} and the generalized form of the HEOM~\cite{IkedaJCP2020}.

First, we expand Eq.~\eqref{eq:CF} in the following form:
\begin{align}
    C(t) = & \sum_{k=1}^{K} e^{-\gamma_k t}
    \{d'_k \cos \omega_k t + d''_k \sin \omega_k t\}  \\
    & + i \sum_{k=1}^{K} e^{-\gamma_k t}
    \{d''_k \cos \omega_kt - d'_k \sin \omega_k t\}  \\
    = & \sum_{k=1}^{K} \{\phi_k(t) + i \psi_k(t)\}\, .
\end{align}
Here, we have introduced the real and imaginary part of the coefficient $d_k$ as $d_k = d'_k + i d''_k$.
Utilizing the superoperator
\begin{align}
    \Theta_k(\mu, \mu';t, s) = 
    & \phi_k(t-s)\frac{-i}{\hbar}V^\times(\mu, \mu';s)  \\
    & + \psi_k(t-s)\frac{1}{\hbar}V^\circ(\mu, \mu';s)
    \label{eq:theta}
\end{align}
for $k = 1, \ldots, K$, we can express the influence functional in Eq.~\eqref{eq:IF2} as 
\begin{align}
    & \mathscr{F}[\mu, \mu';t]  \\
    = & \exp\left[\int_{0}^{t} \!\!dt' \int_{0}^{t'}\!\!dt''
    \frac{-i}{\hbar}V^\times(\mu, \mu';t') \sum_{k=1}^{K}
    \Theta_k(\mu, \mu';t', t'')\right]\, . 
    \label{eq:IF3}
\end{align}
By introducing $\bar{\Theta}_k(\mu, \mu';t, s)$ as
\begin{align}
    \bar{\Theta}_k(\mu, \mu';t, s) = & \bar{\phi}_k(t-s)
    \frac{-i}{\hbar}V^\times(\mu, \mu';s)  \\
    & + \bar{\psi}_k(t-s) \frac{1}{\hbar}V^\circ(\mu, \mu';s)\, ,
\end{align}
where $\bar{\phi}_{k}(t)$ and $\bar{\psi}_{k}(t)$ are given by
\begin{align}
    \bar{\phi}_{k}(t) = &e^{-\gamma_k t}
    \{-d'_k\sin \omega_k t + d''_k \cos \omega_kt\}\, , \\
    \bar{\psi}_{k}(t) = & -e^{-\gamma_k t}
    \{d''_k\sin \omega_kt + d'_k \cos \omega_kt\}\, ,
\end{align}
we obtain the following equation:
\begin{align}
    \begin{aligned}
        \frac{\partial}{\partial t} \Theta_k(\mu, \mu';t, s)
        & \!=\! -\gamma_k \Theta_k(\mu, \mu';t, s)
        \!+\! \omega_k \bar{\Theta}_k(\mu, \mu';t, s)\, , \\
        \frac{\partial}{\partial t} \bar{\Theta}_k(\mu, \mu';t, s) 
        & \!=\! -\gamma_k \bar{\Theta}_k(\mu, \mu';t, s)
        \!-\! \omega_k \Theta_k(\mu, \mu';t, s)\, .
        \end{aligned} \quad
    \label{eq:dTheta}
\end{align}
\begin{widetext}
    Defining the ADO and its time derivative as
    \begin{align}
        \mel{\mu}{\hat{\rho}_{\vec{m}, \vec{n}}(t)}{\mu'}
        = & \int \frac{d^2 \mu_i d^2 \mu'_i}{N(\mu_i)N(\mu'_i)}
        \int_{\mu(0)=\mu_i}^{\mu(t)=\mu}
        \hspace{-5ex}\mathcal{D}[\mu(\cdot)]
        \int_{\mu'(0)=\mu'_i}^{\mu'(t)=\mu'}
        \hspace{-5ex}\mathcal{D}[\mu'(\cdot)]
         \\
        & \times \prod_{k=1}^{K} \frac{1}{\sqrt{m_k!n_k!}}
        \left(\int_{0}^{t} dt'' \Theta_k(\mu, \mu';t, t'')
        \right)^{m_k}
        \left(\int_{0}^{t} dt''
        \bar{\Theta}_k(\mu, \mu';t, t'')\right)^{n_k} 
        \\  
        & \times \exp\left[\frac{i}{\hbar}\int_{0}^{t} dt' L_{S}(\mu;t')
        -\frac{i}{\hbar}\int_{0}^{t} dt' L_{S}(\mu';t')\right]
        \mel{\mu_i}{\hat{\rho}_S(0)}{\mu'_i}
        \mathscr{F}[\mu, \mu'; t]
    \end{align}
    and
    \begin{align}
        & \frac{\partial \hat{\rho}_{\vec{m}, \vec{n}}(t)}
        {\partial t}
        = \int 
        \frac{d^{2}\mu d^{2}\mu'}{N(\mu)N(\mu')} \ket{\mu}
        \lim_{\Delta t \to 0} 
        \frac{\mel{\mu}
        {\hat{\rho}_{\vec{m}, \vec{n}}(t+\Delta t)}{\mu'}
        - \mel{\mu}{\hat{\rho}_{\vec{m}, \vec{n}}(t)}
        {\mu'}}
        {\Delta t}
        \bra{\mu'}\, ,
    \end{align}
    respectively, we obtain the HEOM in the following form through the use of Eq.~\eqref{eq:dTheta}:
    \begin{align}
       \frac{\partial}{\partial t} \hat{\rho}_{\vec{m}, \vec{n}}(t) = &
       -\frac{i}{\hbar} \hat{H}^\times_S(\Omega, \phi;t)
       \hat{\rho}_{\vec{m}, \vec{n}}(t)
       - \sum_{k=1}^{K} (m_k+n_k) \gamma_k \hat{\rho}_{\vec{m}, \vec{n}}(t) \\
       & + \sum_{k=1}^{K} \omega_k
           \Bigl\{\sqrt{m_k(n_k+1)}
           \hat{\rho}_{\vec{m}-\vec{e}_k, \vec{n}+\vec{e}_k}(t)
           - \sqrt{(m_k+1)n_k}
           \hat{\rho}_{\vec{m}+\vec{e}_k, \vec{n}-\vec{e}_k}(t) 
           \Bigr\}
       \\
       & - \frac{i}{\hbar} \hat{V}^\times \sum_{k=1}^{K}
       \sqrt{m_k+1}\hat{\rho}_{\vec{m}+\vec{e}_k, \vec{n}}(t)  \\
       & + \sum_{k=1}^{K} \left[\sqrt{m_k}\left\{
       -\frac{id'_k}{\hbar}\hat{V}^\times
       + \frac{d''_k}{\hbar} \hat{V}^\circ \right\}
       \hat{\rho}_{\vec{m}-\vec{e}_k, \vec{n}}(t)
       + \sqrt{n_k}\left\{
       -\frac{id''_k}{\hbar}\hat{V}^\times
       - \frac{d'_k}{\hbar}\hat{V}^\circ\right\}
       \hat{\rho}_{\vec{m}, \vec{n}-\vec{e}_k}(t) \right].
       \label{eq:HEOMFull}
    \end{align}
\end{widetext}
The vector $\vec{e}_k$ is the unit vector of the $k$th element, and $\hat{\rho}_{\vec{0}, \vec{0}}(t)$ corresponds to the RDO $\hat{\rho}_S(t)$. The symbols $\times$ and $\circ$ denote the commutator and anticommutator respectively, as $\hat{O}_1^\times \hat{O}_2 = \hat{O}_1\hat{O}_2-\hat{O}_2\hat{O}_1$ and $\hat{O}^\circ_1\hat{O}_2 = \hat{O}_1\hat{O}_2+\hat{O}_2\hat{O}_1$.
Note that the last line of Eq.~\eqref{eq:HEOMFull} corresponds to $\Theta(\mu, \mu';t, t)$ and $\bar{\Theta}(\mu, \mu';t, t)$.
Furthermore, the third and fourth line of Eq.~\eqref{eq:HEOMFull} correspond to $\mathcal{L}_k^{+}$ and $\mathcal{L}_k^{-}$ in Eq.~\eqref{eq:HEOM}, respectively.
The superoperator $\mathcal{L}_S$ in Eq.~\eqref{eq:HEOM} is defined as $\mathcal{L}_S = \hat{H}_S^{\times}(\Omega, \phi; t) / \hbar$ in Eq.~\eqref{eq:HEOMFull}.
The dynamics following from Eqs.~\eqref{eq:HEOM} and \eqref{eq:HEOMFull} are same, but Eq.~\eqref{eq:HEOMFull} is computationally more advantageous because of the Hermitian ADOs:
We only need to treat upper (or lower) triangular elements of density matrices.
Computationally, we can exploit this advantage through the use of (generalized) Bloch-vector representation~\cite{LiniovPRE2019}.
\section{Two-time correlation function and depth of HEOM} \label{sec:appCF}
\begin{figure}
    \includegraphics[width=\linewidth]{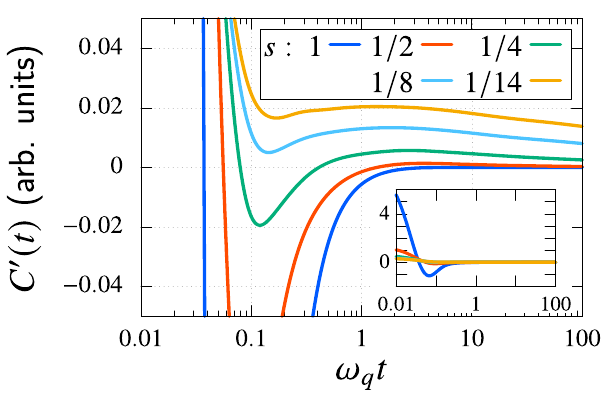}
    \caption{Real part of the two-time correlation function of the reservoir, $C'(t)$, with arbitrary (arb.) units. Only the region around $0$, i.e., $-0.05 \leq C'(t) \leq 0.05$, is plotted to depict the detailed profile. The whole profile is exhibited in the inset.
    \label{fig:CF}}
\end{figure}

Here, we depict the two-time correlation function of the reservoir.
The spectral density is given by Eq.~\eqref{eq:SD}, and the parameter values are the same as in the main text.
The corresponding two-time correlation function is evaluated as Eq.~\eqref{eq:CF} with the aid of the barycentric representation.
The number of modes for the two-time correlation function $K$ is listed in Table~\ref{tbl:param}.
Figure~\ref{fig:CF} displays the dynamics of the real part of the two-time correlation function, $C'(t)$.
In the Ohmic case, the fast decay caused by the large portion of the high-frequency modes $\omega \simeq \omega_c$ is observed in the short-time region, $\omega_q t \leq 0.1$ (inset of Fig.~\ref{fig:CF}), and the function approaches $0$ around the time $\omega_q t \simeq 5$.
In the sub-Ohmic case, the fast decay in the short-time region is suppressed. As the spectral exponent decreases, slower decay in the long-time region is observed. In the case with $s \leq 1/4$, the correlation function takes a finite value even at the time $\omega_q t = 100$.
This slow decay results from the low-frequency modes of the spectral noise power, which is approximated with $1/\omega^{1-s}$.
As discussed in the main text, the slow decay causes non-Markovian effects.

\begin{table}
\caption{The number of modes for the two-time correlation function $K$ and the maximum depth of the hierarchy $\mathcal{N}_{\max}$ for various spectral exponents $s$.
\label{tbl:param}}
\begin{ruledtabular}
\begin{tabular}{cccccc}
   $s$ & $1$ & $1/2$ & $1/4$ & $1/8$ & $1/14$ \\
   \hline
   $K$ & $10$ & $24$ & $18$ & $8$ & $7$ \\
   $\mathcal{N}_{\max}$ & $3$ & $3$ & $4$ & $8$ & $10$ \\
\end{tabular}
\end{ruledtabular}
\end{table}

As implied by the form of the influence functional in Eq.~\eqref{eq:IF1}, the sluggish decay of $C(t)$ allows the system and reservoir to interact many times.
Within the framework of HEOM, the depth of the hierarchy $\mathcal{N}_{\max}$ needs to be large in order to describe these multiple interactions.
For this reason, $\mathcal{N}_{\max}$ increases as the spectral exponent $s$ decreases in Table ~\ref{tbl:param}.
Note that since the multiple interactions cannot be described within the second-order perturbation theory, higher-order terms must be taken into account when one considers $1/f$-type noise with perturbative methods, e.g., time-convolutionless master equations.

Technically, simulations with lager $\mathcal{N}_{\max}$ demand more computational resources, and the cost for the simulations for $s = 1/8$ and $1/14$ is prohibitively expensive.
For this reason, we reduced the number of the elements $K$ with the aid of the method of least squares in those cases.
The values of $K$ for $s = 1/8$ and $1/14$ in Table~\ref{tbl:param} indicate the number of modes of the approximate set.

\section{Dynamics of a single qubit without pulses}
In this appendix, we investigate dynamics of a single qubit without pulses. We consider the same Hamiltonian as in the main text [Eq.~\eqref{eq:H_CL}], with the amplitude $\Omega = 0$. The parameter values of the spectral density are also the same as in the main text, and we used the HEOM to obtain the results.

\subsection{Population relaxation} \label{sec:appRelax}
\begin{figure}
    \includegraphics[width=\linewidth]{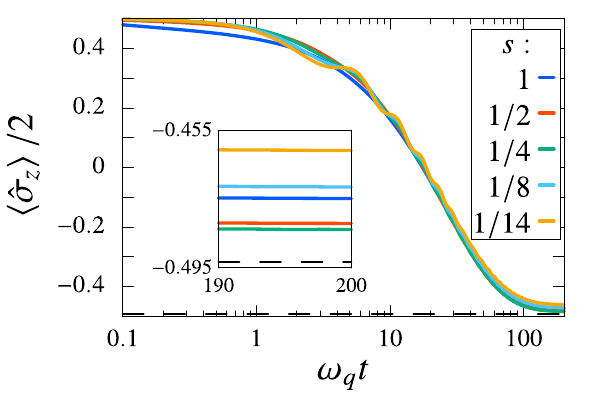}
    \caption{Dynamics of the expectation value $\ev*{\hat{\sigma}_z(t)}/2$ with various spectral densities. We do not consider pulses, which corresponds to $\Omega = 0$, and the excited initial state is adopted ($\hat{\rho}_0 = \hat{\rho}_e$). The dashed-horizontal line indicates the value $\ev*{\hat{\sigma}_z(t)}/2 = -0.5 \tanh(\beta \hbar \omega_q / 2)$, which corresponds to the equilibrium value obtained from the Boltzmann distribution of the bare system. The inset displays the expectation value in the long-time region.
    \label{fig:relax}}
\end{figure}

Here, we focus on the dynamics of the population relaxation.
The excited state $\hat{\rho}_e$ is chosen as the initial state.
Figure~\ref{fig:relax} displays the dynamics of the expectation value $\ev*{\hat{\sigma}_z(t)}/2$ with the Ohmic and sub-Ohmic spectral densities.
In the sub-Ohmic cases, we consider four exponents, $s = 1/2$, $1/4$, $1/8,$ and $1/14$, in the same way as the main text.
As is clear from the inset of Fig.~\ref{fig:relax}, the system reaches the equilibrium state up to the time $\omega_q t \leq 200$. We adopted the RDO and ADOs at the time $\omega_q t = 200$ as the equilibrium initial states, $\hat{\rho}_{eq}$, which were used in the simulations of the pulse sequences in the main text.

Overall, the dynamics of the population relaxation is qualitatively similar irrespective of the spectral exponent $s$. This is because we chose the parameter values such that $S_\beta(\pm\omega_q)$, which corresponds to the decay rate within the Bloch--Redfield theory [cf.\ Eq.~\eqref{eq:Lindblad}], takes the same value regardless of $s$. Conversely, the discrepancy of these dynamics indicates the effects beyond the Born--Markov approximation.

First, we focus on the dynamics in the short-time region, $\omega_q t \ll 1$.
The slower decay is observed as the spectral exponent decreases, which results from the ``universal decoherence''~\cite{TuorilaPRR2019,BraunPRL2001} as follows:
In the short-time region, the contribution of the system Hamiltonian $\hat{H}_S$ to the dynamics is negligibly small.
Because we assume the condition $\hat{H}_S \simeq 0$, the commutation relation $[\hat{H}_S, \hat{V}] = 0$ holds.
With this condition, the time evolution of the expectation value in the short-time region is evaluated as
\begin{align}
    \ev*{\hat{\sigma}_z(t)} 
    = & \exp \left[-4\hbar\int_{0}^{\infty} d\omega J(\omega)
    \coth\frac{\beta\hbar\omega}{2}
    \frac{1-\cos \omega t}{\omega^2}\right]\, .
\end{align}
Here, the initial state $\hat{\rho}_e$ is considered.
This decay in the short-time region is referred to as the universal decoherence.
When we consider further shorter-time region, $\omega_c t \ll 1$, we can approximate the term $(1-\cos \omega t)/ \omega^2$ with $t^2 / 2$, and the decay rate is proportional to $J(\omega) \coth (\beta \hbar \omega /2)$.
We confirmed that the area decreases as the spectral exponent decreases.
Although the region $0.1 < \omega_q t < 1$ in Fig.~\ref{fig:relax} does not fulfill the condition $\omega_c t \ll 1$, which corresponds to the condition $\omega_q t \ll 0.02$ in our case, we found the similar tendency of the decay in this region: the decay is slower as the exponent decreases.

After the universal decoherence, relatively fast decay is observed around the time $\omega_q t \simeq 3$ for the deep sub-Ohmic reservoirs. The subsequent oscillatory behavior is found in the cases for $s = 1/8$ and $1/14$.
These oscillations are caused by the slow dynamics of the two-time correlation function of the reservoir and reflects the non-Markovianity of the reservoir. 

Now, we turn to the analysis of the equilibrium states, the inset of Fig.~\ref{fig:relax}.
The dashed line in Fig.~\ref{fig:relax} is the expectation value for the equilibrium state obtained from the Born--Markov approach [$\ev*{\hat{\sigma}_z(t)}/2 = -0.5 \tanh(\beta \hbar \omega_q / 2)$], and every result deviates from this.
The system--reservoir coupling is ignored for the equilibrium states within the Born--Markov approximation, while included for the total equilibrium states.
This difference leads to the deviation, which is smallest in the case for $s = 1/4$, while largest in the case for $s = 1/14$.

\subsection{Ramsey experiments} \label{sec:appRamsey}
In this section, we illustrate the numerical results of the Ramsey experiments. The initial state is given by $\hat{\rho}_\mathrm{tot}(0) = (\ket{0} + \ket{1})(\bra{0} + \bra{1})/2 \otimes \hat{\rho}_{R, eq}$.

\begin{figure}
    \includegraphics[scale=0.7]{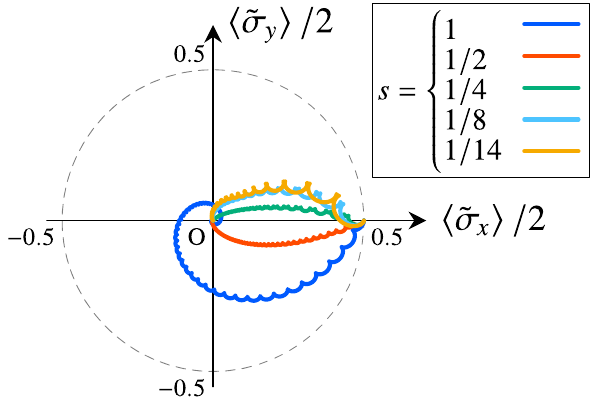}
    \caption{Time evolution of the Bloch vectors projected onto the $\ev*{\tilde{\sigma}_x}$--$\ev*{\tilde{\sigma}_y}$ plane.
    \label{fig:RamseyXY}}
\end{figure}

We depict the time evolution of the Bloch vectors projected onto the $\ev*{\tilde{\sigma}_x}$--$\ev*{\tilde{\sigma}_y}$ plane in Fig.~\ref{fig:RamseyXY}. The initial state corresponds to the point $(0.5, 0)$. Decoherence occurs during the time evolution, and the system reaches the equilibrium state, in which the off-diagonal elements are zero, corresponding to the point $(0, 0)$. Here, the rotating frame $\hat{R}_z(-\omega_q t)\hat{\rho}_S(t)\hat{R}_z(\omega_qt)$ is considered. If the off-diagonal elements, $\ev*{\hat{\sigma}_x}$ and $\ev*{\hat{\sigma}_y}$, oscillate with the frequency $\omega_q$, the projected Bloch vector in the rotating frame is always in the direction of $\ev*{\tilde{\sigma}_x}$.
In the Bloch--Redfield theory, the frequency change from $\omega_q$ because of the Lamb shift, and the projected Bloch vector rotates.
Because the Lamb shift is time independent within this approximation and the length of the vector decreases because of the decoherence, the locus is a spiral. In our case, however, the oscillatory behavior along the spiral is observed. This implies that we cannot express the oscillation behavior with a single frequency.

\begin{figure}
    \includegraphics[width=\linewidth]{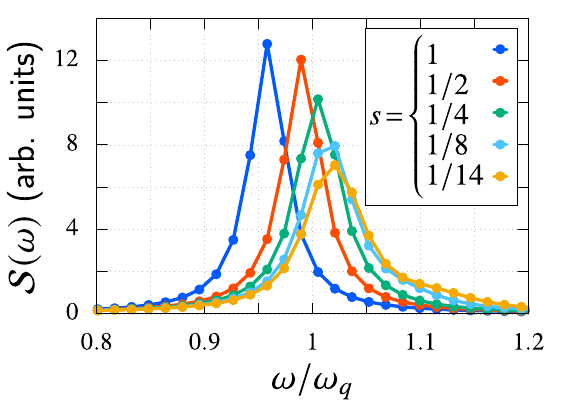}
    \caption{Fourier transform of $\ev*{\hat{\sigma}_x(t)}$ obtained with the Ramsey experiments, $\mathcal{S}(\omega)$, with arbitrary (arb.) units.
    \label{fig:RamseyFFT}}
\end{figure}


Figure~\ref{fig:RamseyFFT} displays the Fourier transform of $\ev*{\hat{\sigma}_x(t)}$ obtained with the Ramsey experiments, which is defined as 
\begin{align}
    \mathcal{S}(\omega) = \mathrm{Re}\left\{\int_{0}^{\infty} dt \frac{\ev*{\hat{\sigma}_x(t)}}{2} e^{-i \omega t}\right\}\, .
    \label{eq:FFT}
\end{align}
The upper bound of the integral is replaced with a sufficiently large value ($t = 400 / \omega_q$ in this case).
The effective frequency of the qubit increases as the spectral exponent decreases.
It is smaller than $\omega_q$ in the cases $s=1$ and $1/2$, while larger in the cases $s = 1/4$, $1/8$, and $1/14$.
This corresponds to the direction of the rotation of the spiral in Fig.~\ref{fig:RamseyXY}.
In addition, the absolute value of the frequency shift reflects the degree of the deviation of the locus from the $\ev*{\tilde{\sigma}_x}$ axis.
As discussed above, even if we choose the effective frequency obtained in Fig.~\ref{fig:RamseyFFT} for $\omega_\mathrm{ex}$, the oscillatory behavior cannot be removed completely.

\begin{figure*}
    \centering
    \includegraphics[width=0.88\linewidth]{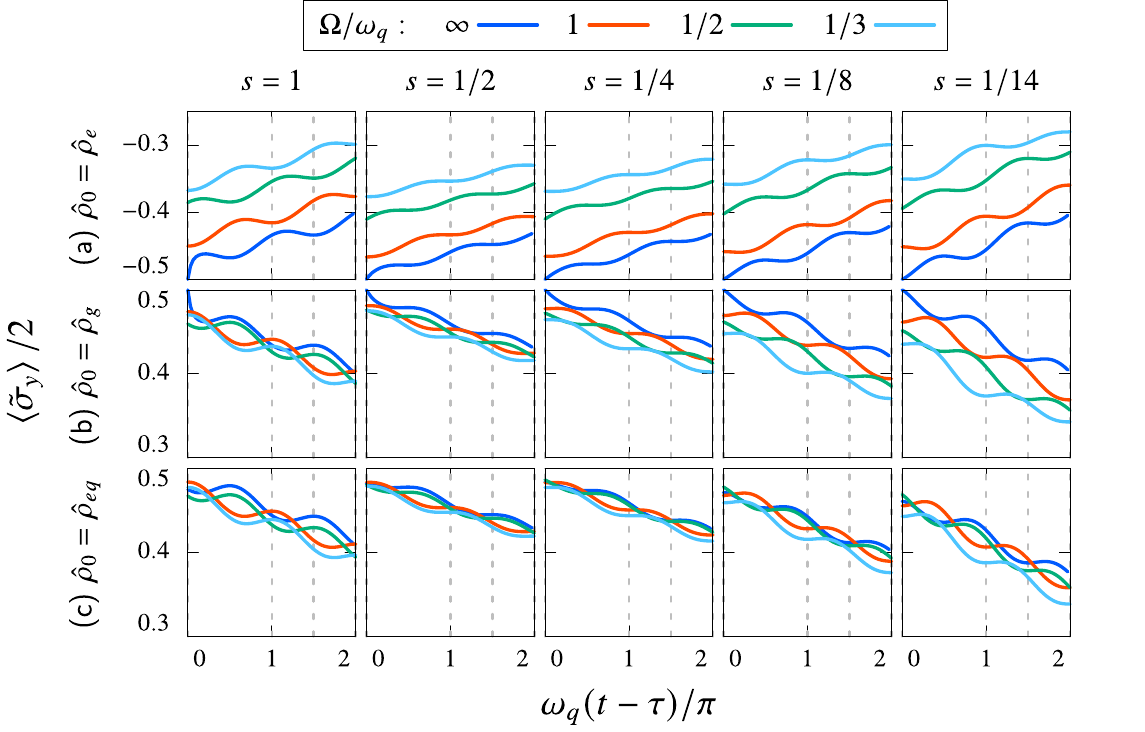}
    \caption{Dynamics of the expectation value $\ev*{\tilde{\sigma}_y(t)}/2$ in the rotating frame during the first idle phase with various initial states $\hat{\rho}_0$ and spectral exponents $s$. The sequence of $R_x(\pi/2)$ gates is considered. The gray-vertical-dashed lines indicate $\Delta t$ in Fig.~\ref{fig:HMap_HPI}. 
    \label{fig:idle_y}}
    \includegraphics[width=0.88\linewidth]{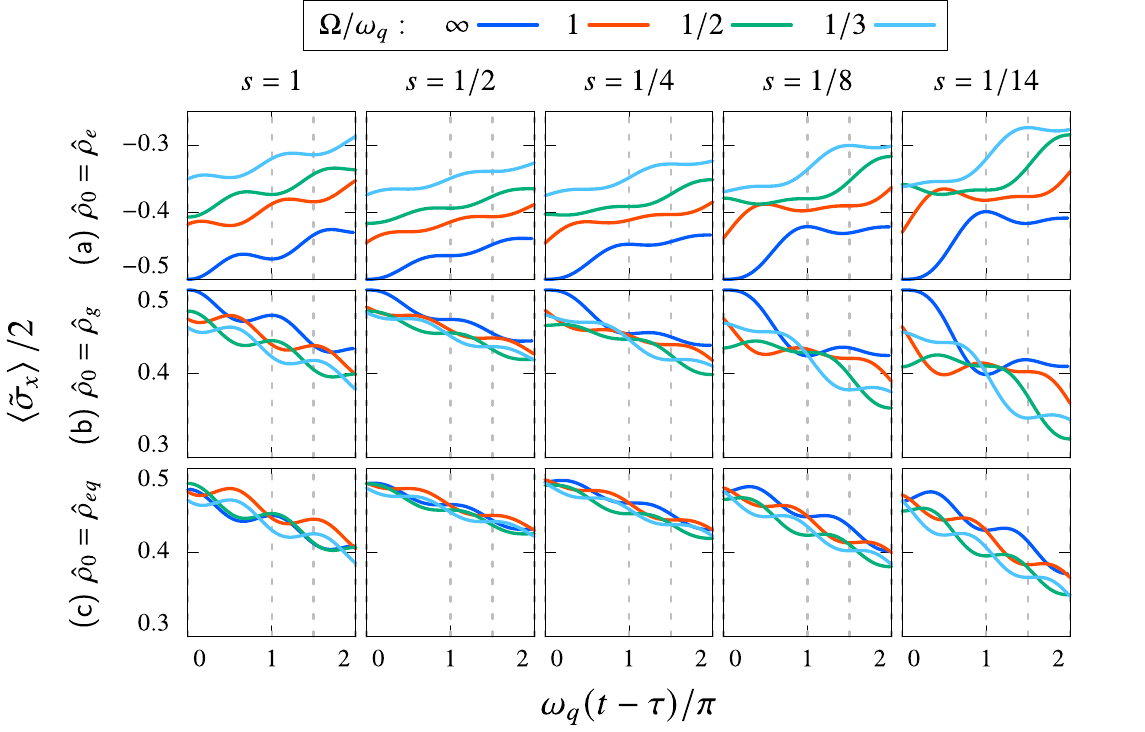}
    \caption{Dynamics of the expectation value $\ev*{\tilde{\sigma}_x(t)}/2$ in the rotating frame during the first idle phase with various initial states $\hat{\rho}_0$ and spectral exponents $s$. The sequence of three $H$ gates is applied. The gray vertical-dashed-lines indicate $\Delta t$ in Fig.~\ref{fig:HMap_H}.
    \label{fig:idle_x}}
\end{figure*}
\section{Detailed dynamics of a qubit during idle phases}
In Sec.~\ref{sec:resPI}, dynamics of the first idle phase after the $R_x(\pi)$-gate application were discussed (Fig.~\ref{fig:idle_z}).
Here, we report other interesting dynamics during idle phases.

\subsection{$R_x(\pi/2)$ gates} \label{sec:appHPI}
Figure~\ref{fig:idle_y} displays dynamics of the $\ev*{\tilde{\sigma}_y}$ element of the Bloch vector during the first idle phase of the $R_x(\pi/2)$-gate sequence.
In contrast to the case of the $R_x(\pi)$ gates (Fig.~\ref{fig:idle_z}), the oscillatory behavior is universally observed.
It is found that this oscillation has a $\pi$ periodicity with respect to $\omega_q \tau$ [$= \pi \omega_q/(2\Omega)$].
The relative angle of $\hat{V} \propto \hat{\sigma}_x$ and $\tilde{\sigma}_y$ determines this periodicity. 
The fast decoherence around the time $t \simeq \tau$ is observed with $\Omega = \infty $ for the reservoir $s = 1$, which is gradually suppressed as the spectral exponent decreases.
High-frequency modes of the spectral density cause steep decay of both population and coherence.
The amplitude of the oscillation tends to be smallest for $s = 1/2$ and $1/4$, while largest for $s = 1$ and $1/14$.
This reflects the effective Larmor frequency obtained from Fig.~\ref{fig:RamseyFFT} and appears to be related to the tendency of the fidelity in terms of $s$ discussed in the main text.

The peculiar oscillation, which was found in the $R_x(\pi)$-gate cases (Sec.~\ref{sec:resPI}), is not observed in Fig.~\ref{fig:idle_y}.
This leads to the emergence of the expected order in all the cases in Figs.~\ref{fig:HMap_HPI}(a)--\ref{fig:HMap_HPI}(c) at $d = 2$.

The time $t = \tau$ corresponds to the end of the first pulse application, $d = 1$.
The absolute value $|\ev*{\tilde{\sigma}_y}|$ mainly contributes to the fidelity and is in general smaller with the initial state $\hat{\rho}_e$ than with the state $\hat{\rho}_g$ and $\hat{\rho}_{eq}$ for a fixed $\Omega$.
As discussed in the main text, the rotation from $(0,0,1)$ to $(0,-1,0)$ is disadvantageous compared to the rotation from $(0,0,-1)$ to $(0,1,0)$.

\subsection{Hadamard ($H$) gates} \label{sec:appH}
Dynamics of the $\ev*{\tilde{\sigma}_x}$ element of the Bloch vector during the first idle phase of the $H$-gate sequence are depicted in Fig.~\ref{fig:idle_x}.
In the cases for $s \leq 1/4$ with $\hat{\rho}_0 = \hat{\rho}_e$ and $\hat{\rho}_g$, periodical oscillations are no longer observed, while oscillatory behavior similar to the $R_x(\pi/2)$-gate case is observed in the other cases.
For the nonperiodical results, one finds intense decay of the expectation value around the time $\omega_q (t - \tau)/ \pi \simeq 0.5$ with the amplitude $\Omega / \omega_q = \infty$ and $1$, while $\omega_q (t-\tau) / \pi \simeq 1.5$ with $\Omega / \omega_q = 1/2$ and $1/3$, which leads to the violation of the expected order defined in the main text.
The slow decay of the two-time correlation function appears to contribute to this behavior.
It is also enhanced in the instantaneous-pulse case:
the slow reestablishment of the system--reservoir correlations contributes.
This argument is also supported by the fact that this behavior is suppressed in the case $\hat{\rho}_0 = \hat{\rho}_{eq}$.

The initial phase of the oscillation behavior is different with $\pi$ from the case of $R_x(\pi/2)$ gates, and the fast decoherence around the time $t = \tau$ with $\Omega = \infty$ for $s = 1$ is not observed because of this difference of the initial phase.

In the same manner as the $R_x(\pi/2)$ gates, the absolute value $|\ev*{\tilde{\sigma}_x}|$ mainly contributes to the fidelity, and it is worse in case $\hat{\rho}_e$ than in the other two cases. The disadvantage of the $\pi/2$-rotation from the excited state is independent of the rotation axis. 

\subsection{Asymptotic behavior with respect to the pulse duration} \label{sec:appAsym}
\begin{figure}[b]
    \includegraphics[width=\linewidth]{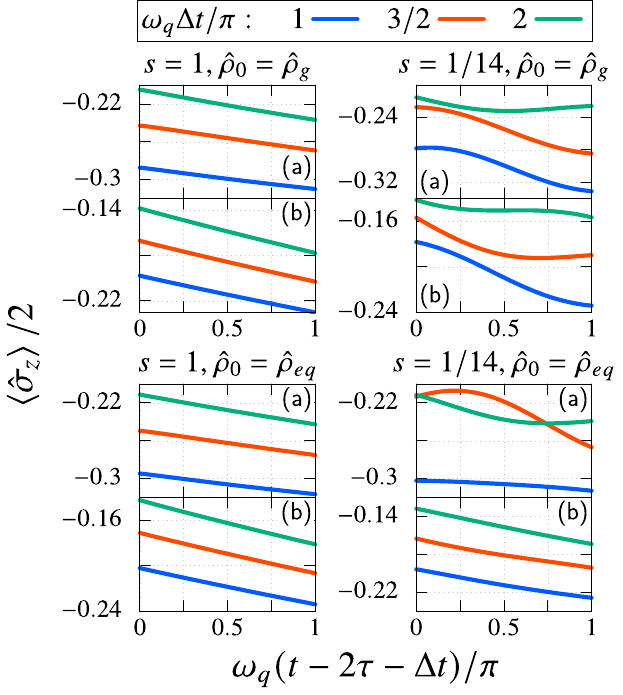}
    \caption{Dynamics of the expectation value $\ev*{\hat{\sigma}_z(t)}/2$ during the second idle phase of the sequence of three $R_x(\pi)$ gates. As representatives, the cases for $s=1$ and $s=1/14$ are depicted. The initial states are given by $\hat{\rho}_0 = \hat{\rho}_g$ and $\hat{\rho}_{eq}$, respectively. The pulse amplitude is chosen as follows: (a) $\Omega/\omega_q = 1$, (b) $\Omega/\omega_q = 1/3$.
    \label{fig:idle_asym}}
\end{figure}

Figure~\ref{fig:idle_asym} displays the dynamics of the $\ev*{\hat{\sigma}_z}$ element of the Bloch vector during the second idle phase of the $R_x(\pi)$-gate sequence.
The cases with $\hat{\rho}_0 = \hat{\rho}_g$ and $\hat{\rho}_{eq}$ for $s=1$ and $1/14$ are depicted as representatives.
We varied the amplitudes and durations of the idle phase, as $\Omega / \omega_q = 1$ and $1/3$ [(a) and (b) in each panel], and $\omega_q \Delta t/\pi= 1$, $3/2$, and $2$, respectively.

The decay is nearly linear irrespective of $\Delta t$ in the Ohmic case ($s = 1$, left panel), and further, the rate is almost independent of $\Delta t$, while similar behavior is only observed in the deep sub-Ohmic case ($s = 1/14$, right panel) with the condition $\Omega/\omega_q = 1/3 $ and $\hat{\rho}_0 = \hat{\rho}_{eq}$.
This indicates that when the pulse duration is sufficiently large and the initial state is prepared into the equilibrium state, the decay with an almost same rate irrespective of $\Delta t$ and $s$ occurs.
From this result, we can interpret this behavior as asymptotic behavior with respect to the pulse duration.
As discussed above, the reconfiguration process is less significant with the initial state $\hat{\rho}_0 = \hat{\rho}_{eq}$ compared to $\hat{\rho}_e$ and $\hat{\rho}_g$. 
In addition, the process is completed in a shorter period of time when the decay of the two-time correlation function is faster.
We found that the asymptotic behavior tends to be observed in the cases where the impact of the reconfiguration is small, and we suggest that the Markovianity  of the reservoir \textcolor{black}{(instantaneous response of the reservoir without memory effects)} contributes to this asymptotic behavior.

Note that during a sequence with the impulsive pulses, the above asymptotic behavior is not observed even in the Ohmic case.
The reconfiguration process does not occur during the pulse application, and hence the dynamics during the second idle phase are directly affected by those during the first idle phase.

Finally, we comprehensively mention the behavior during the second idle phase.
In the case for $s \geq 1/2$, the asymptotic behavior is always observed for the pulse amplitude $\Omega / \omega_q = 1$, $1/2$, and $1/3$, with the initial state $\hat{\rho}_0 = \hat{\rho}_e$, $\hat{\rho}_g$, and $\hat{\rho}_{eq}$, while in the case for $s \leq 1/8$, this behavior is only observed for $\Omega / \omega_q = 1/3$ with $\hat{\rho}_0 = \hat{\rho}_{eq}$.
The case for $s = 1/4$ is intermediate of these two cases:
the asymptotic behavior is observed for $\Omega / \omega_q = 1/3$ with all the initial states.

\bibliography{reference,qubit,dd}

\begin{thebibliography}{66}%
\makeatletter
\providecommand \@ifxundefined [1]{%
 \@ifx{#1\undefined}
}%
\providecommand \@ifnum [1]{%
 \ifnum #1\expandafter \@firstoftwo
 \else \expandafter \@secondoftwo
 \fi
}%
\providecommand \@ifx [1]{%
 \ifx #1\expandafter \@firstoftwo
 \else \expandafter \@secondoftwo
 \fi
}%
\providecommand \natexlab [1]{#1}%
\providecommand \enquote  [1]{``#1''}%
\providecommand \bibnamefont  [1]{#1}%
\providecommand \bibfnamefont [1]{#1}%
\providecommand \citenamefont [1]{#1}%
\providecommand \href@noop [0]{\@secondoftwo}%
\providecommand \href [0]{\begingroup \@sanitize@url \@href}%
\providecommand \@href[1]{\@@startlink{#1}\@@href}%
\providecommand \@@href[1]{\endgroup#1\@@endlink}%
\providecommand \@sanitize@url [0]{\catcode `\\12\catcode `\$12\catcode
  `\&12\catcode `\#12\catcode `\^12\catcode `\_12\catcode `\%12\relax}%
\providecommand \@@startlink[1]{}%
\providecommand \@@endlink[0]{}%
\providecommand \url  [0]{\begingroup\@sanitize@url \@url }%
\providecommand \@url [1]{\endgroup\@href {#1}{\urlprefix }}%
\providecommand \urlprefix  [0]{URL }%
\providecommand \Eprint [0]{\href }%
\providecommand \doibase [0]{https://doi.org/}%
\providecommand \selectlanguage [0]{\@gobble}%
\providecommand \bibinfo  [0]{\@secondoftwo}%
\providecommand \bibfield  [0]{\@secondoftwo}%
\providecommand \translation [1]{[#1]}%
\providecommand \BibitemOpen [0]{}%
\providecommand \bibitemStop [0]{}%
\providecommand \bibitemNoStop [0]{.\EOS\space}%
\providecommand \EOS [0]{\spacefactor3000\relax}%
\providecommand \BibitemShut  [1]{\csname bibitem#1\endcsname}%
\let\auto@bib@innerbib\@empty
\bibitem [{\citenamefont {Place}\ \emph {et~al.}(2021)\citenamefont {Place},
  \citenamefont {Rodgers}, \citenamefont {Mundada}, \citenamefont {Smitham},
  \citenamefont {Fitzpatrick}, \citenamefont {Leng}, \citenamefont {Premkumar},
  \citenamefont {Bryon}, \citenamefont {Vrajitoarea}, \citenamefont {Sussman}
  \emph {et~al.}}]{PlaceNATCOMMUN2021}%
  \BibitemOpen
  \bibfield  {author} {\bibinfo {author} {\bibfnamefont {A.~P.~M.}\
  \bibnamefont {Place}}, \bibinfo {author} {\bibfnamefont {L.~V.~H.}\
  \bibnamefont {Rodgers}}, \bibinfo {author} {\bibfnamefont {P.}~\bibnamefont
  {Mundada}}, \bibinfo {author} {\bibfnamefont {B.~M.}\ \bibnamefont
  {Smitham}}, \bibinfo {author} {\bibfnamefont {M.}~\bibnamefont
  {Fitzpatrick}}, \bibinfo {author} {\bibfnamefont {Z.}~\bibnamefont {Leng}},
  \bibinfo {author} {\bibfnamefont {A.}~\bibnamefont {Premkumar}}, \bibinfo
  {author} {\bibfnamefont {J.}~\bibnamefont {Bryon}}, \bibinfo {author}
  {\bibfnamefont {A.}~\bibnamefont {Vrajitoarea}}, \bibinfo {author}
  {\bibfnamefont {S.}~\bibnamefont {Sussman}}, \emph {et~al.},\ }\bibfield
  {title} {\bibinfo {title} {{New material platform for superconducting
  transmon qubits with coherence times exceeding 0.3 milliseconds}},\ }\href
  {https://doi.org/10.1038/s41467-021-22030-5} {\bibfield  {journal} {\bibinfo
  {journal} {Nat. Commun.}\ }\textbf {\bibinfo {volume} {12}},\ \bibinfo
  {pages} {1779} (\bibinfo {year} {2021})}\BibitemShut {NoStop}%
\bibitem [{\citenamefont {Wang}\ \emph {et~al.}(2022)\citenamefont {Wang},
  \citenamefont {Li}, \citenamefont {Xu}, \citenamefont {Li}, \citenamefont
  {Wang}, \citenamefont {Yang}, \citenamefont {Mi}, \citenamefont {Liang},
  \citenamefont {Su}, \citenamefont {Yang} \emph {et~al.}}]{WangNPJQI2022}%
  \BibitemOpen
  \bibfield  {author} {\bibinfo {author} {\bibfnamefont {C.}~\bibnamefont
  {Wang}}, \bibinfo {author} {\bibfnamefont {X.}~\bibnamefont {Li}}, \bibinfo
  {author} {\bibfnamefont {H.}~\bibnamefont {Xu}}, \bibinfo {author}
  {\bibfnamefont {Z.}~\bibnamefont {Li}}, \bibinfo {author} {\bibfnamefont
  {J.}~\bibnamefont {Wang}}, \bibinfo {author} {\bibfnamefont {Z.}~\bibnamefont
  {Yang}}, \bibinfo {author} {\bibfnamefont {Z.}~\bibnamefont {Mi}}, \bibinfo
  {author} {\bibfnamefont {X.}~\bibnamefont {Liang}}, \bibinfo {author}
  {\bibfnamefont {T.}~\bibnamefont {Su}}, \bibinfo {author} {\bibfnamefont
  {C.}~\bibnamefont {Yang}}, \emph {et~al.},\ }\bibfield  {title} {\bibinfo
  {title} {{Towards practical quantum computers: transmon qubit with a lifetime
  approaching 0.5 milliseconds}},\ }\href
  {https://doi.org/10.1038/s41534-021-00510-2} {\bibfield  {journal} {\bibinfo
  {journal} {npj Quantum Inf.}\ }\textbf {\bibinfo {volume} {8}},\ \bibinfo
  {pages} {3} (\bibinfo {year} {2022})}\BibitemShut {NoStop}%
\bibitem [{\citenamefont {Neg\^{\i}rneac}\ \emph {et~al.}(2021)\citenamefont
  {Neg\^{\i}rneac}, \citenamefont {Ali}, \citenamefont {Muthusubramanian},
  \citenamefont {Battistel}, \citenamefont {Sagastizabal}, \citenamefont
  {Moreira}, \citenamefont {Marques}, \citenamefont {Vlothuizen}, \citenamefont
  {Beekman}, \citenamefont {Zachariadis} \emph {et~al.}}]{NegirneacPRL2021}%
  \BibitemOpen
  \bibfield  {author} {\bibinfo {author} {\bibfnamefont {V.}~\bibnamefont
  {Neg\^{\i}rneac}}, \bibinfo {author} {\bibfnamefont {H.}~\bibnamefont {Ali}},
  \bibinfo {author} {\bibfnamefont {N.}~\bibnamefont {Muthusubramanian}},
  \bibinfo {author} {\bibfnamefont {F.}~\bibnamefont {Battistel}}, \bibinfo
  {author} {\bibfnamefont {R.}~\bibnamefont {Sagastizabal}}, \bibinfo {author}
  {\bibfnamefont {M.~S.}\ \bibnamefont {Moreira}}, \bibinfo {author}
  {\bibfnamefont {J.~F.}\ \bibnamefont {Marques}}, \bibinfo {author}
  {\bibfnamefont {W.~J.}\ \bibnamefont {Vlothuizen}}, \bibinfo {author}
  {\bibfnamefont {M.}~\bibnamefont {Beekman}}, \bibinfo {author} {\bibfnamefont
  {C.}~\bibnamefont {Zachariadis}}, \emph {et~al.},\ }\bibfield  {title}
  {\bibinfo {title} {{High-Fidelity Controlled-$Z$ Gate with Maximal
  Intermediate Leakage Operating at the Speed Limit in a Superconducting
  Quantum Processor}},\ }\href {https://doi.org/10.1103/PhysRevLett.126.220502}
  {\bibfield  {journal} {\bibinfo  {journal} {Phys. Rev. Lett.}\ }\textbf
  {\bibinfo {volume} {126}},\ \bibinfo {pages} {220502} (\bibinfo {year}
  {2021})}\BibitemShut {NoStop}%
\bibitem [{\citenamefont {Sung}\ \emph {et~al.}(2021)\citenamefont {Sung},
  \citenamefont {Ding}, \citenamefont {Braum\"uller}, \citenamefont
  {Veps\"al\"ainen}, \citenamefont {Kannan}, \citenamefont {Kjaergaard},
  \citenamefont {Greene}, \citenamefont {Samach}, \citenamefont {McNally},
  \citenamefont {Kim} \emph {et~al.}}]{SungPRX2021}%
  \BibitemOpen
  \bibfield  {author} {\bibinfo {author} {\bibfnamefont {Y.}~\bibnamefont
  {Sung}}, \bibinfo {author} {\bibfnamefont {L.}~\bibnamefont {Ding}}, \bibinfo
  {author} {\bibfnamefont {J.}~\bibnamefont {Braum\"uller}}, \bibinfo {author}
  {\bibfnamefont {A.}~\bibnamefont {Veps\"al\"ainen}}, \bibinfo {author}
  {\bibfnamefont {B.}~\bibnamefont {Kannan}}, \bibinfo {author} {\bibfnamefont
  {M.}~\bibnamefont {Kjaergaard}}, \bibinfo {author} {\bibfnamefont
  {A.}~\bibnamefont {Greene}}, \bibinfo {author} {\bibfnamefont {G.~O.}\
  \bibnamefont {Samach}}, \bibinfo {author} {\bibfnamefont {C.}~\bibnamefont
  {McNally}}, \bibinfo {author} {\bibfnamefont {D.}~\bibnamefont {Kim}}, \emph
  {et~al.},\ }\bibfield  {title} {\bibinfo {title} {{Realization of
  High-Fidelity CZ and $ZZ$-Free iSWAP Gates with a Tunable Coupler}},\ }\href
  {https://doi.org/10.1103/PhysRevX.11.021058} {\bibfield  {journal} {\bibinfo
  {journal} {Phys. Rev. X}\ }\textbf {\bibinfo {volume} {11}},\ \bibinfo
  {pages} {021058} (\bibinfo {year} {2021})}\BibitemShut {NoStop}%
\bibitem [{\citenamefont {Kandala}\ \emph {et~al.}(2021)\citenamefont
  {Kandala}, \citenamefont {Wei}, \citenamefont {Srinivasan}, \citenamefont
  {Magesan}, \citenamefont {Carnevale}, \citenamefont {Keefe}, \citenamefont
  {Klaus}, \citenamefont {Dial},\ and\ \citenamefont {McKay}}]{KandalaPRL2021}%
  \BibitemOpen
  \bibfield  {author} {\bibinfo {author} {\bibfnamefont {A.}~\bibnamefont
  {Kandala}}, \bibinfo {author} {\bibfnamefont {K.~X.}\ \bibnamefont {Wei}},
  \bibinfo {author} {\bibfnamefont {S.}~\bibnamefont {Srinivasan}}, \bibinfo
  {author} {\bibfnamefont {E.}~\bibnamefont {Magesan}}, \bibinfo {author}
  {\bibfnamefont {S.}~\bibnamefont {Carnevale}}, \bibinfo {author}
  {\bibfnamefont {G.~A.}\ \bibnamefont {Keefe}}, \bibinfo {author}
  {\bibfnamefont {D.}~\bibnamefont {Klaus}}, \bibinfo {author} {\bibfnamefont
  {O.}~\bibnamefont {Dial}},\ and\ \bibinfo {author} {\bibfnamefont {D.~C.}\
  \bibnamefont {McKay}},\ }\bibfield  {title} {\bibinfo {title} {{Demonstration
  of a High-Fidelity \textsc{cnot} Gate for Fixed-Frequency Transmons with
  Engineered $ZZ$ Suppression}},\ }\href
  {https://doi.org/10.1103/PhysRevLett.127.130501} {\bibfield  {journal}
  {\bibinfo  {journal} {Phys. Rev. Lett.}\ }\textbf {\bibinfo {volume} {127}},\
  \bibinfo {pages} {130501} (\bibinfo {year} {2021})}\BibitemShut {NoStop}%
\bibitem [{\citenamefont {Arute}\ \emph {et~al.}(2019)\citenamefont {Arute},
  \citenamefont {Arya}, \citenamefont {Babbush}, \citenamefont {Bacon},
  \citenamefont {Bardin}, \citenamefont {Barends}, \citenamefont {Biswas},
  \citenamefont {Boixo}, \citenamefont {Brandao}, \citenamefont {Buell} \emph
  {et~al.}}]{GoogleNATURE2019}%
  \BibitemOpen
  \bibfield  {author} {\bibinfo {author} {\bibfnamefont {F.}~\bibnamefont
  {Arute}}, \bibinfo {author} {\bibfnamefont {K.}~\bibnamefont {Arya}},
  \bibinfo {author} {\bibfnamefont {R.}~\bibnamefont {Babbush}}, \bibinfo
  {author} {\bibfnamefont {D.}~\bibnamefont {Bacon}}, \bibinfo {author}
  {\bibfnamefont {J.~C.}\ \bibnamefont {Bardin}}, \bibinfo {author}
  {\bibfnamefont {R.}~\bibnamefont {Barends}}, \bibinfo {author} {\bibfnamefont
  {R.}~\bibnamefont {Biswas}}, \bibinfo {author} {\bibfnamefont
  {S.}~\bibnamefont {Boixo}}, \bibinfo {author} {\bibfnamefont {F.~G. S.~L.}\
  \bibnamefont {Brandao}}, \bibinfo {author} {\bibfnamefont {D.~A.}\
  \bibnamefont {Buell}}, \emph {et~al.},\ }\bibfield  {title} {\bibinfo {title}
  {{Quantum supremacy using a programmable superconducting processor}},\ }\href
  {https://doi.org/10.1038/s41586-019-1666-5} {\bibfield  {journal} {\bibinfo
  {journal} {{Nature (London)}}\ }\textbf {\bibinfo {volume} {574}},\ \bibinfo
  {pages} {505} (\bibinfo {year} {2019})}\BibitemShut {NoStop}%
\bibitem [{\citenamefont {{Google Quantum AI}}(2023)}]{GoogleNATURE2023}%
  \BibitemOpen
  \bibfield  {author} {\bibinfo {author} {\bibnamefont {{Google Quantum AI}}},\
  }\bibfield  {title} {\bibinfo {title} {{Suppressing quantum errors by scaling
  a surface code logical qubit}},\ }\href
  {https://doi.org/10.1038/s41586-022-05434-1} {\bibfield  {journal} {\bibinfo
  {journal} {{Nature (London)}}\ }\textbf {\bibinfo {volume} {614}},\ \bibinfo
  {pages} {676} (\bibinfo {year} {2023})}\BibitemShut {NoStop}%
\bibitem [{\citenamefont {Kim}\ \emph {et~al.}(2023)\citenamefont {Kim},
  \citenamefont {Eddins}, \citenamefont {Anand}, \citenamefont {Wei},
  \citenamefont {van~den Berg}, \citenamefont {Rosenblatt}, \citenamefont
  {Nayfeh}, \citenamefont {Wu}, \citenamefont {Zaletel}, \citenamefont {Temme}
  \emph {et~al.}}]{IBMNATURE2023}%
  \BibitemOpen
  \bibfield  {author} {\bibinfo {author} {\bibfnamefont {Y.}~\bibnamefont
  {Kim}}, \bibinfo {author} {\bibfnamefont {A.}~\bibnamefont {Eddins}},
  \bibinfo {author} {\bibfnamefont {S.}~\bibnamefont {Anand}}, \bibinfo
  {author} {\bibfnamefont {K.~X.}\ \bibnamefont {Wei}}, \bibinfo {author}
  {\bibfnamefont {E.}~\bibnamefont {van~den Berg}}, \bibinfo {author}
  {\bibfnamefont {S.}~\bibnamefont {Rosenblatt}}, \bibinfo {author}
  {\bibfnamefont {H.}~\bibnamefont {Nayfeh}}, \bibinfo {author} {\bibfnamefont
  {Y.}~\bibnamefont {Wu}}, \bibinfo {author} {\bibfnamefont {M.}~\bibnamefont
  {Zaletel}}, \bibinfo {author} {\bibfnamefont {K.}~\bibnamefont {Temme}},
  \emph {et~al.},\ }\bibfield  {title} {\bibinfo {title} {{Evidence for the
  utility of quantum computing before fault tolerance}},\ }\href
  {https://doi.org/10.1038/s41586-023-06096-3} {\bibfield  {journal} {\bibinfo
  {journal} {{Nature (London)}}\ }\textbf {\bibinfo {volume} {618}},\ \bibinfo
  {pages} {500} (\bibinfo {year} {2023})}\BibitemShut {NoStop}%
\bibitem [{\citenamefont {Havl{\'i}{\v{c}}ek}\ \emph
  {et~al.}(2019)\citenamefont {Havl{\'i}{\v{c}}ek}, \citenamefont
  {C{\'o}rcoles}, \citenamefont {Temme}, \citenamefont {Harrow}, \citenamefont
  {Kandala}, \citenamefont {Chow},\ and\ \citenamefont
  {Gambetta}}]{HavlicekNATURE2019}%
  \BibitemOpen
  \bibfield  {author} {\bibinfo {author} {\bibfnamefont {V.}~\bibnamefont
  {Havl{\'i}{\v{c}}ek}}, \bibinfo {author} {\bibfnamefont {A.~D.}\ \bibnamefont
  {C{\'o}rcoles}}, \bibinfo {author} {\bibfnamefont {K.}~\bibnamefont {Temme}},
  \bibinfo {author} {\bibfnamefont {A.~W.}\ \bibnamefont {Harrow}}, \bibinfo
  {author} {\bibfnamefont {A.}~\bibnamefont {Kandala}}, \bibinfo {author}
  {\bibfnamefont {J.~M.}\ \bibnamefont {Chow}},\ and\ \bibinfo {author}
  {\bibfnamefont {J.~M.}\ \bibnamefont {Gambetta}},\ }\bibfield  {title}
  {\bibinfo {title} {{Supervised learning with quantum-enhanced feature
  spaces}},\ }\href {https://doi.org/10.1038/s41586-019-0980-2} {\bibfield
  {journal} {\bibinfo  {journal} {{Nature (London)}}\ }\textbf {\bibinfo
  {volume} {567}},\ \bibinfo {pages} {209} (\bibinfo {year}
  {2019})}\BibitemShut {NoStop}%
\bibitem [{\citenamefont {Kandala}\ \emph {et~al.}(2017)\citenamefont
  {Kandala}, \citenamefont {Mezzacapo}, \citenamefont {Temme}, \citenamefont
  {Takita}, \citenamefont {Brink}, \citenamefont {Chow},\ and\ \citenamefont
  {Gambetta}}]{KandalaNATURE2017}%
  \BibitemOpen
  \bibfield  {author} {\bibinfo {author} {\bibfnamefont {A.}~\bibnamefont
  {Kandala}}, \bibinfo {author} {\bibfnamefont {A.}~\bibnamefont {Mezzacapo}},
  \bibinfo {author} {\bibfnamefont {K.}~\bibnamefont {Temme}}, \bibinfo
  {author} {\bibfnamefont {M.}~\bibnamefont {Takita}}, \bibinfo {author}
  {\bibfnamefont {M.}~\bibnamefont {Brink}}, \bibinfo {author} {\bibfnamefont
  {J.~M.}\ \bibnamefont {Chow}},\ and\ \bibinfo {author} {\bibfnamefont
  {J.~M.}\ \bibnamefont {Gambetta}},\ }\bibfield  {title} {\bibinfo {title}
  {{Hardware-efficient variational quantum eigensolver for small molecules and
  quantum magnets}},\ }\href {https://doi.org/10.1038/nature23879} {\bibfield
  {journal} {\bibinfo  {journal} {{Nature (London)}}\ }\textbf {\bibinfo
  {volume} {549}},\ \bibinfo {pages} {242} (\bibinfo {year}
  {2017})}\BibitemShut {NoStop}%
\bibitem [{\citenamefont {Rossmannek}\ \emph {et~al.}(2023)\citenamefont
  {Rossmannek}, \citenamefont {Pavo\v{s}evi\'{c}}, \citenamefont {Rubio},\ and\
  \citenamefont {Tavernelli}}]{RossmannekJPCL2023}%
  \BibitemOpen
  \bibfield  {author} {\bibinfo {author} {\bibfnamefont {M.}~\bibnamefont
  {Rossmannek}}, \bibinfo {author} {\bibfnamefont {F.}~\bibnamefont
  {Pavo\v{s}evi\'{c}}}, \bibinfo {author} {\bibfnamefont {A.}~\bibnamefont
  {Rubio}},\ and\ \bibinfo {author} {\bibfnamefont {I.}~\bibnamefont
  {Tavernelli}},\ }\bibfield  {title} {\bibinfo {title} {{Quantum Embedding
  Method for the Simulation of Strongly Correlated Systems on Quantum
  Computers}},\ }\href {https://doi.org/10.1021/acs.jpclett.3c00330} {\bibfield
   {journal} {\bibinfo  {journal} {J. Phys. Chem. Lett.}\ }\textbf {\bibinfo
  {volume} {14}},\ \bibinfo {pages} {3491} (\bibinfo {year}
  {2023})}\BibitemShut {NoStop}%
\bibitem [{\citenamefont {Willsch}\ \emph {et~al.}(2020)\citenamefont
  {Willsch}, \citenamefont {Willsch}, \citenamefont {Jin}, \citenamefont
  {De~Raedt},\ and\ \citenamefont {Michielsen}}]{WillschQIP2020}%
  \BibitemOpen
  \bibfield  {author} {\bibinfo {author} {\bibfnamefont {M.}~\bibnamefont
  {Willsch}}, \bibinfo {author} {\bibfnamefont {D.}~\bibnamefont {Willsch}},
  \bibinfo {author} {\bibfnamefont {F.}~\bibnamefont {Jin}}, \bibinfo {author}
  {\bibfnamefont {H.}~\bibnamefont {De~Raedt}},\ and\ \bibinfo {author}
  {\bibfnamefont {K.}~\bibnamefont {Michielsen}},\ }\bibfield  {title}
  {\bibinfo {title} {{Benchmarking the quantum approximate optimization
  algorithm}},\ }\href {https://doi.org/10.1007/s11128-020-02692-8} {\bibfield
  {journal} {\bibinfo  {journal} {Quantum Inf. Process.}\ }\textbf {\bibinfo
  {volume} {19}},\ \bibinfo {pages} {197} (\bibinfo {year} {2020})}\BibitemShut
  {NoStop}%
\bibitem [{\citenamefont {Garc{\'i}a-P{\'e}rez}\ \emph
  {et~al.}(2020)\citenamefont {Garc{\'i}a-P{\'e}rez}, \citenamefont {Rossi},\
  and\ \citenamefont {Maniscalco}}]{PerezNPJQI2020}%
  \BibitemOpen
  \bibfield  {author} {\bibinfo {author} {\bibfnamefont {G.}~\bibnamefont
  {Garc{\'i}a-P{\'e}rez}}, \bibinfo {author} {\bibfnamefont {M.~A.~C.}\
  \bibnamefont {Rossi}},\ and\ \bibinfo {author} {\bibfnamefont
  {S.}~\bibnamefont {Maniscalco}},\ }\bibfield  {title} {\bibinfo {title} {{IBM
  Q Experience as a versatile experimental testbed for simulating open quantum
  systems}},\ }\href {https://doi.org/10.1038/s41534-019-0235-y} {\bibfield
  {journal} {\bibinfo  {journal} {npj Quantum Inf.}\ }\textbf {\bibinfo
  {volume} {6}},\ \bibinfo {pages} {1} (\bibinfo {year} {2020})}\BibitemShut
  {NoStop}%
\bibitem [{\citenamefont {Farrell}\ \emph {et~al.}(2024)\citenamefont
  {Farrell}, \citenamefont {Illa}, \citenamefont {Ciavarella},\ and\
  \citenamefont {Savage}}]{FarrellPRD2024}%
  \BibitemOpen
  \bibfield  {author} {\bibinfo {author} {\bibfnamefont {R.~C.}\ \bibnamefont
  {Farrell}}, \bibinfo {author} {\bibfnamefont {M.}~\bibnamefont {Illa}},
  \bibinfo {author} {\bibfnamefont {A.~N.}\ \bibnamefont {Ciavarella}},\ and\
  \bibinfo {author} {\bibfnamefont {M.~J.}\ \bibnamefont {Savage}},\ }\bibfield
   {title} {\bibinfo {title} {{Quantum simulations of hadron dynamics in the
  Schwinger model using 112 qubits}},\ }\href
  {https://doi.org/10.1103/PhysRevD.109.114510} {\bibfield  {journal} {\bibinfo
   {journal} {Phys. Rev. D}\ }\textbf {\bibinfo {volume} {109}},\ \bibinfo
  {pages} {114510} (\bibinfo {year} {2024})}\BibitemShut {NoStop}%
\bibitem [{\citenamefont {Devitt}\ \emph {et~al.}(2013)\citenamefont {Devitt},
  \citenamefont {Munro},\ and\ \citenamefont {Nemoto}}]{DevittRPP2013}%
  \BibitemOpen
  \bibfield  {author} {\bibinfo {author} {\bibfnamefont {S.~J.}\ \bibnamefont
  {Devitt}}, \bibinfo {author} {\bibfnamefont {W.~J.}\ \bibnamefont {Munro}},\
  and\ \bibinfo {author} {\bibfnamefont {K.}~\bibnamefont {Nemoto}},\
  }\bibfield  {title} {\bibinfo {title} {{Quantum error correction for
  beginners}},\ }\href {https://doi.org/10.1088/0034-4885/76/7/076001}
  {\bibfield  {journal} {\bibinfo  {journal} {Rep. Prog. Phys.}\ }\textbf
  {\bibinfo {volume} {76}},\ \bibinfo {pages} {076001} (\bibinfo {year}
  {2013})}\BibitemShut {NoStop}%
\bibitem [{\citenamefont {Cai}\ \emph {et~al.}(2023)\citenamefont {Cai},
  \citenamefont {Babbush}, \citenamefont {Benjamin}, \citenamefont {Endo},
  \citenamefont {Huggins}, \citenamefont {Li}, \citenamefont {McClean},\ and\
  \citenamefont {O'Brien}}]{CaiRMP2023}%
  \BibitemOpen
  \bibfield  {author} {\bibinfo {author} {\bibfnamefont {Z.}~\bibnamefont
  {Cai}}, \bibinfo {author} {\bibfnamefont {R.}~\bibnamefont {Babbush}},
  \bibinfo {author} {\bibfnamefont {S.~C.}\ \bibnamefont {Benjamin}}, \bibinfo
  {author} {\bibfnamefont {S.}~\bibnamefont {Endo}}, \bibinfo {author}
  {\bibfnamefont {W.~J.}\ \bibnamefont {Huggins}}, \bibinfo {author}
  {\bibfnamefont {Y.}~\bibnamefont {Li}}, \bibinfo {author} {\bibfnamefont
  {J.~R.}\ \bibnamefont {McClean}},\ and\ \bibinfo {author} {\bibfnamefont
  {T.~E.}\ \bibnamefont {O'Brien}},\ }\bibfield  {title} {\bibinfo {title}
  {{Quantum error mitigation}},\ }\href
  {https://doi.org/10.1103/RevModPhys.95.045005} {\bibfield  {journal}
  {\bibinfo  {journal} {Rev. Mod. Phys.}\ }\textbf {\bibinfo {volume} {95}},\
  \bibinfo {pages} {045005} (\bibinfo {year} {2023})}\BibitemShut {NoStop}%
\bibitem [{\citenamefont {Gul\'acsi}\ and\ \citenamefont
  {Burkard}(2023)}]{GulasciPRB2022}%
  \BibitemOpen
  \bibfield  {author} {\bibinfo {author} {\bibfnamefont {B.}~\bibnamefont
  {Gul\'acsi}}\ and\ \bibinfo {author} {\bibfnamefont {G.}~\bibnamefont
  {Burkard}},\ }\bibfield  {title} {\bibinfo {title} {{Signatures of
  non-Markovianity of a superconducting qubit}},\ }\href
  {https://doi.org/10.1103/PhysRevB.107.174511} {\bibfield  {journal} {\bibinfo
   {journal} {Phys. Rev. B}\ }\textbf {\bibinfo {volume} {107}},\ \bibinfo
  {pages} {174511} (\bibinfo {year} {2023})}\BibitemShut {NoStop}%
\bibitem [{\citenamefont {Papi\v{c}}\ \emph {et~al.}()\citenamefont
  {Papi\v{c}}, \citenamefont {Auer},\ and\ \citenamefont
  {de~Vega}}]{PapicARXIV2023}%
  \BibitemOpen
  \bibfield  {author} {\bibinfo {author} {\bibfnamefont {M.}~\bibnamefont
  {Papi\v{c}}}, \bibinfo {author} {\bibfnamefont {A.}~\bibnamefont {Auer}},\
  and\ \bibinfo {author} {\bibfnamefont {I.}~\bibnamefont {de~Vega}},\
  }\href@noop {} {\bibinfo {title} {{Fast Estimation of Physical Error
  Contributions of Quantum Gates}}},\ \Eprint
  {https://arxiv.org/abs/2305.08916} {arXiv:2305.08916 [quant-ph]} \BibitemShut
  {NoStop}%
\bibitem [{\citenamefont {Rist{\`e}}\ \emph {et~al.}(2013)\citenamefont
  {Rist{\`e}}, \citenamefont {Bultink}, \citenamefont {Tiggelman},
  \citenamefont {Schouten}, \citenamefont {Lehnert},\ and\ \citenamefont
  {DiCarlo}}]{RisteNATCOMMUN2013}%
  \BibitemOpen
  \bibfield  {author} {\bibinfo {author} {\bibfnamefont {D.}~\bibnamefont
  {Rist{\`e}}}, \bibinfo {author} {\bibfnamefont {C.~C.}\ \bibnamefont
  {Bultink}}, \bibinfo {author} {\bibfnamefont {M.~J.}\ \bibnamefont
  {Tiggelman}}, \bibinfo {author} {\bibfnamefont {R.~N.}\ \bibnamefont
  {Schouten}}, \bibinfo {author} {\bibfnamefont {K.~W.}\ \bibnamefont
  {Lehnert}},\ and\ \bibinfo {author} {\bibfnamefont {L.}~\bibnamefont
  {DiCarlo}},\ }\bibfield  {title} {\bibinfo {title} {{Millisecond
  charge-parity fluctuations and induced decoherence in a superconducting
  transmon qubit}},\ }\href {https://doi.org/10.1038/ncomms2936} {\bibfield
  {journal} {\bibinfo  {journal} {Nat. Commun.}\ }\textbf {\bibinfo {volume}
  {4}},\ \bibinfo {pages} {1913} (\bibinfo {year} {2013})}\BibitemShut
  {NoStop}%
\bibitem [{\citenamefont {Cardani}\ \emph {et~al.}(2021)\citenamefont
  {Cardani}, \citenamefont {Valenti}, \citenamefont {Casali}, \citenamefont
  {Catelani}, \citenamefont {Charpentier}, \citenamefont {Clemenza},
  \citenamefont {Colantoni}, \citenamefont {Cruciani}, \citenamefont
  {D'Imperio}, \citenamefont {Gironi} \emph {et~al.}}]{CardaniNATCOMMUN2021}%
  \BibitemOpen
  \bibfield  {author} {\bibinfo {author} {\bibfnamefont {L.}~\bibnamefont
  {Cardani}}, \bibinfo {author} {\bibfnamefont {F.}~\bibnamefont {Valenti}},
  \bibinfo {author} {\bibfnamefont {N.}~\bibnamefont {Casali}}, \bibinfo
  {author} {\bibfnamefont {G.}~\bibnamefont {Catelani}}, \bibinfo {author}
  {\bibfnamefont {T.}~\bibnamefont {Charpentier}}, \bibinfo {author}
  {\bibfnamefont {M.}~\bibnamefont {Clemenza}}, \bibinfo {author}
  {\bibfnamefont {I.}~\bibnamefont {Colantoni}}, \bibinfo {author}
  {\bibfnamefont {A.}~\bibnamefont {Cruciani}}, \bibinfo {author}
  {\bibfnamefont {G.}~\bibnamefont {D'Imperio}}, \bibinfo {author}
  {\bibfnamefont {L.}~\bibnamefont {Gironi}}, \emph {et~al.},\ }\bibfield
  {title} {\bibinfo {title} {{Reducing the impact of radioactivity on quantum
  circuits in a deep-underground facility}},\ }\href
  {https://doi.org/10.1038/s41467-021-23032-z} {\bibfield  {journal} {\bibinfo
  {journal} {Nat. Commun.}\ }\textbf {\bibinfo {volume} {12}},\ \bibinfo
  {pages} {2733} (\bibinfo {year} {2021})}\BibitemShut {NoStop}%
\bibitem [{\citenamefont {Pan}\ \emph {et~al.}(2022)\citenamefont {Pan},
  \citenamefont {Zhou}, \citenamefont {Yuan}, \citenamefont {Nie},
  \citenamefont {Wei}, \citenamefont {Zhang}, \citenamefont {Li}, \citenamefont
  {Liu}, \citenamefont {Jiang}, \citenamefont {Catelani} \emph
  {et~al.}}]{PanNATCOMMUN2022}%
  \BibitemOpen
  \bibfield  {author} {\bibinfo {author} {\bibfnamefont {X.}~\bibnamefont
  {Pan}}, \bibinfo {author} {\bibfnamefont {Y.}~\bibnamefont {Zhou}}, \bibinfo
  {author} {\bibfnamefont {H.}~\bibnamefont {Yuan}}, \bibinfo {author}
  {\bibfnamefont {L.}~\bibnamefont {Nie}}, \bibinfo {author} {\bibfnamefont
  {W.}~\bibnamefont {Wei}}, \bibinfo {author} {\bibfnamefont {L.}~\bibnamefont
  {Zhang}}, \bibinfo {author} {\bibfnamefont {J.}~\bibnamefont {Li}}, \bibinfo
  {author} {\bibfnamefont {S.}~\bibnamefont {Liu}}, \bibinfo {author}
  {\bibfnamefont {Z.~H.}\ \bibnamefont {Jiang}}, \bibinfo {author}
  {\bibfnamefont {G.}~\bibnamefont {Catelani}}, \emph {et~al.},\ }\bibfield
  {title} {\bibinfo {title} {{Engineering superconducting qubits to reduce
  quasiparticles and charge noise}},\ }\href
  {https://doi.org/10.1038/s41467-022-34727-2} {\bibfield  {journal} {\bibinfo
  {journal} {Nat. Commun.}\ }\textbf {\bibinfo {volume} {13}},\ \bibinfo
  {pages} {7196} (\bibinfo {year} {2022})}\BibitemShut {NoStop}%
\bibitem [{\citenamefont {Tuorila}\ \emph {et~al.}(2019)\citenamefont
  {Tuorila}, \citenamefont {Stockburger}, \citenamefont {Ala-Nissila},
  \citenamefont {Ankerhold},\ and\ \citenamefont
  {M\"ott\"onen}}]{TuorilaPRR2019}%
  \BibitemOpen
  \bibfield  {author} {\bibinfo {author} {\bibfnamefont {J.}~\bibnamefont
  {Tuorila}}, \bibinfo {author} {\bibfnamefont {J.}~\bibnamefont
  {Stockburger}}, \bibinfo {author} {\bibfnamefont {T.}~\bibnamefont
  {Ala-Nissila}}, \bibinfo {author} {\bibfnamefont {J.}~\bibnamefont
  {Ankerhold}},\ and\ \bibinfo {author} {\bibfnamefont {M.}~\bibnamefont
  {M\"ott\"onen}},\ }\bibfield  {title} {\bibinfo {title} {{System-environment
  correlations in qubit initialization and control}},\ }\href
  {https://doi.org/10.1103/PhysRevResearch.1.013004} {\bibfield  {journal}
  {\bibinfo  {journal} {Phys. Rev. Research}\ }\textbf {\bibinfo {volume}
  {1}},\ \bibinfo {pages} {013004} (\bibinfo {year} {2019})}\BibitemShut
  {NoStop}%
\bibitem [{\citenamefont {Babu}\ \emph {et~al.}(2021)\citenamefont {Babu},
  \citenamefont {Tuorila},\ and\ \citenamefont {Ala-Nissila}}]{BabuNPJQI2021}%
  \BibitemOpen
  \bibfield  {author} {\bibinfo {author} {\bibfnamefont {A.~P.}\ \bibnamefont
  {Babu}}, \bibinfo {author} {\bibfnamefont {J.}~\bibnamefont {Tuorila}},\ and\
  \bibinfo {author} {\bibfnamefont {T.}~\bibnamefont {Ala-Nissila}},\
  }\bibfield  {title} {\bibinfo {title} {{State leakage during fast decay and
  control of a superconducting transmon qubit}},\ }\href
  {https://doi.org/10.1038/s41534-020-00357-z} {\bibfield  {journal} {\bibinfo
  {journal} {npj Quantum Inf.}\ }\textbf {\bibinfo {volume} {7}},\ \bibinfo
  {pages} {30} (\bibinfo {year} {2021})}\BibitemShut {NoStop}%
\bibitem [{\citenamefont {Nakamura}\ and\ \citenamefont
  {Ankerhold}(2024)}]{NakamuraPRB2024}%
  \BibitemOpen
  \bibfield  {author} {\bibinfo {author} {\bibfnamefont {K.}~\bibnamefont
  {Nakamura}}\ and\ \bibinfo {author} {\bibfnamefont {J.}~\bibnamefont
  {Ankerhold}},\ }\bibfield  {title} {\bibinfo {title} {{Qubit dynamics beyond
  Lindblad: Non-Markovianity versus rotating wave approximation}},\ }\href
  {https://doi.org/10.1103/PhysRevB.109.014315} {\bibfield  {journal} {\bibinfo
   {journal} {Phys. Rev. B}\ }\textbf {\bibinfo {volume} {109}},\ \bibinfo
  {pages} {014315} (\bibinfo {year} {2024})}\BibitemShut {NoStop}%
\bibitem [{\citenamefont {Babu}\ \emph {et~al.}(2023)\citenamefont {Babu},
  \citenamefont {Orell}, \citenamefont {Vadimov}, \citenamefont {Teixeira},
  \citenamefont {M\"ott\"onen},\ and\ \citenamefont {Silveri}}]{BabuPRR2023}%
  \BibitemOpen
  \bibfield  {author} {\bibinfo {author} {\bibfnamefont {A.~P.}\ \bibnamefont
  {Babu}}, \bibinfo {author} {\bibfnamefont {T.}~\bibnamefont {Orell}},
  \bibinfo {author} {\bibfnamefont {V.}~\bibnamefont {Vadimov}}, \bibinfo
  {author} {\bibfnamefont {W.}~\bibnamefont {Teixeira}}, \bibinfo {author}
  {\bibfnamefont {M.}~\bibnamefont {M\"ott\"onen}},\ and\ \bibinfo {author}
  {\bibfnamefont {M.}~\bibnamefont {Silveri}},\ }\bibfield  {title} {\bibinfo
  {title} {{Quantum error correction under numerically exact
  open-quantum-system dynamics}},\ }\href
  {https://doi.org/10.1103/PhysRevResearch.5.043161} {\bibfield  {journal}
  {\bibinfo  {journal} {Phys. Rev. Research}\ }\textbf {\bibinfo {volume}
  {5}},\ \bibinfo {pages} {043161} (\bibinfo {year} {2023})}\BibitemShut
  {NoStop}%
\bibitem [{\citenamefont {Figueroa-Romero}\ \emph {et~al.}(2024)\citenamefont
  {Figueroa-Romero}, \citenamefont {Papi\v{c}}, \citenamefont {Auer},
  \citenamefont {Hsieh}, \citenamefont {Modi},\ and\ \citenamefont
  {de~Vega}}]{RomeroQST2024}%
  \BibitemOpen
  \bibfield  {author} {\bibinfo {author} {\bibfnamefont {P.}~\bibnamefont
  {Figueroa-Romero}}, \bibinfo {author} {\bibfnamefont {M.}~\bibnamefont
  {Papi\v{c}}}, \bibinfo {author} {\bibfnamefont {A.}~\bibnamefont {Auer}},
  \bibinfo {author} {\bibfnamefont {M.-H.}\ \bibnamefont {Hsieh}}, \bibinfo
  {author} {\bibfnamefont {K.}~\bibnamefont {Modi}},\ and\ \bibinfo {author}
  {\bibfnamefont {I.}~\bibnamefont {de~Vega}},\ }\bibfield  {title} {\bibinfo
  {title} {{Operational Markovianization in randomized benchmarking}},\ }\href
  {https://doi.org/10.1088/2058-9565/ad3f44} {\bibfield  {journal} {\bibinfo
  {journal} {Quantum Sci. Technol.}\ }\textbf {\bibinfo {volume} {9}},\
  \bibinfo {pages} {035020} (\bibinfo {year} {2024})}\BibitemShut {NoStop}%
\bibitem [{\citenamefont {Agarwal}\ \emph {et~al.}()\citenamefont {Agarwal},
  \citenamefont {Lindoy}, \citenamefont {Lall}, \citenamefont {Jamet},\ and\
  \citenamefont {Rungger}}]{AgarwalARXIV2023}%
  \BibitemOpen
  \bibfield  {author} {\bibinfo {author} {\bibfnamefont {A.}~\bibnamefont
  {Agarwal}}, \bibinfo {author} {\bibfnamefont {L.~P.}\ \bibnamefont {Lindoy}},
  \bibinfo {author} {\bibfnamefont {D.}~\bibnamefont {Lall}}, \bibinfo {author}
  {\bibfnamefont {F.}~\bibnamefont {Jamet}},\ and\ \bibinfo {author}
  {\bibfnamefont {I.}~\bibnamefont {Rungger}},\ }\href@noop {} {\bibinfo
  {title} {{Modelling non-Markovian noise in driven superconducting qubits}}},\
  \Eprint {https://arxiv.org/abs/2306.13021} {arXiv:2306.13021 [quant-ph]}
  \BibitemShut {NoStop}%
\bibitem [{\citenamefont {Lorenzo}\ \emph {et~al.}(2011)\citenamefont
  {Lorenzo}, \citenamefont {Plastina},\ and\ \citenamefont
  {Paternostro}}]{LorenzoPRA2011}%
  \BibitemOpen
  \bibfield  {author} {\bibinfo {author} {\bibfnamefont {S.}~\bibnamefont
  {Lorenzo}}, \bibinfo {author} {\bibfnamefont {F.}~\bibnamefont {Plastina}},\
  and\ \bibinfo {author} {\bibfnamefont {M.}~\bibnamefont {Paternostro}},\
  }\bibfield  {title} {\bibinfo {title} {{Role of environmental correlations in
  the non-Markovian dynamics of a spin system}},\ }\href
  {https://doi.org/10.1103/PhysRevA.84.032124} {\bibfield  {journal} {\bibinfo
  {journal} {Phys. Rev. A}\ }\textbf {\bibinfo {volume} {84}},\ \bibinfo
  {pages} {032124} (\bibinfo {year} {2011})}\BibitemShut {NoStop}%
\bibitem [{\citenamefont {Laine}\ \emph {et~al.}(2012)\citenamefont {Laine},
  \citenamefont {Breuer}, \citenamefont {Piilo}, \citenamefont {Li},\ and\
  \citenamefont {Guo}}]{LainePRL2012}%
  \BibitemOpen
  \bibfield  {author} {\bibinfo {author} {\bibfnamefont {E.-M.}\ \bibnamefont
  {Laine}}, \bibinfo {author} {\bibfnamefont {H.-P.}\ \bibnamefont {Breuer}},
  \bibinfo {author} {\bibfnamefont {J.}~\bibnamefont {Piilo}}, \bibinfo
  {author} {\bibfnamefont {C.-F.}\ \bibnamefont {Li}},\ and\ \bibinfo {author}
  {\bibfnamefont {G.-C.}\ \bibnamefont {Guo}},\ }\bibfield  {title} {\bibinfo
  {title} {{Nonlocal Memory Effects in the Dynamics of Open Quantum Systems}},\
  }\href {https://doi.org/10.1103/PhysRevLett.108.210402} {\bibfield  {journal}
  {\bibinfo  {journal} {Phys. Rev. Lett.}\ }\textbf {\bibinfo {volume} {108}},\
  \bibinfo {pages} {210402} (\bibinfo {year} {2012})}\BibitemShut {NoStop}%
\bibitem [{\citenamefont {Fanchini}\ \emph {et~al.}(2014)\citenamefont
  {Fanchini}, \citenamefont {Karpat}, \citenamefont {\c{C}akmak}, \citenamefont
  {Castelano}, \citenamefont {Aguilar}, \citenamefont {Far\'{\i}as},
  \citenamefont {Walborn}, \citenamefont {Ribeiro},\ and\ \citenamefont
  {de~Oliveira}}]{FanchiniPRL2014}%
  \BibitemOpen
  \bibfield  {author} {\bibinfo {author} {\bibfnamefont {F.~F.}\ \bibnamefont
  {Fanchini}}, \bibinfo {author} {\bibfnamefont {G.}~\bibnamefont {Karpat}},
  \bibinfo {author} {\bibfnamefont {B.}~\bibnamefont {\c{C}akmak}}, \bibinfo
  {author} {\bibfnamefont {L.~K.}\ \bibnamefont {Castelano}}, \bibinfo {author}
  {\bibfnamefont {G.~H.}\ \bibnamefont {Aguilar}}, \bibinfo {author}
  {\bibfnamefont {O.~J.}\ \bibnamefont {Far\'{\i}as}}, \bibinfo {author}
  {\bibfnamefont {S.~P.}\ \bibnamefont {Walborn}}, \bibinfo {author}
  {\bibfnamefont {P.~H.~S.}\ \bibnamefont {Ribeiro}},\ and\ \bibinfo {author}
  {\bibfnamefont {M.~C.}\ \bibnamefont {de~Oliveira}},\ }\bibfield  {title}
  {\bibinfo {title} {{Non-Markovianity through Accessible Information}},\
  }\href {https://doi.org/10.1103/PhysRevLett.112.210402} {\bibfield  {journal}
  {\bibinfo  {journal} {Phys. Rev. Lett.}\ }\textbf {\bibinfo {volume} {112}},\
  \bibinfo {pages} {210402} (\bibinfo {year} {2014})}\BibitemShut {NoStop}%
\bibitem [{\citenamefont {Siltanen}\ \emph {et~al.}(2021)\citenamefont
  {Siltanen}, \citenamefont {Kuusela},\ and\ \citenamefont
  {Piilo}}]{SiltanenPRA2021}%
  \BibitemOpen
  \bibfield  {author} {\bibinfo {author} {\bibfnamefont {O.}~\bibnamefont
  {Siltanen}}, \bibinfo {author} {\bibfnamefont {T.}~\bibnamefont {Kuusela}},\
  and\ \bibinfo {author} {\bibfnamefont {J.}~\bibnamefont {Piilo}},\ }\bibfield
   {title} {\bibinfo {title} {{Interferometric approach to open quantum systems
  and non-Markovian dynamics}},\ }\href
  {https://doi.org/10.1103/PhysRevA.103.032223} {\bibfield  {journal} {\bibinfo
   {journal} {Phys. Rev. A}\ }\textbf {\bibinfo {volume} {103}},\ \bibinfo
  {pages} {032223} (\bibinfo {year} {2021})}\BibitemShut {NoStop}%
\bibitem [{\citenamefont {Xu}\ \emph {et~al.}(2022)\citenamefont {Xu},
  \citenamefont {Yan}, \citenamefont {Shi}, \citenamefont {Ankerhold},\ and\
  \citenamefont {Stockburger}}]{XuPRL2022}%
  \BibitemOpen
  \bibfield  {author} {\bibinfo {author} {\bibfnamefont {M.}~\bibnamefont
  {Xu}}, \bibinfo {author} {\bibfnamefont {Y.}~\bibnamefont {Yan}}, \bibinfo
  {author} {\bibfnamefont {Q.}~\bibnamefont {Shi}}, \bibinfo {author}
  {\bibfnamefont {J.}~\bibnamefont {Ankerhold}},\ and\ \bibinfo {author}
  {\bibfnamefont {J.~T.}\ \bibnamefont {Stockburger}},\ }\bibfield  {title}
  {\bibinfo {title} {{Taming Quantum Noise for Efficient Low Temperature
  Simulations of Open Quantum Systems}},\ }\href
  {https://doi.org/10.1103/PhysRevLett.129.230601} {\bibfield  {journal}
  {\bibinfo  {journal} {Phys. Rev. Lett.}\ }\textbf {\bibinfo {volume} {129}},\
  \bibinfo {pages} {230601} (\bibinfo {year} {2022})}\BibitemShut {NoStop}%
\bibitem [{\citenamefont {Blais}\ \emph {et~al.}(2021)\citenamefont {Blais},
  \citenamefont {Grimsmo}, \citenamefont {Girvin},\ and\ \citenamefont
  {Wallraff}}]{BlaisRMP2021}%
  \BibitemOpen
  \bibfield  {author} {\bibinfo {author} {\bibfnamefont {A.}~\bibnamefont
  {Blais}}, \bibinfo {author} {\bibfnamefont {A.~L.}\ \bibnamefont {Grimsmo}},
  \bibinfo {author} {\bibfnamefont {S.~M.}\ \bibnamefont {Girvin}},\ and\
  \bibinfo {author} {\bibfnamefont {A.}~\bibnamefont {Wallraff}},\ }\bibfield
  {title} {\bibinfo {title} {{Circuit quantum electrodynamics}},\ }\href
  {https://doi.org/10.1103/RevModPhys.93.025005} {\bibfield  {journal}
  {\bibinfo  {journal} {Rev. Mod. Phys.}\ }\textbf {\bibinfo {volume} {93}},\
  \bibinfo {pages} {025005} (\bibinfo {year} {2021})}\BibitemShut {NoStop}%
\bibitem [{\citenamefont {Breuer}\ and\ \citenamefont
  {Petruccione}(2002)}]{Breuer2002}%
  \BibitemOpen
  \bibfield  {author} {\bibinfo {author} {\bibfnamefont {H.-P.}\ \bibnamefont
  {Breuer}}\ and\ \bibinfo {author} {\bibfnamefont {F.}~\bibnamefont
  {Petruccione}},\ }\href@noop {} {\emph {\bibinfo {title} {{The Theory of Open
  Quantum Systems}}}}\ (\bibinfo  {publisher} {Oxford University Press},\
  \bibinfo {address} {Oxford},\ \bibinfo {year} {2002})\BibitemShut {NoStop}%
\bibitem [{\citenamefont {Weiss}(2012)}]{Weiss2012}%
  \BibitemOpen
  \bibfield  {author} {\bibinfo {author} {\bibfnamefont {U.}~\bibnamefont
  {Weiss}},\ }\href@noop {} {\emph {\bibinfo {title} {{Quantum Dissipative
  Systems}}}},\ \bibinfo {edition} {4th}\ ed.\ (\bibinfo  {publisher} {World
  Scientific},\ \bibinfo {address} {Singapore},\ \bibinfo {year}
  {2012})\BibitemShut {NoStop}%
\bibitem [{\citenamefont {Feynman}\ and\ \citenamefont
  {Vernon~Jr.}(1963)}]{FeynmanANNPHYS1963}%
  \BibitemOpen
  \bibfield  {author} {\bibinfo {author} {\bibfnamefont {R.~P.}\ \bibnamefont
  {Feynman}}\ and\ \bibinfo {author} {\bibfnamefont {F.~L.}\ \bibnamefont
  {Vernon~Jr.}},\ }\bibfield  {title} {\bibinfo {title} {{The theory of a
  general quantum system interacting with a linear dissipative system}},\
  }\href {https://doi.org/https://doi.org/10.1016/0003-4916(63)90068-X}
  {\bibfield  {journal} {\bibinfo  {journal} {Ann. Phys.}\ }\textbf {\bibinfo
  {volume} {24}},\ \bibinfo {pages} {118} (\bibinfo {year} {1963})}\BibitemShut
  {NoStop}%
\bibitem [{\citenamefont {Makri}(1999)}]{MakriJPCB1999}%
  \BibitemOpen
  \bibfield  {author} {\bibinfo {author} {\bibfnamefont {N.}~\bibnamefont
  {Makri}},\ }\bibfield  {title} {\bibinfo {title} {{The Linear Response
  Approximation and Its Lowest Order Corrections: An Influence Functional
  Approach}},\ }\href {https://doi.org/10.1021/jp9847540} {\bibfield  {journal}
  {\bibinfo  {journal} {J. Phys. Chem. B}\ }\textbf {\bibinfo {volume} {103}},\
  \bibinfo {pages} {2823} (\bibinfo {year} {1999})}\BibitemShut {NoStop}%
\bibitem [{\citenamefont {Sza\'{n}kowski}\ \emph {et~al.}(2017)\citenamefont
  {Sza\'{n}kowski}, \citenamefont {Ramon}, \citenamefont {Krzywda},
  \citenamefont {Kwiatkowski},\ and\ \citenamefont
  {Cywi\'{n}ski}}]{SzankowskiJPHYS2017}%
  \BibitemOpen
  \bibfield  {author} {\bibinfo {author} {\bibfnamefont {P.}~\bibnamefont
  {Sza\'{n}kowski}}, \bibinfo {author} {\bibfnamefont {G.}~\bibnamefont
  {Ramon}}, \bibinfo {author} {\bibfnamefont {J.}~\bibnamefont {Krzywda}},
  \bibinfo {author} {\bibfnamefont {D.}~\bibnamefont {Kwiatkowski}},\ and\
  \bibinfo {author} {\bibfnamefont {{\L}.}~\bibnamefont {Cywi\'{n}ski}},\
  }\bibfield  {title} {\bibinfo {title} {{Environmental noise spectroscopy with
  qubits subjected to dynamical decoupling}},\ }\href
  {https://doi.org/10.1088/1361-648X/aa7648} {\bibfield  {journal} {\bibinfo
  {journal} {J. Phys.: Condens. Matter}\ }\textbf {\bibinfo {volume} {29}},\
  \bibinfo {pages} {333001} (\bibinfo {year} {2017})}\BibitemShut {NoStop}%
\bibitem [{\citenamefont {Veps{\"a}l{\"a}inen}\ \emph
  {et~al.}(2020)\citenamefont {Veps{\"a}l{\"a}inen}, \citenamefont {Karamlou},
  \citenamefont {Orrell}, \citenamefont {Dogra}, \citenamefont {Loer},
  \citenamefont {Vasconcelos}, \citenamefont {Kim}, \citenamefont {Melville},
  \citenamefont {Niedzielski}, \citenamefont {Yoder} \emph
  {et~al.}}]{VepsalainenNATURE2020}%
  \BibitemOpen
  \bibfield  {author} {\bibinfo {author} {\bibfnamefont {A.~P.}\ \bibnamefont
  {Veps{\"a}l{\"a}inen}}, \bibinfo {author} {\bibfnamefont {A.~H.}\
  \bibnamefont {Karamlou}}, \bibinfo {author} {\bibfnamefont {J.~L.}\
  \bibnamefont {Orrell}}, \bibinfo {author} {\bibfnamefont {A.~S.}\
  \bibnamefont {Dogra}}, \bibinfo {author} {\bibfnamefont {B.}~\bibnamefont
  {Loer}}, \bibinfo {author} {\bibfnamefont {F.}~\bibnamefont {Vasconcelos}},
  \bibinfo {author} {\bibfnamefont {D.~K.}\ \bibnamefont {Kim}}, \bibinfo
  {author} {\bibfnamefont {A.~J.}\ \bibnamefont {Melville}}, \bibinfo {author}
  {\bibfnamefont {B.~M.}\ \bibnamefont {Niedzielski}}, \bibinfo {author}
  {\bibfnamefont {J.~L.}\ \bibnamefont {Yoder}}, \emph {et~al.},\ }\bibfield
  {title} {\bibinfo {title} {{Impact of ionizing radiation on superconducting
  qubit coherence}},\ }\href {https://doi.org/10.1038/s41586-020-2619-8}
  {\bibfield  {journal} {\bibinfo  {journal} {{Nature (London)}}\ }\textbf
  {\bibinfo {volume} {584}},\ \bibinfo {pages} {551} (\bibinfo {year}
  {2020})}\BibitemShut {NoStop}%
\bibitem [{\citenamefont {Ithier}\ \emph {et~al.}(2005)\citenamefont {Ithier},
  \citenamefont {Collin}, \citenamefont {Joyez}, \citenamefont {Meeson},
  \citenamefont {Vion}, \citenamefont {Esteve}, \citenamefont {Chiarello},
  \citenamefont {Shnirman}, \citenamefont {Makhlin}, \citenamefont {Schriefl}
  \emph {et~al.}}]{IthierPRB2005}%
  \BibitemOpen
  \bibfield  {author} {\bibinfo {author} {\bibfnamefont {G.}~\bibnamefont
  {Ithier}}, \bibinfo {author} {\bibfnamefont {E.}~\bibnamefont {Collin}},
  \bibinfo {author} {\bibfnamefont {P.}~\bibnamefont {Joyez}}, \bibinfo
  {author} {\bibfnamefont {P.~J.}\ \bibnamefont {Meeson}}, \bibinfo {author}
  {\bibfnamefont {D.}~\bibnamefont {Vion}}, \bibinfo {author} {\bibfnamefont
  {D.}~\bibnamefont {Esteve}}, \bibinfo {author} {\bibfnamefont
  {F.}~\bibnamefont {Chiarello}}, \bibinfo {author} {\bibfnamefont
  {A.}~\bibnamefont {Shnirman}}, \bibinfo {author} {\bibfnamefont
  {Y.}~\bibnamefont {Makhlin}}, \bibinfo {author} {\bibfnamefont
  {J.}~\bibnamefont {Schriefl}}, \emph {et~al.},\ }\bibfield  {title} {\bibinfo
  {title} {{Decoherence in a superconducting quantum bit circuit}},\ }\href
  {https://doi.org/10.1103/PhysRevB.72.134519} {\bibfield  {journal} {\bibinfo
  {journal} {Phys. Rev. B}\ }\textbf {\bibinfo {volume} {72}},\ \bibinfo
  {pages} {134519} (\bibinfo {year} {2005})}\BibitemShut {NoStop}%
\bibitem [{\citenamefont {Tanimura}(2015)}]{Tanimura2015}%
  \BibitemOpen
  \bibfield  {author} {\bibinfo {author} {\bibfnamefont {Y.}~\bibnamefont
  {Tanimura}},\ }\bibfield  {title} {\bibinfo {title} {{Real-time and
  imaginary-time quantum hierarchal Fokker--Planck equations}},\ }\href
  {https://doi.org/10.1063/1.4916647} {\bibfield  {journal} {\bibinfo
  {journal} {J. Chem. Phys.}\ }\textbf {\bibinfo {volume} {142}},\ \bibinfo
  {pages} {144110} (\bibinfo {year} {2015})}\BibitemShut {NoStop}%
\bibitem [{\citenamefont {Tanimura}(2020)}]{tanimuraJCP20}%
  \BibitemOpen
  \bibfield  {author} {\bibinfo {author} {\bibfnamefont {Y.}~\bibnamefont
  {Tanimura}},\ }\bibfield  {title} {\bibinfo {title} {{Numerically ``exact''
  approach to open quantum dynamics: The hierarchical equations of motion
  ({HEOM})}},\ }\href {https://doi.org/10.1063/5.0011599} {\bibfield  {journal}
  {\bibinfo  {journal} {J. Chem. Phys.}\ }\textbf {\bibinfo {volume} {153}},\
  \bibinfo {pages} {020901} (\bibinfo {year} {2020})}\BibitemShut {NoStop}%
\bibitem [{\citenamefont {Xu}\ \emph {et~al.}()\citenamefont {Xu},
  \citenamefont {Vadimov}, \citenamefont {Krug}, \citenamefont {Stockburger},\
  and\ \citenamefont {Ankerhold}}]{Xu2023}%
  \BibitemOpen
  \bibfield  {author} {\bibinfo {author} {\bibfnamefont {M.}~\bibnamefont
  {Xu}}, \bibinfo {author} {\bibfnamefont {V.}~\bibnamefont {Vadimov}},
  \bibinfo {author} {\bibfnamefont {M.}~\bibnamefont {Krug}}, \bibinfo {author}
  {\bibfnamefont {J.~T.}\ \bibnamefont {Stockburger}},\ and\ \bibinfo {author}
  {\bibfnamefont {J.}~\bibnamefont {Ankerhold}},\ }\href@noop {} {\bibinfo
  {title} {{A Universal Framework for Quantum Dissipation:Minimally Extended
  State Space and Exact Time-Local Dynamics}}},\ \Eprint
  {https://arxiv.org/abs/2307.16790} {arXiv:2307.16790 [quant-ph]} \BibitemShut
  {NoStop}%
\bibitem [{\citenamefont {Bylander}\ \emph {et~al.}(2011)\citenamefont
  {Bylander}, \citenamefont {Gustavsson}, \citenamefont {Yan}, \citenamefont
  {Yoshihara}, \citenamefont {Harrabi}, \citenamefont {Fitch}, \citenamefont
  {Cory}, \citenamefont {Nakamura}, \citenamefont {Tsai},\ and\ \citenamefont
  {Oliver}}]{BylanderNP2011}%
  \BibitemOpen
  \bibfield  {author} {\bibinfo {author} {\bibfnamefont {J.}~\bibnamefont
  {Bylander}}, \bibinfo {author} {\bibfnamefont {S.}~\bibnamefont
  {Gustavsson}}, \bibinfo {author} {\bibfnamefont {F.}~\bibnamefont {Yan}},
  \bibinfo {author} {\bibfnamefont {F.}~\bibnamefont {Yoshihara}}, \bibinfo
  {author} {\bibfnamefont {K.}~\bibnamefont {Harrabi}}, \bibinfo {author}
  {\bibfnamefont {G.}~\bibnamefont {Fitch}}, \bibinfo {author} {\bibfnamefont
  {D.~G.}\ \bibnamefont {Cory}}, \bibinfo {author} {\bibfnamefont
  {Y.}~\bibnamefont {Nakamura}}, \bibinfo {author} {\bibfnamefont {J.-S.}\
  \bibnamefont {Tsai}},\ and\ \bibinfo {author} {\bibfnamefont {W.~D.}\
  \bibnamefont {Oliver}},\ }\bibfield  {title} {\bibinfo {title} {{Noise
  spectroscopy through dynamical decoupling with a superconducting flux
  qubit}},\ }\href {https://doi.org/10.1038/nphys1994} {\bibfield  {journal}
  {\bibinfo  {journal} {Nat. Phys.}\ }\textbf {\bibinfo {volume} {7}},\
  \bibinfo {pages} {565} (\bibinfo {year} {2011})}\BibitemShut {NoStop}%
\bibitem [{\citenamefont {Barone}\ and\ \citenamefont
  {Patern\`{o}}(1982)}]{Barone1982}%
  \BibitemOpen
  \bibfield  {author} {\bibinfo {author} {\bibfnamefont {A.}~\bibnamefont
  {Barone}}\ and\ \bibinfo {author} {\bibfnamefont {G.}~\bibnamefont
  {Patern\`{o}}},\ }\href@noop {} {\emph {\bibinfo {title} {{Physics and
  Applications of the Josephson Effect}}}}\ (\bibinfo  {publisher} {John Wiley
  \& Sons},\ \bibinfo {address} {New York},\ \bibinfo {year}
  {1982})\BibitemShut {NoStop}%
\bibitem [{\citenamefont {Wendin}\ and\ \citenamefont
  {Shumeiko}()}]{WendinARXIV2005}%
  \BibitemOpen
  \bibfield  {author} {\bibinfo {author} {\bibfnamefont {G.}~\bibnamefont
  {Wendin}}\ and\ \bibinfo {author} {\bibfnamefont {V.~S.}\ \bibnamefont
  {Shumeiko}},\ }\href@noop {} {\bibinfo {title} {{Superconducting Quantum
  Circuits, Qubits and Computing}}},\ \Eprint
  {https://arxiv.org/abs/cond-mat/0508729} {arXiv:cond-mat/0508729
  [cond-mat.supr-con]} \BibitemShut {NoStop}%
\bibitem [{\citenamefont {Machlup}(1954)}]{MachlupJAP1954}%
  \BibitemOpen
  \bibfield  {author} {\bibinfo {author} {\bibfnamefont {S.}~\bibnamefont
  {Machlup}},\ }\bibfield  {title} {\bibinfo {title} {{Noise in Semiconductors:
  Spectrum of a Two‐Parameter Random Signal}},\ }\href
  {https://doi.org/10.1063/1.1721637} {\bibfield  {journal} {\bibinfo
  {journal} {J. Appl. Phys.}\ }\textbf {\bibinfo {volume} {25}},\ \bibinfo
  {pages} {341} (\bibinfo {year} {1954})}\BibitemShut {NoStop}%
\bibitem [{\citenamefont {Paladino}\ \emph {et~al.}(2014)\citenamefont
  {Paladino}, \citenamefont {Galperin}, \citenamefont {Falci},\ and\
  \citenamefont {Altshuler}}]{PaladinoRMP2014}%
  \BibitemOpen
  \bibfield  {author} {\bibinfo {author} {\bibfnamefont {E.}~\bibnamefont
  {Paladino}}, \bibinfo {author} {\bibfnamefont {Y.~M.}\ \bibnamefont
  {Galperin}}, \bibinfo {author} {\bibfnamefont {G.}~\bibnamefont {Falci}},\
  and\ \bibinfo {author} {\bibfnamefont {B.~L.}\ \bibnamefont {Altshuler}},\
  }\bibfield  {title} {\bibinfo {title} {{$1/f$ noise: Implications for
  solid-state quantum information}},\ }\href
  {https://doi.org/10.1103/RevModPhys.86.361} {\bibfield  {journal} {\bibinfo
  {journal} {Rev. Mod. Phys.}\ }\textbf {\bibinfo {volume} {86}},\ \bibinfo
  {pages} {361} (\bibinfo {year} {2014})}\BibitemShut {NoStop}%
\bibitem [{\citenamefont {M\"uller}\ \emph {et~al.}(2019)\citenamefont
  {M\"uller}, \citenamefont {Cole},\ and\ \citenamefont
  {Lisenfeld}}]{MullerRPP2019}%
  \BibitemOpen
  \bibfield  {author} {\bibinfo {author} {\bibfnamefont {C.}~\bibnamefont
  {M\"uller}}, \bibinfo {author} {\bibfnamefont {J.~H.}\ \bibnamefont {Cole}},\
  and\ \bibinfo {author} {\bibfnamefont {J.}~\bibnamefont {Lisenfeld}},\
  }\bibfield  {title} {\bibinfo {title} {{Towards understanding
  two-level-systems in amorphous solids: insights from quantum circuits}},\
  }\href {https://doi.org/10.1088/1361-6633/ab3a7e} {\bibfield  {journal}
  {\bibinfo  {journal} {Rep. Prog. Phys.}\ }\textbf {\bibinfo {volume} {82}},\
  \bibinfo {pages} {124501} (\bibinfo {year} {2019})}\BibitemShut {NoStop}%
\bibitem [{\citenamefont {Glazman}\ and\ \citenamefont
  {Catelani}(2021)}]{GlazmanSPPLN2021}%
  \BibitemOpen
  \bibfield  {author} {\bibinfo {author} {\bibfnamefont {L.~I.}\ \bibnamefont
  {Glazman}}\ and\ \bibinfo {author} {\bibfnamefont {G.}~\bibnamefont
  {Catelani}},\ }\bibfield  {title} {\bibinfo {title} {{Bogoliubov
  quasiparticles in superconducting qubits}},\ }\bibfield  {journal} {\bibinfo
  {journal} {SciPost Phys. Lect. Notes}\ }\textbf {\bibinfo {volume} {31}},\
  \href {https://doi.org/10.21468/SciPostPhysLectNotes.31}
  {10.21468/SciPostPhysLectNotes.31} (\bibinfo {year} {2021})\BibitemShut
  {NoStop}%
\bibitem [{\citenamefont {Eroms}\ \emph {et~al.}(2006)\citenamefont {Eroms},
  \citenamefont {van Schaarenburg}, \citenamefont {Driessen}, \citenamefont
  {Plantenberg}, \citenamefont {Huizinga}, \citenamefont {Schouten},
  \citenamefont {Verbruggen}, \citenamefont {Harmans},\ and\ \citenamefont
  {Mooij}}]{EromsAPL2006}%
  \BibitemOpen
  \bibfield  {author} {\bibinfo {author} {\bibfnamefont {J.}~\bibnamefont
  {Eroms}}, \bibinfo {author} {\bibfnamefont {L.~C.}\ \bibnamefont {van
  Schaarenburg}}, \bibinfo {author} {\bibfnamefont {E.~F.~C.}\ \bibnamefont
  {Driessen}}, \bibinfo {author} {\bibfnamefont {J.~H.}\ \bibnamefont
  {Plantenberg}}, \bibinfo {author} {\bibfnamefont {C.~M.}\ \bibnamefont
  {Huizinga}}, \bibinfo {author} {\bibfnamefont {R.~N.}\ \bibnamefont
  {Schouten}}, \bibinfo {author} {\bibfnamefont {A.~H.}\ \bibnamefont
  {Verbruggen}}, \bibinfo {author} {\bibfnamefont {C.~J. P.~M.}\ \bibnamefont
  {Harmans}},\ and\ \bibinfo {author} {\bibfnamefont {J.~E.}\ \bibnamefont
  {Mooij}},\ }\bibfield  {title} {\bibinfo {title} {{Low-frequency noise in
  Josephson junctions for superconducting qubits}},\ }\href
  {https://doi.org/10.1063/1.2357010} {\bibfield  {journal} {\bibinfo
  {journal} {Appl. Phys. Lett.}\ }\textbf {\bibinfo {volume} {89}},\ \bibinfo
  {pages} {122516} (\bibinfo {year} {2006})}\BibitemShut {NoStop}%
\bibitem [{\citenamefont {Nugroho}\ \emph {et~al.}(2013)\citenamefont
  {Nugroho}, \citenamefont {Orlyanchik},\ and\ \citenamefont
  {Van~Harlingen}}]{NugrohoAPL2013}%
  \BibitemOpen
  \bibfield  {author} {\bibinfo {author} {\bibfnamefont {C.~D.}\ \bibnamefont
  {Nugroho}}, \bibinfo {author} {\bibfnamefont {V.}~\bibnamefont
  {Orlyanchik}},\ and\ \bibinfo {author} {\bibfnamefont {D.~J.}\ \bibnamefont
  {Van~Harlingen}},\ }\bibfield  {title} {\bibinfo {title} {{Low frequency
  resistance and critical current fluctuations in Al-based Josephson
  junctions}},\ }\href {https://doi.org/10.1063/1.4801521} {\bibfield
  {journal} {\bibinfo  {journal} {Appl. Phys. Lett.}\ }\textbf {\bibinfo
  {volume} {102}},\ \bibinfo {pages} {142602} (\bibinfo {year}
  {2013})}\BibitemShut {NoStop}%
\bibitem [{\citenamefont {Kenyon}\ \emph {et~al.}(2000)\citenamefont {Kenyon},
  \citenamefont {Lobb},\ and\ \citenamefont {Wellstood}}]{KenyonJAP2000}%
  \BibitemOpen
  \bibfield  {author} {\bibinfo {author} {\bibfnamefont {M.}~\bibnamefont
  {Kenyon}}, \bibinfo {author} {\bibfnamefont {C.~J.}\ \bibnamefont {Lobb}},\
  and\ \bibinfo {author} {\bibfnamefont {F.~C.}\ \bibnamefont {Wellstood}},\
  }\bibfield  {title} {\bibinfo {title} {{Temperature dependence of
  low-frequency noise in Al–Al$_2$O$_3$–Al single-electron transistors}},\
  }\href {https://doi.org/10.1063/1.1312846} {\bibfield  {journal} {\bibinfo
  {journal} {J. Appl. Phys.}\ }\textbf {\bibinfo {volume} {88}},\ \bibinfo
  {pages} {6536} (\bibinfo {year} {2000})}\BibitemShut {NoStop}%
\bibitem [{\citenamefont {Wellstood}\ \emph {et~al.}(2004)\citenamefont
  {Wellstood}, \citenamefont {Urbina},\ and\ \citenamefont
  {Clarke}}]{WellstoodAPL2004}%
  \BibitemOpen
  \bibfield  {author} {\bibinfo {author} {\bibfnamefont {F.~C.}\ \bibnamefont
  {Wellstood}}, \bibinfo {author} {\bibfnamefont {C.}~\bibnamefont {Urbina}},\
  and\ \bibinfo {author} {\bibfnamefont {J.}~\bibnamefont {Clarke}},\
  }\bibfield  {title} {\bibinfo {title} {{Flicker ($1/f$) noise in the critical
  current of Josephson junctions at $0.09$--$4.2$K}},\ }\href
  {https://doi.org/10.1063/1.1826236} {\bibfield  {journal} {\bibinfo
  {journal} {Appl. Phys. Lett.}\ }\textbf {\bibinfo {volume} {85}},\ \bibinfo
  {pages} {5296} (\bibinfo {year} {2004})}\BibitemShut {NoStop}%
\bibitem [{\citenamefont {Tanimura}(2014)}]{Tanimura2014}%
  \BibitemOpen
  \bibfield  {author} {\bibinfo {author} {\bibfnamefont {Y.}~\bibnamefont
  {Tanimura}},\ }\bibfield  {title} {\bibinfo {title} {{Reduced hierarchical
  equations of motion in real and imaginary time: Correlated initial states and
  thermodynamic quantities}},\ }\href {https://doi.org/10.1063/1.4890441}
  {\bibfield  {journal} {\bibinfo  {journal} {J. Chem. Phys.}\ }\textbf
  {\bibinfo {volume} {141}},\ \bibinfo {pages} {044114} (\bibinfo {year}
  {2014})}\BibitemShut {NoStop}%
\bibitem [{\citenamefont {McKay}\ \emph {et~al.}(2017)\citenamefont {McKay},
  \citenamefont {Wood}, \citenamefont {Sheldon}, \citenamefont {Chow},\ and\
  \citenamefont {Gambetta}}]{McKayPRA2017}%
  \BibitemOpen
  \bibfield  {author} {\bibinfo {author} {\bibfnamefont {D.~C.}\ \bibnamefont
  {McKay}}, \bibinfo {author} {\bibfnamefont {C.~J.}\ \bibnamefont {Wood}},
  \bibinfo {author} {\bibfnamefont {S.}~\bibnamefont {Sheldon}}, \bibinfo
  {author} {\bibfnamefont {J.~M.}\ \bibnamefont {Chow}},\ and\ \bibinfo
  {author} {\bibfnamefont {J.~M.}\ \bibnamefont {Gambetta}},\ }\bibfield
  {title} {\bibinfo {title} {{Efficient $Z$ gates for quantum computing}},\
  }\href {https://doi.org/10.1103/PhysRevA.96.022330} {\bibfield  {journal}
  {\bibinfo  {journal} {Phys. Rev. A}\ }\textbf {\bibinfo {volume} {96}},\
  \bibinfo {pages} {022330} (\bibinfo {year} {2017})}\BibitemShut {NoStop}%
\bibitem [{\citenamefont {Biercuk}\ \emph {et~al.}(2009)\citenamefont
  {Biercuk}, \citenamefont {Uys}, \citenamefont {VanDevender}, \citenamefont
  {Shiga}, \citenamefont {Itano},\ and\ \citenamefont
  {Bollinger}}]{BiercukNATURE2009}%
  \BibitemOpen
  \bibfield  {author} {\bibinfo {author} {\bibfnamefont {M.~J.}\ \bibnamefont
  {Biercuk}}, \bibinfo {author} {\bibfnamefont {H.}~\bibnamefont {Uys}},
  \bibinfo {author} {\bibfnamefont {A.~P.}\ \bibnamefont {VanDevender}},
  \bibinfo {author} {\bibfnamefont {N.}~\bibnamefont {Shiga}}, \bibinfo
  {author} {\bibfnamefont {W.~M.}\ \bibnamefont {Itano}},\ and\ \bibinfo
  {author} {\bibfnamefont {J.~J.}\ \bibnamefont {Bollinger}},\ }\bibfield
  {title} {\bibinfo {title} {{Optimized dynamical decoupling in a model quantum
  memory}},\ }\href {https://doi.org/10.1038/nature07951} {\bibfield  {journal}
  {\bibinfo  {journal} {{Nature (London)}}\ }\textbf {\bibinfo {volume}
  {458}},\ \bibinfo {pages} {996} (\bibinfo {year} {2009})}\BibitemShut
  {NoStop}%
\bibitem [{\citenamefont {Gaikwad}\ \emph {et~al.}(2024)\citenamefont
  {Gaikwad}, \citenamefont {Kowsari}, \citenamefont {Brame}, \citenamefont
  {Song}, \citenamefont {Zhang}, \citenamefont {Esposito}, \citenamefont
  {Ranadive}, \citenamefont {Cappelli}, \citenamefont {Roch}, \citenamefont
  {Levenson-Falk} \emph {et~al.}}]{GaikwadPRL2024}%
  \BibitemOpen
  \bibfield  {author} {\bibinfo {author} {\bibfnamefont {C.}~\bibnamefont
  {Gaikwad}}, \bibinfo {author} {\bibfnamefont {D.}~\bibnamefont {Kowsari}},
  \bibinfo {author} {\bibfnamefont {C.}~\bibnamefont {Brame}}, \bibinfo
  {author} {\bibfnamefont {X.}~\bibnamefont {Song}}, \bibinfo {author}
  {\bibfnamefont {H.}~\bibnamefont {Zhang}}, \bibinfo {author} {\bibfnamefont
  {M.}~\bibnamefont {Esposito}}, \bibinfo {author} {\bibfnamefont
  {A.}~\bibnamefont {Ranadive}}, \bibinfo {author} {\bibfnamefont
  {G.}~\bibnamefont {Cappelli}}, \bibinfo {author} {\bibfnamefont
  {N.}~\bibnamefont {Roch}}, \bibinfo {author} {\bibfnamefont {E.~M.}\
  \bibnamefont {Levenson-Falk}}, \emph {et~al.},\ }\bibfield  {title} {\bibinfo
  {title} {{Entanglement Assisted Probe of the Non-Markovian to Markovian
  Transition in Open Quantum System Dynamics}},\ }\href
  {https://doi.org/10.1103/PhysRevLett.132.200401} {\bibfield  {journal}
  {\bibinfo  {journal} {Phys. Rev. Lett.}\ }\textbf {\bibinfo {volume} {132}},\
  \bibinfo {pages} {200401} (\bibinfo {year} {2024})}\BibitemShut {NoStop}%
\bibitem [{\citenamefont {Alicki}\ and\ \citenamefont
  {Lendi}(1987)}]{Alicki1987}%
  \BibitemOpen
  \bibfield  {author} {\bibinfo {author} {\bibfnamefont {R.}~\bibnamefont
  {Alicki}}\ and\ \bibinfo {author} {\bibfnamefont {K.}~\bibnamefont {Lendi}},\
  }\href@noop {} {\emph {\bibinfo {title} {{Quantum Dynamical Semigroups and
  Applications}}}}\ (\bibinfo  {publisher} {Springer-Verlag},\ \bibinfo
  {address} {Berlin},\ \bibinfo {year} {1987})\BibitemShut {NoStop}%
\bibitem [{\citenamefont {Rivas}\ and\ \citenamefont
  {Huelga}(2012)}]{Rivas2012}%
  \BibitemOpen
  \bibfield  {author} {\bibinfo {author} {\bibfnamefont {{\'A}.}~\bibnamefont
  {Rivas}}\ and\ \bibinfo {author} {\bibfnamefont {S.~F.}\ \bibnamefont
  {Huelga}},\ }\href@noop {} {\emph {\bibinfo {title} {{Open Quantum Systems:
  An Introduction}}}}\ (\bibinfo  {publisher} {Springer},\ \bibinfo {address}
  {Heidelberg},\ \bibinfo {year} {2012})\BibitemShut {NoStop}%
\bibitem [{\citenamefont {Motzoi}\ \emph {et~al.}(2009)\citenamefont {Motzoi},
  \citenamefont {Gambetta}, \citenamefont {Rebentrost},\ and\ \citenamefont
  {Wilhelm}}]{MotzoiPRL2009}%
  \BibitemOpen
  \bibfield  {author} {\bibinfo {author} {\bibfnamefont {F.}~\bibnamefont
  {Motzoi}}, \bibinfo {author} {\bibfnamefont {J.~M.}\ \bibnamefont
  {Gambetta}}, \bibinfo {author} {\bibfnamefont {P.}~\bibnamefont
  {Rebentrost}},\ and\ \bibinfo {author} {\bibfnamefont {F.~K.}\ \bibnamefont
  {Wilhelm}},\ }\bibfield  {title} {\bibinfo {title} {{Simple Pulses for
  Elimination of Leakage in Weakly Nonlinear Qubits}},\ }\href
  {https://doi.org/10.1103/PhysRevLett.103.110501} {\bibfield  {journal}
  {\bibinfo  {journal} {Phys. Rev. Lett.}\ }\textbf {\bibinfo {volume} {103}},\
  \bibinfo {pages} {110501} (\bibinfo {year} {2009})}\BibitemShut {NoStop}%
\bibitem [{\citenamefont {Radcliffe}(1971)}]{SpinCoherent}%
  \BibitemOpen
  \bibfield  {author} {\bibinfo {author} {\bibfnamefont {J.~M.}\ \bibnamefont
  {Radcliffe}},\ }\bibfield  {title} {\bibinfo {title} {{Some properties of
  coherent spin states}},\ }\href {https://doi.org/10.1088/0305-4470/4/3/009}
  {\bibfield  {journal} {\bibinfo  {journal} {J. Phys. A}\ }\textbf {\bibinfo
  {volume} {4}},\ \bibinfo {pages} {313} (\bibinfo {year} {1971})}\BibitemShut
  {NoStop}%
\bibitem [{\citenamefont {Nakamura}\ and\ \citenamefont
  {Tanimura}(2018)}]{Nakamura18PRA}%
  \BibitemOpen
  \bibfield  {author} {\bibinfo {author} {\bibfnamefont {K.}~\bibnamefont
  {Nakamura}}\ and\ \bibinfo {author} {\bibfnamefont {Y.}~\bibnamefont
  {Tanimura}},\ }\bibfield  {title} {\bibinfo {title} {{Hierarchical
  Schr\"odinger equations of motion for open quantum dynamics}},\ }\href
  {https://doi.org/10.1103/PhysRevA.98.012109} {\bibfield  {journal} {\bibinfo
  {journal} {Phys. Rev. A}\ }\textbf {\bibinfo {volume} {98}},\ \bibinfo
  {pages} {012109} (\bibinfo {year} {2018})}\BibitemShut {NoStop}%
\bibitem [{\citenamefont {Ikeda}\ and\ \citenamefont
  {Scholes}(2020)}]{IkedaJCP2020}%
  \BibitemOpen
  \bibfield  {author} {\bibinfo {author} {\bibfnamefont {T.}~\bibnamefont
  {Ikeda}}\ and\ \bibinfo {author} {\bibfnamefont {G.~D.}\ \bibnamefont
  {Scholes}},\ }\bibfield  {title} {\bibinfo {title} {{Generalization of the
  hierarchical equations of motion theory for efficient calculations with
  arbitrary correlation functions}},\ }\href
  {https://doi.org/10.1063/5.0007327} {\bibfield  {journal} {\bibinfo
  {journal} {J. Chem. Phys.}\ }\textbf {\bibinfo {volume} {152}},\ \bibinfo
  {pages} {204101} (\bibinfo {year} {2020})}\BibitemShut {NoStop}%
\bibitem [{\citenamefont {Liniov}\ \emph {et~al.}(2019)\citenamefont {Liniov},
  \citenamefont {Meyerov}, \citenamefont {Kozinov}, \citenamefont {Volokitin},
  \citenamefont {Yusipov}, \citenamefont {Ivanchenko},\ and\ \citenamefont
  {Denisov}}]{LiniovPRE2019}%
  \BibitemOpen
  \bibfield  {author} {\bibinfo {author} {\bibfnamefont {A.}~\bibnamefont
  {Liniov}}, \bibinfo {author} {\bibfnamefont {I.}~\bibnamefont {Meyerov}},
  \bibinfo {author} {\bibfnamefont {E.}~\bibnamefont {Kozinov}}, \bibinfo
  {author} {\bibfnamefont {V.}~\bibnamefont {Volokitin}}, \bibinfo {author}
  {\bibfnamefont {I.}~\bibnamefont {Yusipov}}, \bibinfo {author} {\bibfnamefont
  {M.}~\bibnamefont {Ivanchenko}},\ and\ \bibinfo {author} {\bibfnamefont
  {S.}~\bibnamefont {Denisov}},\ }\bibfield  {title} {\bibinfo {title}
  {{Unfolding a quantum master equation into a system of real-valued equations:
  Computationally effective expansion over the basis of $\mathrm{SU}(N)$
  generators}},\ }\href {https://doi.org/10.1103/PhysRevE.100.053305}
  {\bibfield  {journal} {\bibinfo  {journal} {Phys. Rev. E}\ }\textbf {\bibinfo
  {volume} {100}},\ \bibinfo {pages} {053305} (\bibinfo {year}
  {2019})}\BibitemShut {NoStop}%
\bibitem [{\citenamefont {Braun}\ \emph {et~al.}(2001)\citenamefont {Braun},
  \citenamefont {Haake},\ and\ \citenamefont {Strunz}}]{BraunPRL2001}%
  \BibitemOpen
  \bibfield  {author} {\bibinfo {author} {\bibfnamefont {D.}~\bibnamefont
  {Braun}}, \bibinfo {author} {\bibfnamefont {F.}~\bibnamefont {Haake}},\ and\
  \bibinfo {author} {\bibfnamefont {W.~T.}\ \bibnamefont {Strunz}},\ }\bibfield
   {title} {\bibinfo {title} {{Universality of Decoherence}},\ }\href
  {https://doi.org/10.1103/PhysRevLett.86.2913} {\bibfield  {journal} {\bibinfo
   {journal} {Phys. Rev. Lett.}\ }\textbf {\bibinfo {volume} {86}},\ \bibinfo
  {pages} {2913} (\bibinfo {year} {2001})}\BibitemShut {NoStop}%
\end{thebibliography}%
\end{document}